\documentclass[aps,twocolumn,prc,superscriptaddress,showpacs,nofootinbib,floatfix,amssymb,amsfonts,amsmath]{revtex4-1}

\usepackage{graphicx}
\usepackage{dcolumn}
\usepackage{bm}

\usepackage{amsmath}    
\usepackage{amsfonts}   
\usepackage{amssymb}
\usepackage{graphicx}   
\usepackage{longtable}

\begin{document}
\title{Octupole deformation in the ground states of even-even nuclei:
a global analysis within the covariant density functional theory.}

\author{S.\ E.\ Agbemava}
\affiliation{Department of Physics and Astronomy, Mississippi
State University, MS 39762}

\author{A.\ V.\ Afanasjev}
\affiliation{Department of Physics and Astronomy, Mississippi
State University, MS 39762}

\author{P. Ring}
\affiliation{Fakult\"at f\"ur Physik, Technische Universit\"at M\"unchen,
 D-85748 Garching, Germany}

\date{\today}

\begin{abstract}
A systematic investigation of octupole deformed nuclei is presented
for even-even systems with $Z\leq 106$ located between the two-proton and 
two-neutron drip lines.
For this study we use five most up-to-date covariant energy
density functionals of different types, with a non-linear meson coupling,
with density dependent meson couplings, and with density-dependent zero-range
interactions. Pairing correlations are treated within relativistic
Hartree-Bogoliubov (RHB) theory based on an effective separable particle-particle
interaction of finite range. This allows us to assess theoretical uncertainties
within the present covariant models for the prediction of physical observables
relevant for octupole deformed nuclei.
In addition, a detailed comparison with the predictions of non-relativistic
models is performed. A new region of octupole deformation, centered around
$Z\sim 98, N\sim  196$ is predicted for the first time. In terms of its size in
the $(Z,N)$ plane and the impact of octupole deformation on binding
energies this region is similar to the best known region of octupole deformed nuclei
centered at $Z\sim 90, N\sim 136$. For the later island of octupole deformed
nuclei, the calculations suggest substantial increase of its size as compared
with available experimental data.
\end{abstract}


\maketitle

\section{Introduction}

   Reflection asymmetric (or octupole deformed) shapes represent an
interesting example of symmetry breaking of the nuclear mean
field. The physics of such shapes in the normal deformed minimum (both
in non-rotating and rotating systems) has been been extensively studied
in the 80ies and 90ies of the last century (see the review in Ref.\
\cite{BN.96}). Reflection asymmetric shapes are also present for
large deformations at the outer fission barriers in the actinides, superheavy
nuclei and nuclei important in the r-process of nucleosynthesis
\cite{BN.96,AAR.12,ELLMR.12}. At present, there is a revival of
the interest to the study of such shapes. It is seen in a substantial
number of theoretical
\cite{WMZ.10,GPZ.10,RB.11,MDSSL.12,JBJ.12,WLX.12,RR.12,ZLEZ.12,NVL.13,NVNL.14,CGCT.15,NRR.15,WYLX.15,YZL.15}
and experimental
\cite{Ba145.12,Ba147.13,220Rn-224Ra.13,Ra221.13,Pu240.13,BE3-A224.14,Ce147.15,Pa229.15,Nd147.15}
studies of octupole correlations and octupole deformed nuclei in the normal deformed
minimum. Moreover, the attempts to understand microscopically the fission
process, cluster radioactivity and the stability of superheavy elements
\cite{AAR.10,AAR.12,PNLV.12,LZZ.14,WE.12,DGD.08,WR.11,RR.14,SBN.13,SMBDNS.13,MSI.09,KJS.10}
as well as renewed interest to experimental studies of fission
\cite{And.10,238-U-HD,AHD.13}  created a
substantial interest in octupole deformed shapes at large deformations.

  The existence of octupole deformed shapes is dictated by the underlying
shell structure. Strong octupole coupling exists for particle numbers
associated
with a large $\Delta N=1$ interaction between intruder orbitals
with $(l,j)$ and normal-parity orbitals with $(l-3,j-3)$ \cite{BN.96}.
For normal deformed nuclei  not far away from beta stability the tendency
towards octupole deformation or strong octupole correlations occurs just
above closed shells at particle numbers near 34  (the coupling between the
1$g_{9/2}$ and 2$p_{3/2}$ orbitals), 56 (the coupling  between the 1$h_{11/2}$ and 2$d_{5/2}$
orbitals), 88 (the coupling  between the 1$i_{13/2}$ and 2$f_{7/2}$ orbitals)
and 134 (the coupling  between the 1$j_{15/2}$ and 2$g_{9/2}$ orbitals) \cite{BN.96}.

 Some of the studies of the octupole shapes have been performed in the
framework of covariant density functional theory (CDFT)~\cite{VALR.05}.
Built on Lorentz covariance and the Dirac equation, CDFT provides a
natural incorporation of spin degrees of freedom \cite{Rei.89,Ring1996_PPNP37-193}
and a good parameter free description of spin-orbit splittings
\cite{Ring1996_PPNP37-193,BRRMG.99,LA.11}, which have an essential
influence on the underlying shell structure. In CDFT the time-odd
components of the mean fields are given by the spatial components
of the Lorentz vectors. Therefore, because of Lorentz invariance,
these fields are coupled with the same constants as the time-like
components \cite{AA.10} which are fitted in time-even systems
to ground state properties of finite nuclei.

 The first investigation of the role of octupole deformation
in the CDFT framework has been performed in Ref.\ \cite{RMRG.95}.
In this work, the occurrence of stable octupole deformation in
the ground states of the Ra isotopes and the impact of octupole 
deformation on fission barriers of the $^{226}$Ra, $^{232}$Th and 
$^{240}$Pu nuclei has been studied with the covariant energy 
density functionals (CEDFs) NL1, NLSH
and PL-40.  However, because of some deficiencies these
functionals are no longer in use.  During the last ten years some extra
calculations for the ground states of octupole deformed nuclei have
been performed in the Ra \cite{GMT.07,NVNL.14}, Th \cite{NVL.13,NVNL.14},
Ba \cite{ZLZ.10,NVNL.14} and Sm \cite{ZLZM.10,NVNL.14} isotope
chains. The choice of these nuclei have been motivated by the results of
the analysis of experimental data performed in non-relativistic theories.

 \begin{figure*}
  \includegraphics[angle=0,width=5.9cm]{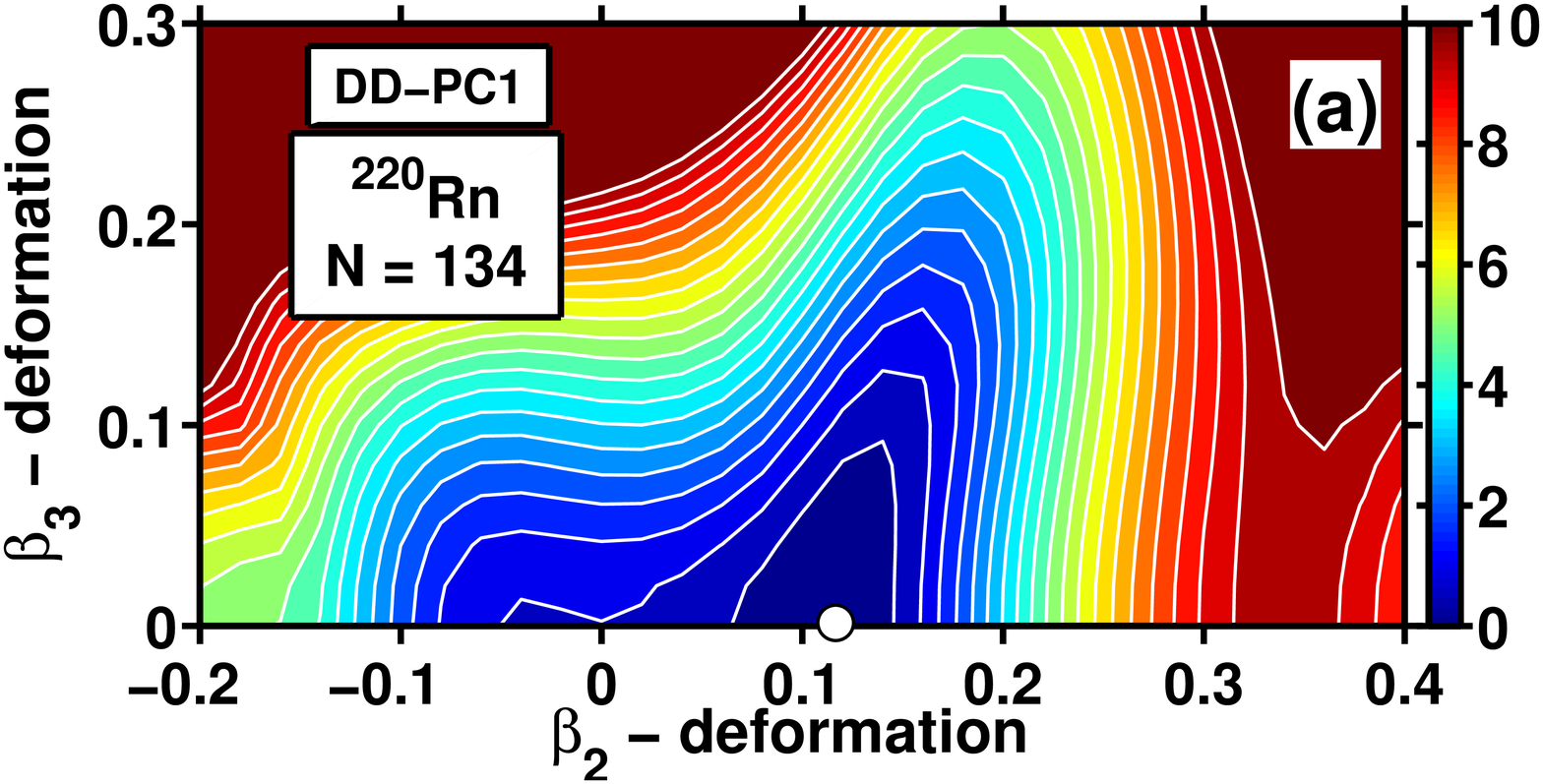}
  \includegraphics[angle=0,width=5.9cm]{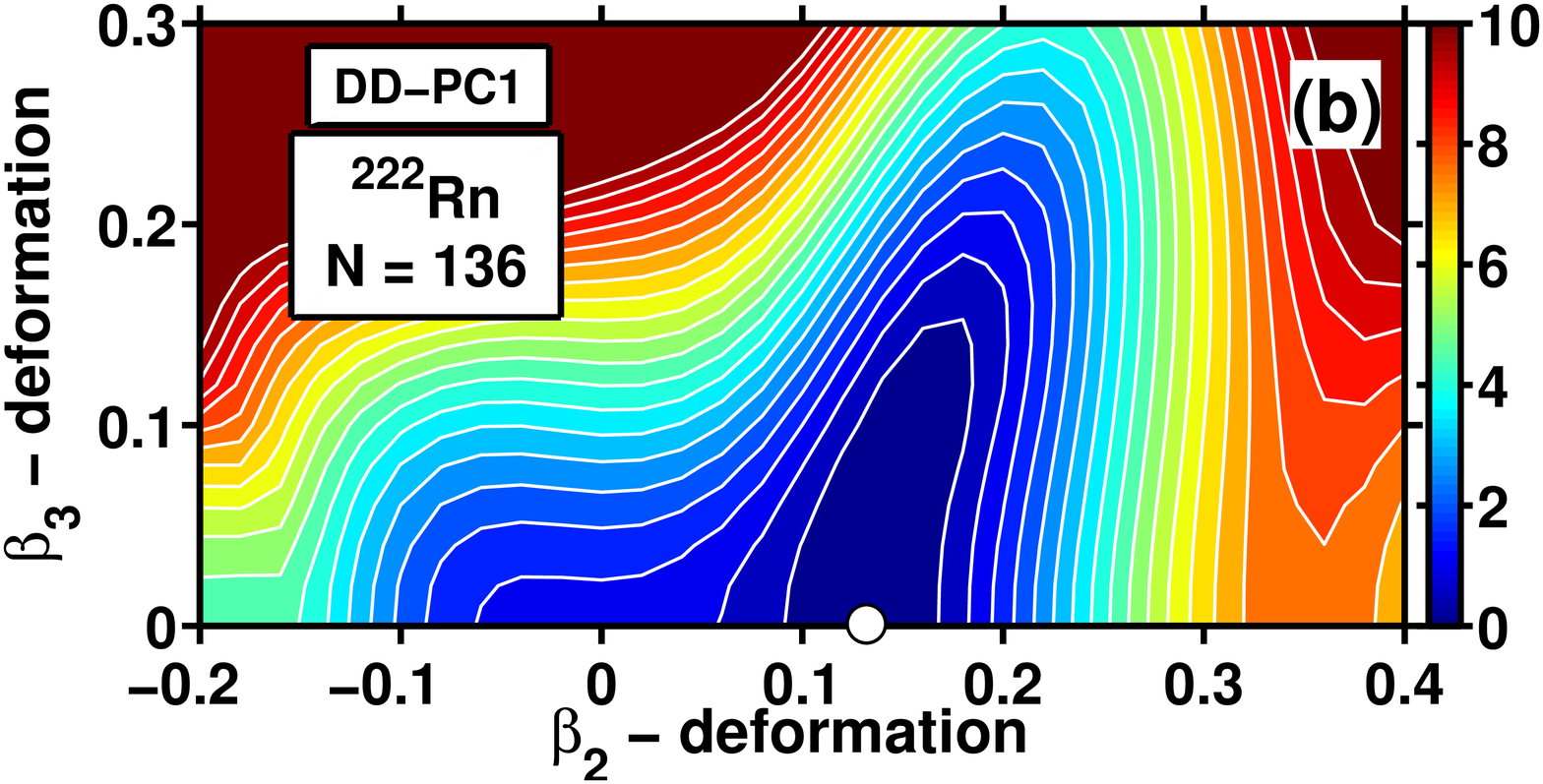}
  \includegraphics[angle=0,width=5.9cm]{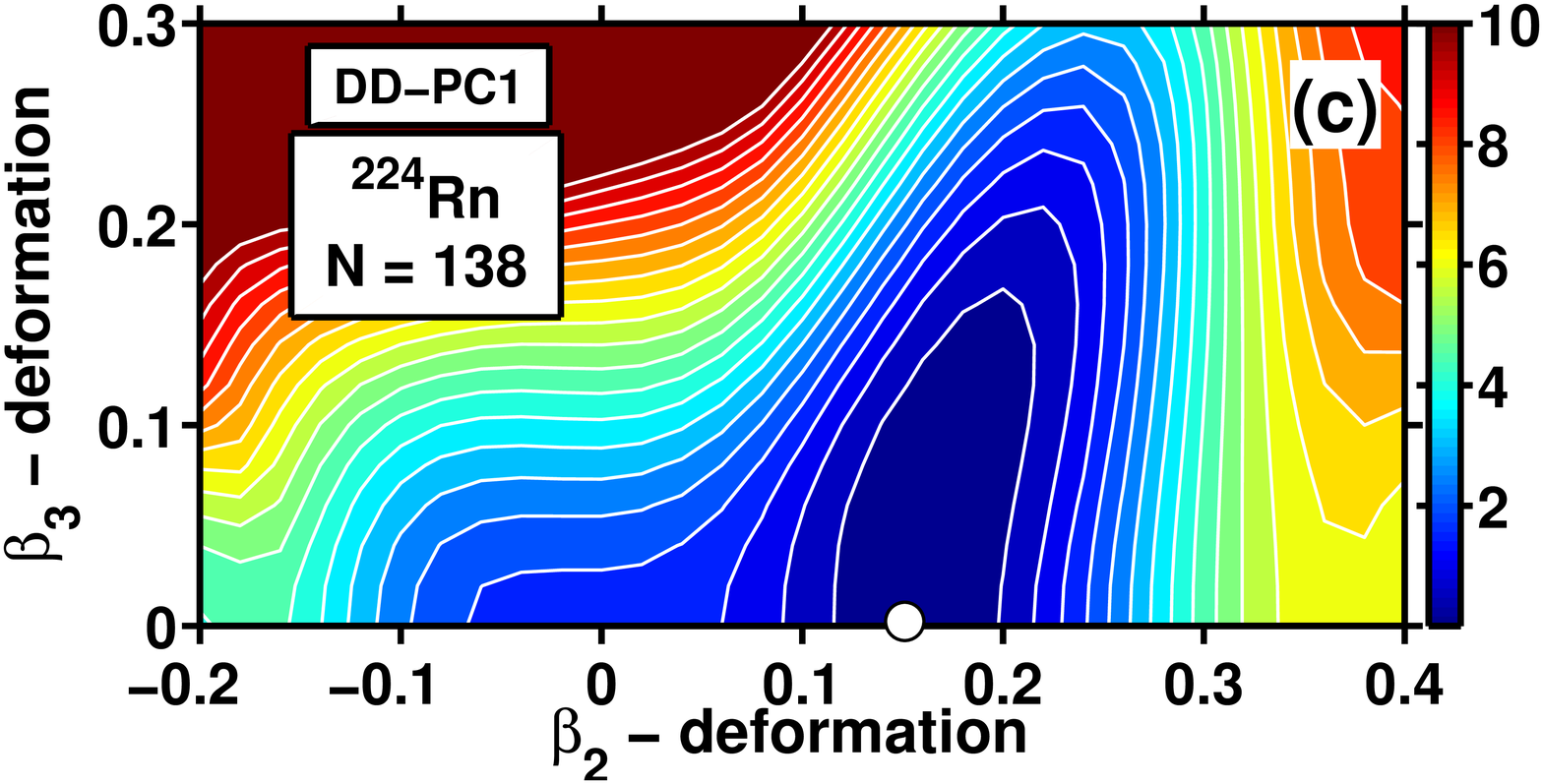}
  \caption{(Color online) Potential energy surfaces of the Rn isotopes
          in the $(\beta_2,\beta_3)$ plane  calculated
          with the CEDF DD-PC1. The white circle indicates the global
          minimum. Equipotential lines are shown in steps of 0.5
          MeV. The neutron number $N$ is shown in each panel in
          order to make the comparison between different isotones
          easier.}
\label{Rn_DD-PC1}
\end{figure*}

  However, a number of questions are left beyond the scope of these
investigations. First of them is related to a global survey of octupole
deformed and octupole  soft nuclei in the CDFT framework across the full
nuclear landscape. As mentioned above, existing CDFT studies are very
limited in scope. On the contrary, more systematic surveys of octupole
deformed nuclei exist in non-relativistic models \cite{MBCOISI.08,RB.11}.
The results of the macroscopic+microscopic
(MM) approach of Ref.\ \cite{MBCOISI.08} cover all nuclei with $Z\leq 108$
from the proton-drip line up to $N=160$. On the contrary, the non-relativistic
Hartree-Fock-Bogoliubov (HFB) studies with the Gogny force \cite{RB.11} 
cover only regions of known nuclei.

   The second question is related to the estimate of theoretical
uncertainties in the description of octupole deformed nuclei. 
The importance of such estimate become clear in the light of recent 
publications \cite{RN.10,DNR.14,AARR.14,AARR.15,AANR.15}. However, 
theoretical uncertainties in the description of octupole deformed 
nuclei have not been studied so far. Such an
estimate is not possible based on the results of previous studies
since they were performed either with only one functional (PK1 in
Ref.\ \cite{ZLZM.10} for Sm isotopes, NL1 and NL3 in Ref.\ \cite{GMT.07}
for $^{226}$Ra, DD-PC1 in \cite{NVL.13,NVNL.14} for Th, Ra, Sm, and Ba
isotopes) or the studies for a given nucleus or isotope chain have
been performed in different frameworks (relativistic mean field plus
BCS (RMF+BCS) in Refs.\ \cite{GMT.07,ZLZ.10,ZLZM.10} versus relativistic 
Hartree Bogoliubov (RHB) in Refs.\ \cite{NVL.13,NVNL.14}) using different 
prescriptions for the pairing interaction. In addition, the choice of 
nuclei in these studies is quite limited (not exceeding six nuclei 
per isotope chain).

  To address these two questions, we have performed a global survey of all
even-even $Z\leq 106$ nuclei located between the two-proton and 
two-neutron drip lines employing the DD-PC1 \cite{DD-PC1} and NL3* 
\cite{NL3*} CEDFs. Additional studies with the DD-ME2 \cite{DD-ME2},
PC-PK1 \cite{PC-PK1} and DD-ME$\delta$ \cite{DD-MEdelta} functionals 
are performed in the known regions of octupole deformed nuclei and their 
vicinity. This allows to estimate theoretical uncertainties in the description of physical
observables.  In addition, the results of our investigation are consistently
compared with the ones obtained in the HFB approach with the Gogny forces and,
in particular, with the MM results presented in Ref.\ \cite{MBCOISI.08}. This
investigation is a continuation of our previous efforts to understand the
accuracy and theoretical uncertainties (and their sources) in the description
of the ground state observables \cite{AARR.14}, the extension of the nuclear landscape
\cite{AARR.13,AARR.14,AARR.15} and the properties of superheavy nuclei
\cite{AANR.15}.

 The paper is organized as follows. Section \ref{theory_details}
describes the details of the solutions of the relativistic
Hartree-Bogoliubov equations. The analysis of octupole deformation
in the actinides with $Z=86-106, N\sim 136$ is presented in Sec.\
\ref{actin}. Sec.\ \ref{lanth} contains the discussion of octupole
deformation in the $A\sim 146$ mass region. The impact of pairing
strength on the relative energies of quadrupole and octupole minima
is discussed in Sec.\ \ref{sec-pairing-impact}. Sec.\ \ref{rotation}
is devoted to the discussion of the impact of the softness of potential
energy surfaces on the rotational properties of actinides. A global
analysis of octupole deformation covering the full nuclear landscape for
nuclei up to $Z=106$ is presented in Sec.\ \ref{glob-an}. Finally,
Sec. \ref{conclusions} summarizes the results of our work.

\section{The details of the theoretical calculations}
\label{theory_details}

 The calculations have been performed in the
Relativistic-Hartree-Bogoliubov (RHB) approach for
which a new parallel computer code
RHB-OCT has been developed using as a basis the octupole deformed
RMF+BCS code DOZ developed in Ref.\ \cite{AAR.12}. Only axial reflection
asymmetric shapes are considered in the RHB-OCT code. The parallel version
allows simultaneous calculations for a significant number of nuclei and
deformation points in each nucleus.

The calculations in the RHB-OCT code perform the variation of the function
\begin{eqnarray}
E_{RHB} + \sum_{\lambda=2,3} C_{\lambda 0}
(\langle\hat{Q}_{\lambda 0}\rangle-q_{\lambda 0})^2
\label{constr}
\end{eqnarray}
employing the method of quadratic constraints. Here $E_{RHB}$ is the
total energy (see Ref.\ \cite{AARR.14} for more details of its
definition) and $\langle\hat{Q}_{\lambda 0}\rangle$ denote the expectation
value of the quadrupole ($\hat{Q}_{20}$) and octupole ($\hat{Q}_{30}$)
moments which are defined as
\begin{eqnarray}
\hat{Q}_{20}&=&2z^2-x^2-y^2,\\
\hat{Q}_{30}&=&z(2z^2-3x^2-3y^2).
\end{eqnarray}
$C_{20}$ and  $C_{30}$ in Eq.\ (\ref{constr}) are corresponding stiffness
constants \cite{RS.80} and $q_{20}$ and $q_{30}$ are constrained values of the
quadrupole and octupole moments. In order to provide the convergence to the
exact value of the desired multipole moment we use the method suggested in
Ref.~\cite{BFH.05}. Here the quantity $q_{\lambda 0}$ is replaced by the parameter
$q_{\lambda 0}^{eff}$, which is automatically modified during the iteration in such
a way that we obtain $\langle\hat{Q}_{\lambda 0}\rangle = q_{\lambda 0}$ for the
converged solution. This method works well in our constrained  calculations.
We also fix the (average) center-of-mass of the nucleus at the origin with
the constraint
\begin{eqnarray}
<\hat{Q}_{10}>=0
\end{eqnarray}
on the center-of-mass operator $\hat{Q}_{10}$ in order to avoid
a spurious motion of the center-of-mass.

  The charge quadrupole and octupole moments are defined as
\begin{eqnarray}
Q_{20} &=& \int d^3r \rho({\bm r})\,(2z^2-r^2_\perp), \\
Q_{30} &=& \int d^3r \rho({\bm r})\,z(2z^2-3r^2_\perp)
\end{eqnarray}
with $r^2_\perp=x^2+y^2$. In principle these values can be directly
compared with experimental data. However, it is more convenient to
transform these quantities into dimensionless deformation
parameters $\beta_2$ and $\beta_3$ using the relations
\begin{eqnarray}
Q_{20}&=&\sqrt{\frac{16\pi}{5}} \frac{3}{4\pi} Z R_0^2 \beta_2,
\label{beta2_def} \\
Q_{30}&=&\sqrt{\frac{16\pi}{7}}\frac{3}{4\pi} Z R_0^3 \beta_3
\label{beta4_def}
\end{eqnarray}
where $R_0=1.2 A^{1/3}$. These deformation parameters are more
frequently used in experimental works than quadrupole and octupole
moments. In addition, the potential energy surfaces (PES) are plotted in
this manuscript in the ($\beta_2,\beta_3$) deformation plane.

In order to avoid the uncertainties connected with the definition of
the size of the pairing window\ \cite{KALR.10}, we use the separable form of the
finite range Gogny pairing interaction introduced by Tian et al \cite{TMR.09}.
Its matrix elements in $r$-space have the form
\begin{eqnarray}
\label{Eq:TMR}
V({\bm r}_1,{\bm r}_2,{\bm r}_1',{\bm r}_2') &=& \nonumber \\
= - f G \delta({\bm R}-&\bm{R'}&)P(r) P(r') \frac{1}{2}(1-P^{\sigma})
\label{TMR}
\end{eqnarray}
with ${\bm R}=({\bm r}_1+{\bm r}_2)/2$ and ${\bm r}={\bm r}_1-{\bm r}_2$
being the center of mass and relative coordinates. The form factor
$P(r)$ is of Gaussian shape
\begin{eqnarray}
P(r)=\frac{1}{(4 \pi a^2)^{3/2}}e^{-r^2/4a^2}
\end{eqnarray}
The two parameters $G=738$ fm$^3$ and $a=0.636$ fm of this interaction
are the same for protons and neutrons and have been derived in Ref.\
\cite{TMR.09} by a mapping of the $^1$S$_0$ pairing gap of infinite
nuclear matter to that of the Gogny force D1S~\cite{D1S}.

  The scaling factor $f$ in Eq.~(\ref{TMR}) is determined by a fine
tuning of the pairing strength in a comparison between experimental
moments of inertia and those obtained in cranked RHB calculations
with the CEDF NL3* (see Ref.\ \cite{AARR.14} for details). It is
fixed at $f=1.0$ in the $Z\geq 88$ actinides and superheavy
nuclei, at $f=1.075$ in the $56\leq Z \leq 76$ and at $f=1.12$ in
the $Z\leq 44$ nuclei. Between these regions, i.e. for $44\leq Z \leq 56$
and for $76\leq Z \leq 88$, the scaling factor $f$ gradually changes
with $Z$ in a linear interpolation. The weak dependence of the scaling
factor $f$ on the CEDF has been seen in the studies of pairing and
rotational properties in the actinides in Refs.\ \cite{A250,AO.13}
and pairing gaps in spherical nuclei in Ref.\ \cite{AARR.14}. Thus,
the same scaling factor $f$ as defined above for the CEDF NL3* is
used in the calculations with DD-PC1, DD-ME2 and DD-ME$\delta$.
Considering the global character of this study, this is a reasonable
choice.

 The truncation of the basis is performed in such a way that all states
belonging to the major shells up to $N_F=16$ fermionic shells for the
Dirac spinors and up to $N_B=20$ bosonic shells for the meson fields
are taken into account (for details see Ref.\ \cite{GRT.90}).
Considering that the calculations are performed
in the vicinity of the normal deformed minimum, this truncation of
the basis provides sufficient numerical accuracy. The potential energy surfaces
are calculated in constrained calculations in the ($\beta_2,\beta_3$)
plane for the $\beta_2$ values ranging from $-0.2$ up to 0.4 and for
the $\beta_3$ values ranging from 0.0 up to 0.3 with a deformation step
of 0.02 in each direction. The energies of the local minima are defined in
unconstrained calculations.

 \begin{figure*}
  \includegraphics[angle=0,width=5.9cm]{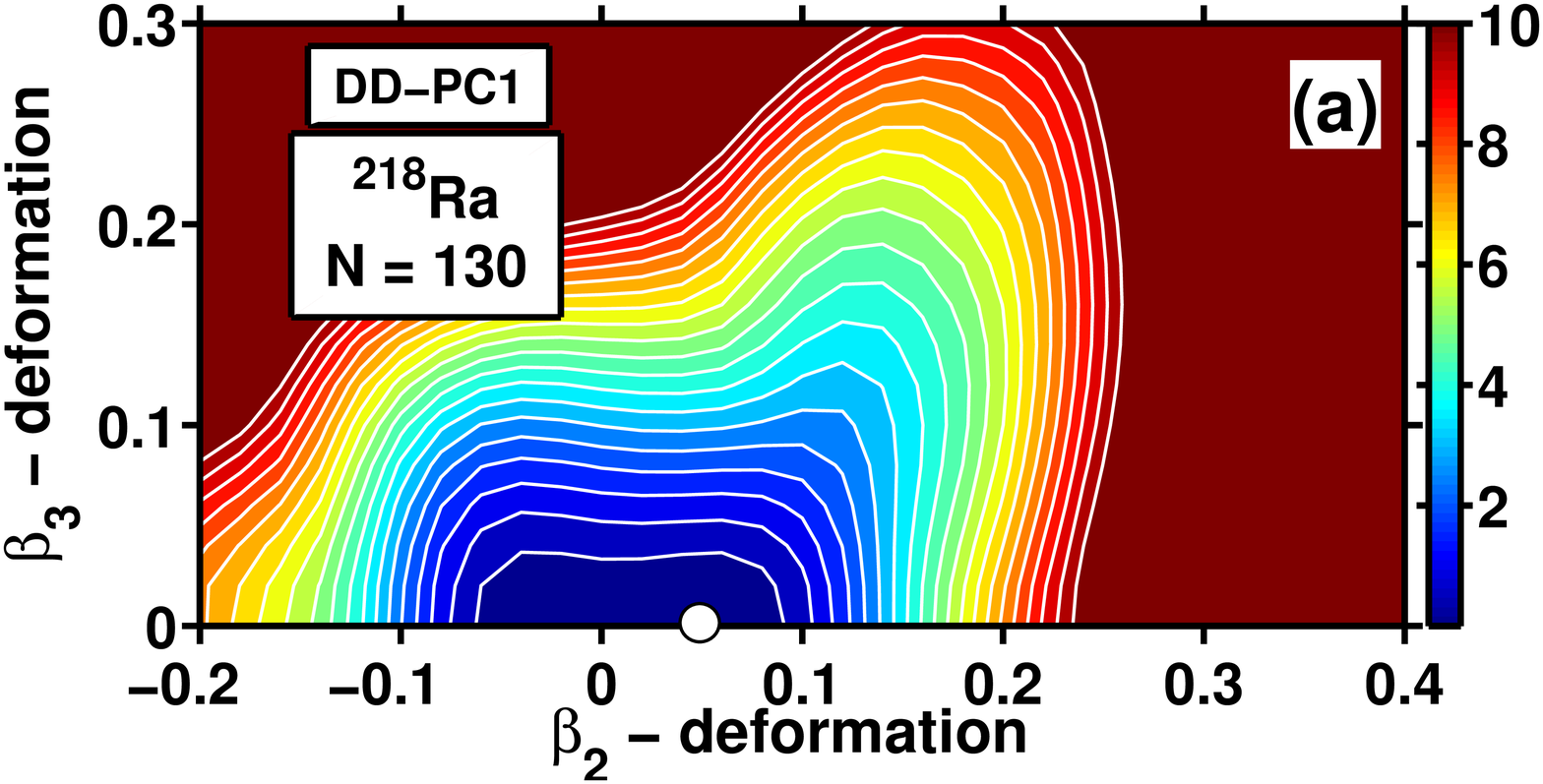}
  \includegraphics[angle=0,width=5.9cm]{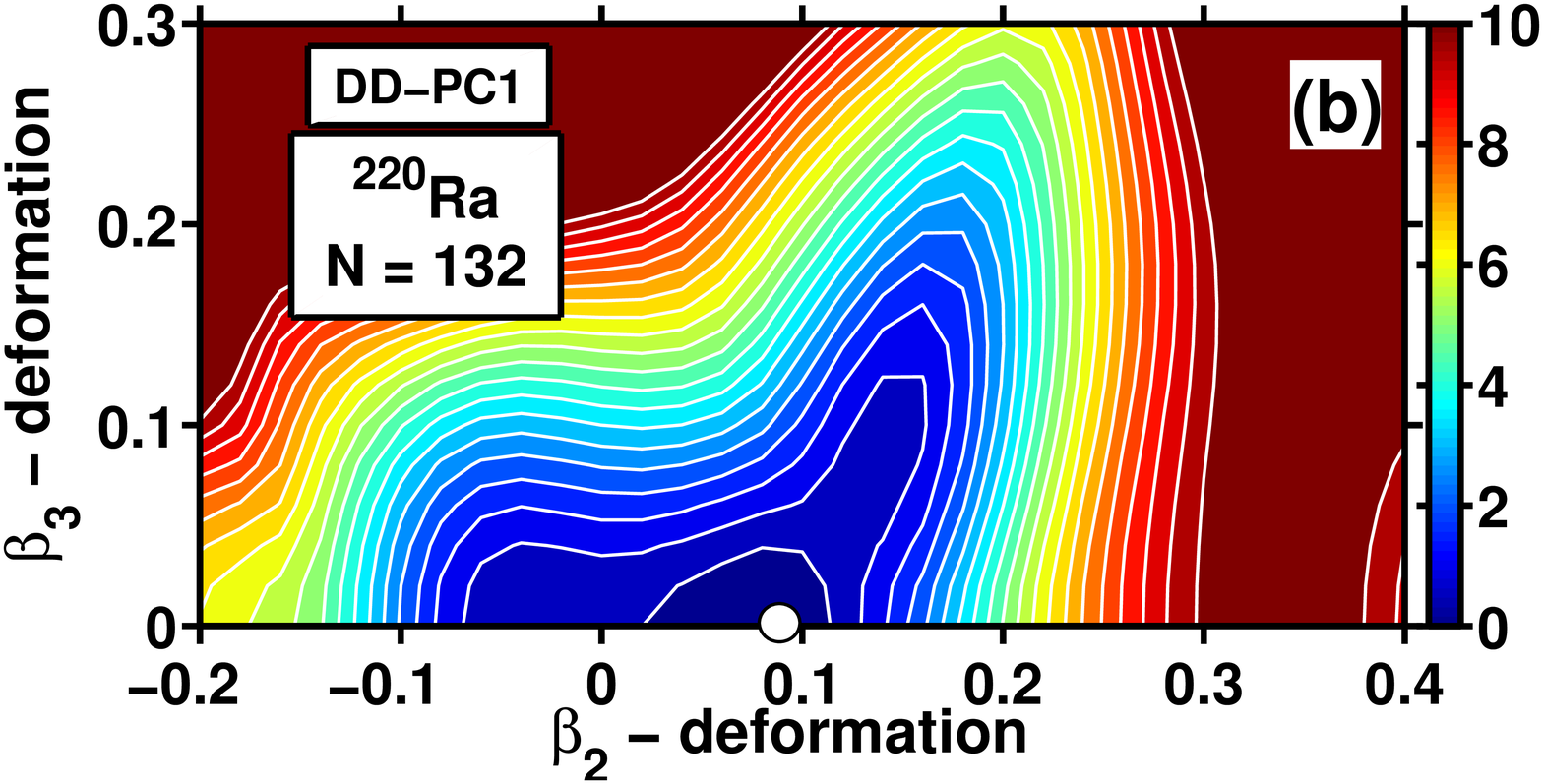}
  \includegraphics[angle=0,width=5.9cm]{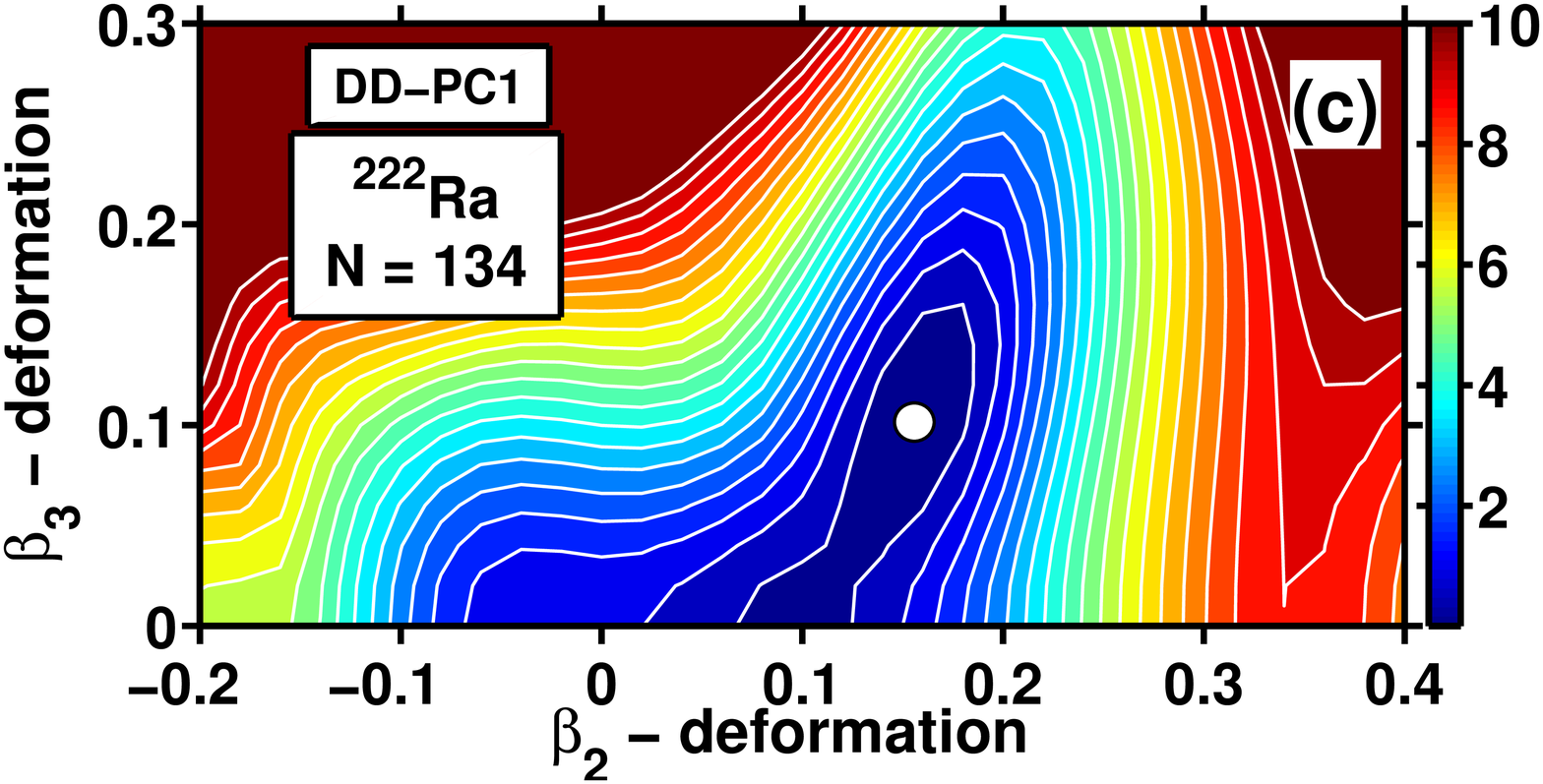}
  \includegraphics[angle=0,width=5.9cm]{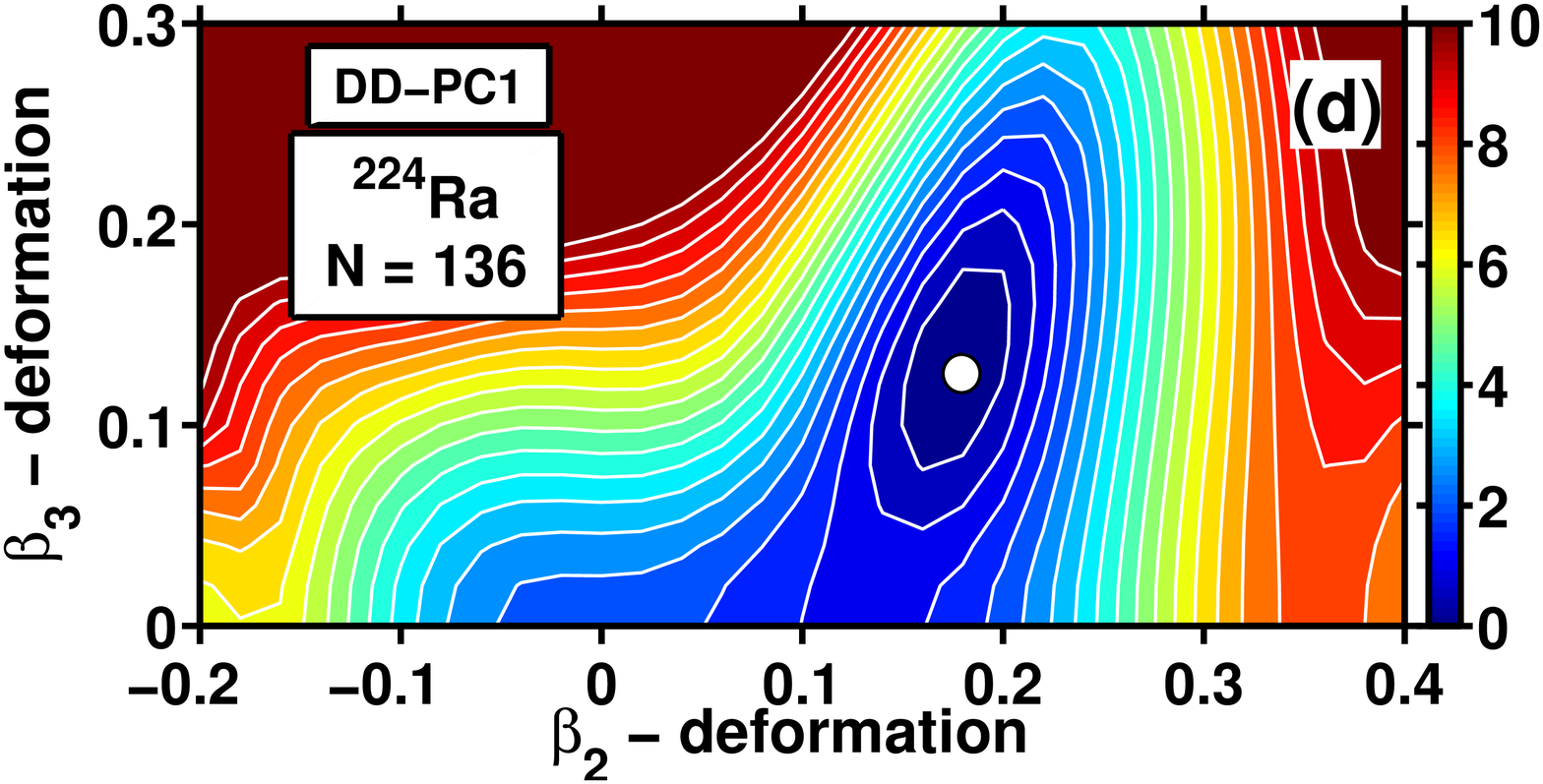}
  \includegraphics[angle=0,width=5.9cm]{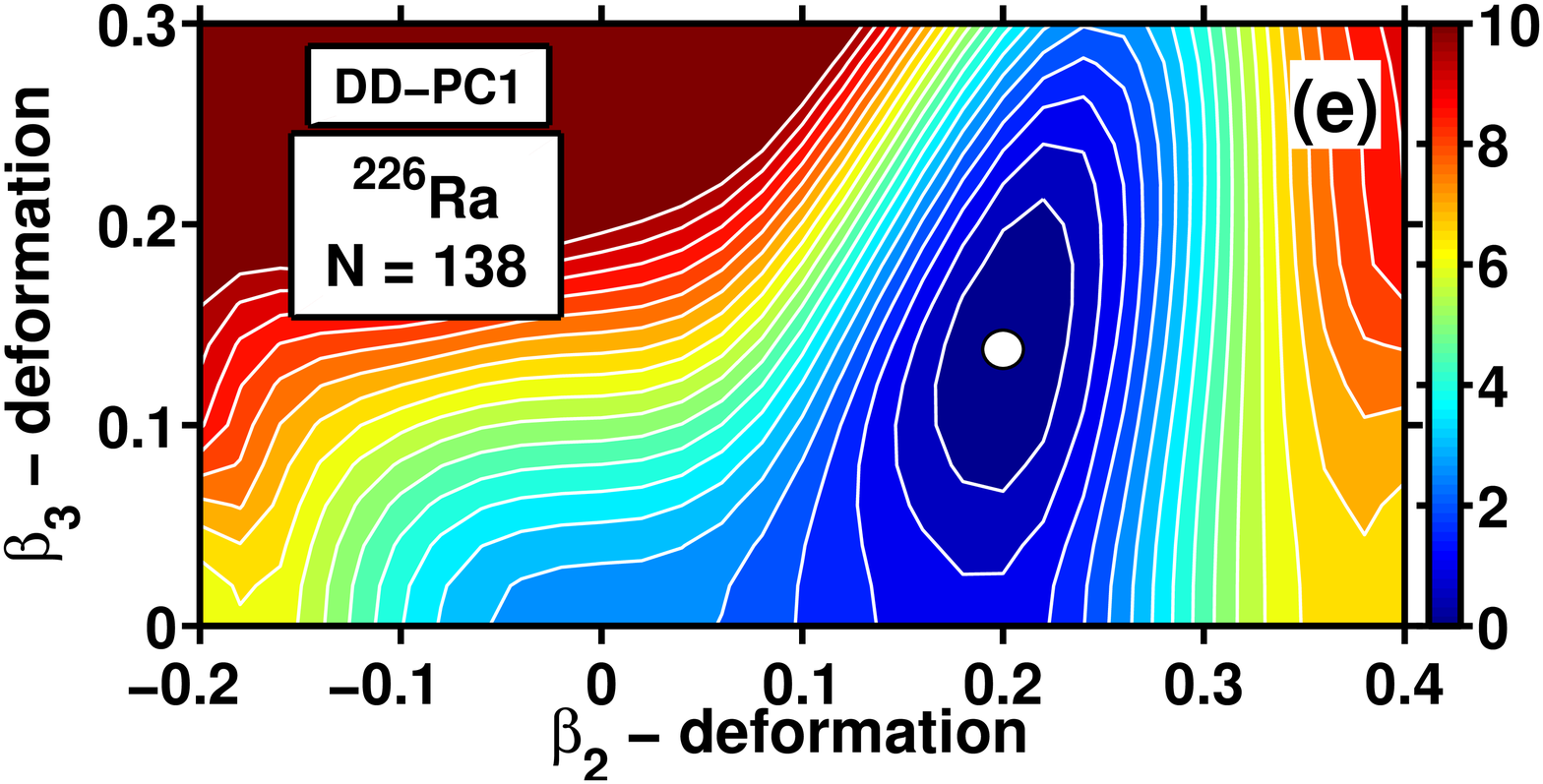}
  \includegraphics[angle=0,width=5.9cm]{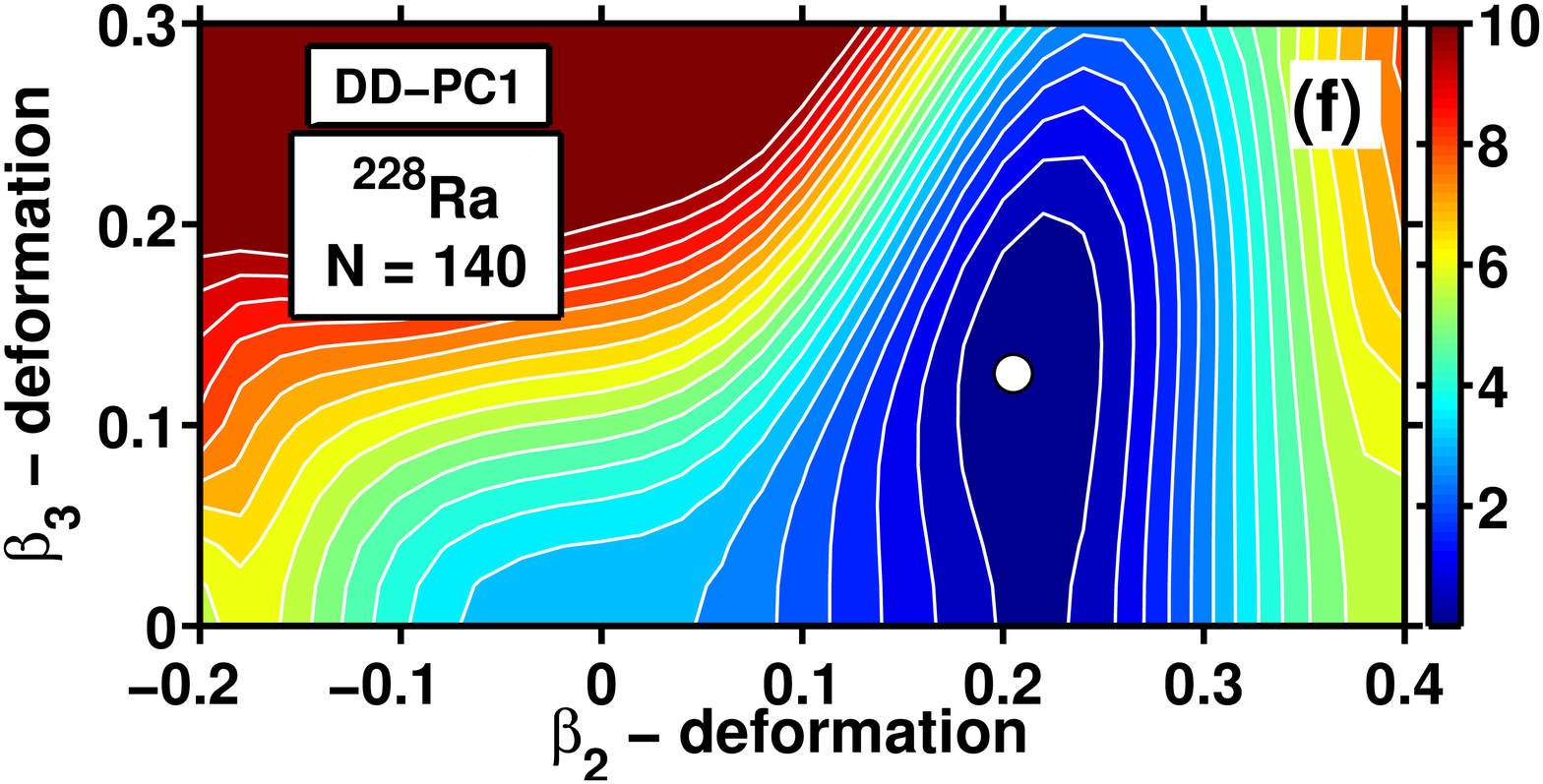}
  \includegraphics[angle=0,width=5.9cm]{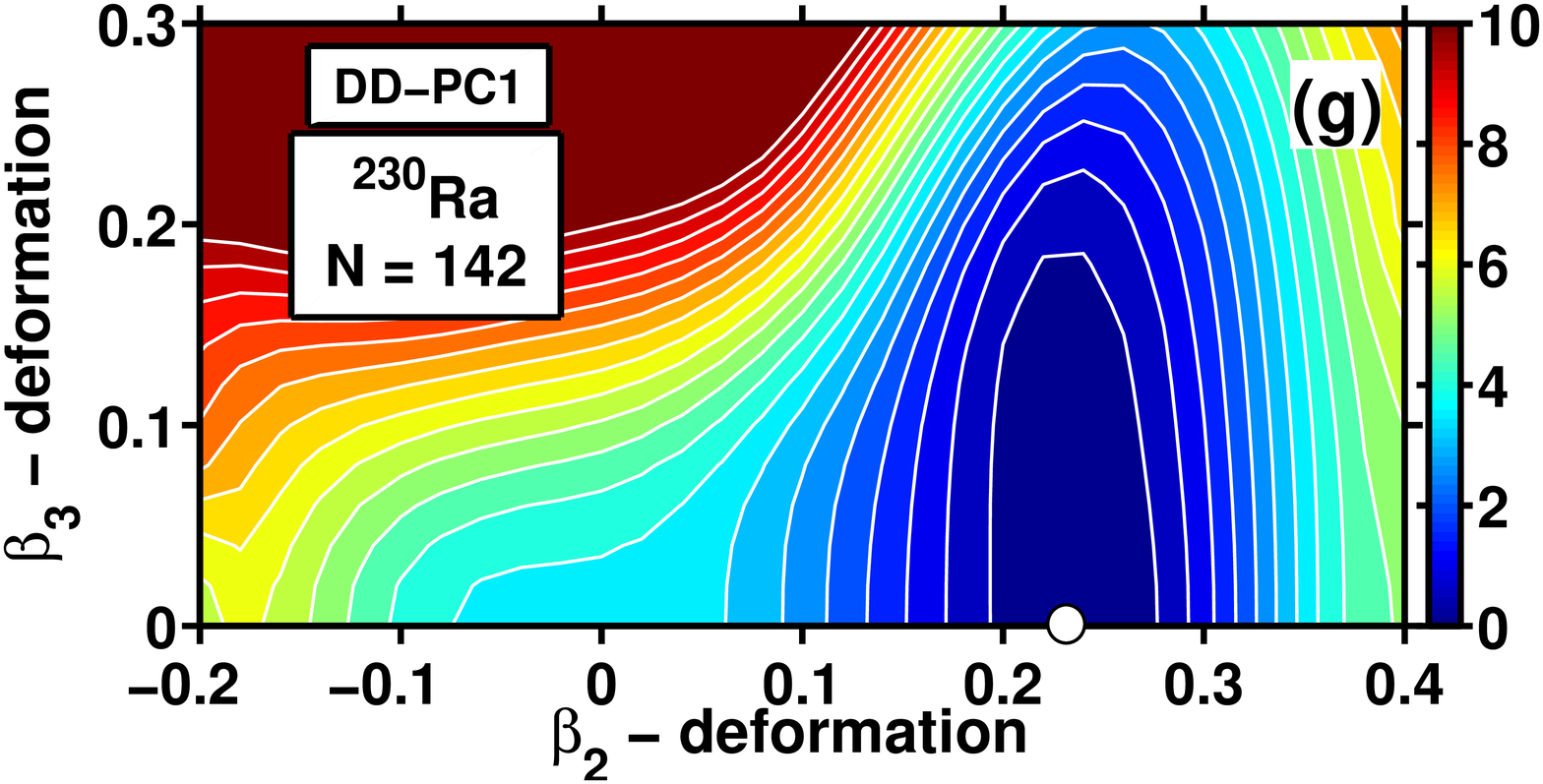}
   \caption{(Color online) The same as Fig.\ \ref{Rn_DD-PC1},
           but for the Ra isotopes.}
\label{Ra_DD-PC1}
\end{figure*}

   The effect of octupole deformation can be quantitatively
characterized by the quantity $\Delta E_{oct}$ defined as
\begin{eqnarray}
\Delta E_{oct} = E^{oct}(\beta_2, \beta_3) - E^{quad}(\beta'_2,\beta'_3=0)
\end{eqnarray}
where $E^{oct}(\beta_2, \beta_3)$ and $E^{quad}(\beta'_2, \beta'_3=0)$
are the binding energies of the nucleus in two local minima of
potential energy surface; the first minimum corresponds to octupole
deformed shapes and second one to the shapes with no octupole
deformation. The quantity  $|\Delta E_{oct}|$ represents the gain of
binding due to octupole deformation. It is also an indicator of
the stability of the octupole deformed shapes. Large $|\Delta E_{oct}|$
values are typical for well pronounced octupole minima in the
PES; for such systems the stabilization of static octupole deformation
is likely. On the contrary, small $|\Delta E_{oct}|$ values are characteristic
for soft (in octupole direction) PES typical for octupole vibrations. In 
such systems  beyond mean field effects can play an important role.
They have profound effect on the spectroscopy of the nuclei, 
in particular, on the E1 and enhanced E3 transition strengths 
\cite{RRS.12,YZL.15,ZYLMR.16}, and on the energy splittings of the 
positive and negative parity branches of alternating parity rotational bands 
\cite{GER.98,YZL.15}. On the other hand, octupole beyond mean field 
correlations do not affect in a significant way the trends and 
systematics of binding energies \cite{Robledo.15}.

\section{Octupole deformation in actinides}
\label{actin}

  Several studies of the octupole deformation in the ground states of
actinides and its impact on spectroscopic properties of these nuclei
have been performed so far in the CDFT framework. The first relativistic 
study of
octupole shapes in the ground states of atomic nuclei has been
performed twenty years ago in Ref.\ \cite{RMRG.95}; in this manuscript
radium isotopes have been investigated in the RMF+BCS approach using
monopole pairing with constant pairing gap and the CEDFs NL1, NL-SH and PL-40.
Shape evolution from spherical to octupole-deformed shapes has
been studied in even-even Th isotopes in the RMF+BCS framework in
Ref.\ \cite{GPZ.10} using monopole pairing with constant pairing gap
and the NL3* and PK1 functionals. Octupole deformed shapes in $^{226}$Ra
have been investigated earlier in Ref.\ \cite{GMT.07} within the same
approach but with NL1 and NL3 functionals. The potential energy
surfaces and octupole deformations of the ground states of even-even
$^{222-232}$Th and $^{218-228}$Ra nuclei have been studied in  the RHB framework
with the DD-PC1 functional and separable pairing forces in Refs.\
\cite{NVL.13,NVNL.14}. The mapping of these potential energy surfaces
onto an equivalent Hamiltonian of the $sdf$ interacting boson model (IBM)
allowed to determine its parameters. Then, the resulting IBM Hamiltonian
was used to calculate excitation spectra and transition rates for positive-
and negative-parity states in these nuclei \cite{NVL.13,NVNL.14}. Recently,
first generator coordinate method studies taking into account dynamical
correlations and quadrupole-octupole shape fluctuations have been undertaken
in $^{224}$Ra employing the PC-PK1 functional \cite{YZL.15}. They reveal
rotation-induced octupole shape stabilization.

\begin{figure*}
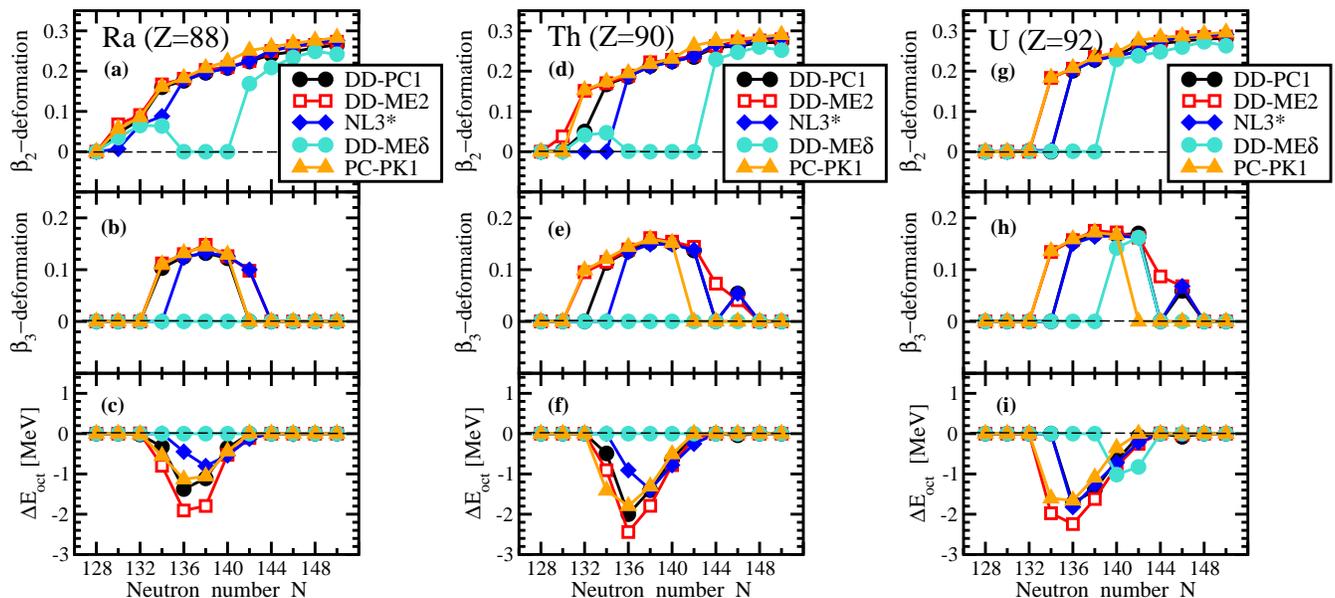

  \includegraphics[angle=0,width=5.8cm]{fig-3-a-new.eps}
  \includegraphics[angle=0,width=5.8cm]{fig-3-b-new.eps}
  \includegraphics[angle=0,width=5.8cm]{fig-3-c-new.eps}
  \caption{(Color online) The calculated equilibrium quadrupole $\beta_{2}$
   (panels (a), (d) and (g)) and octupole  $\beta_{3}$ (panels (b), (e) and (h)) 
   deformations as well as 
   the $\Delta E^{oct}$ quantities (panels (c), (f) and (i)). The
   results of the RHB calculations with five indicated functionals are presented
   for the Ra, Th and U isotope chains.}
\label{massdependence}
\end{figure*}

  It is clear that these studies were quite limited in scope and the
selection of nuclei was guided by the previous studies in non-relativistic
frameworks. A global review of octupole deformed nuclei in this mass
region paints a much richer picture. Our RHB calculations indicate that
not only Ra and Th nuclei (as suggested by previous studies) can have
either stable octupole deformation or be octupole soft, but also U, Pu, Cm,
Cf, Fm, No and Sg nuclei possess these properties. The potential
energy surfaces obtained in the RHB calculations with the DD-PC1 functional
are shown in Figs.\ \ref{Rn_DD-PC1}, \ref{Ra_DD-PC1}, \ref{Th_DD-PC1}, \ref{U_DD-PC1},
\ref{Pu_DD-PC1}, \ref{Cm_DD-PC1}, \ref{Cf_DD-PC1}, and \ref{Fm_DD-PC1}
below for the nuclei in the Rn, Ra, Th, U, Pu, Cm, Cf and Fm isotope chains.
According to previous global surveys of the performance of the state-of-the-art
CEDFs presented in Refs.\ \cite{AARR.14,AANR.15},  this is one of the best CEDFs.
Neutron number dependencies of calculated equilibrium quadrupole and octupole
deformations as well as the gains in binding due to octupole deformation for
these isotope chains are presented for five CEDFs in Figs.\ \ref{massdependence},
\ref{massdependence-Pu-etc} and \ref{massdependence-Fm} below. For simplicity
of the discussion of the results and their comparison with available experimental
data and other theoretical approaches, they are discussed on the
``chain-by-chain'' basis in the next subsections.

\subsection{Discussion: theory versus experiment}

\subsubsection{Rn isotopes}

  No octupole deformation is predicted in the Rn isotopes of interest (see
Table \ref{table-global}). The potential energy surfaces are soft in octupole
direction for the $N=136, 138$ $^{222,224}$Rn nuclei (Fig.\ \ref{Rn_DD-PC1}).
This is in agreement with the analysis of experimental data presented in Refs.\
\cite{Rn-Ra-Th.99,220Rn-224Ra.13} which strongly suggests that the $^{218-222}$Rn
isotopes behave like octupole vibrators, thus supporting the presence of a 
considerable octupole softness of the potential energy surfaces.

  The MM calculations of Ref.\ \cite{NORDLMR.84} with a Woods Saxon potential
suggest that only the $^{222-224}$Rn isotopes with $N=136-138$ have non-zero octupole
deformation. However, the gain in binding due to octupole deformation is rather
small ($\sim 100$ keV). A wider range of octupole deformed Rn isotopes with 
$N=132-140$ (with a maximum value of $|\Delta E^{oct}|=0.85$ MeV at $N=134$) and 
$N=146$ is predicted in Ref.\ \cite{MBCOISI.08} in MM calculations with a folded 
Yukawa potential (see Table \ref{table-global}). Ref.\ \cite{RB.13} shows 
that at least $^{220}$Rn has non-zero octupole deformation in the ground state in 
the HFB calculations with the D1S and D1M Gogny forces. However, this is octupole 
soft nucleus with a relative small gain in binding due to octupole deformation 
($|\Delta E^{oct}|<0.25$ MeV).

\subsubsection{Ra isotopes}

 Potential energy surfaces of the Ra isotopes are shown in Fig.\
\ref{Ra_DD-PC1}. Weakly deformed minima with $\beta_3=0.0$ are 
the lowest in energy in the $^{218,220}$Ra nuclei with $N=130, 132$. The
increase of neutron number leads to the formation of an octupole
minimum which becomes pronounced at $N=136, 138$. At higher neutron
numbers the potential energy surfaces become soft in octupole direction.

\begin{figure*}
  \includegraphics[angle=0,width=5.9cm]{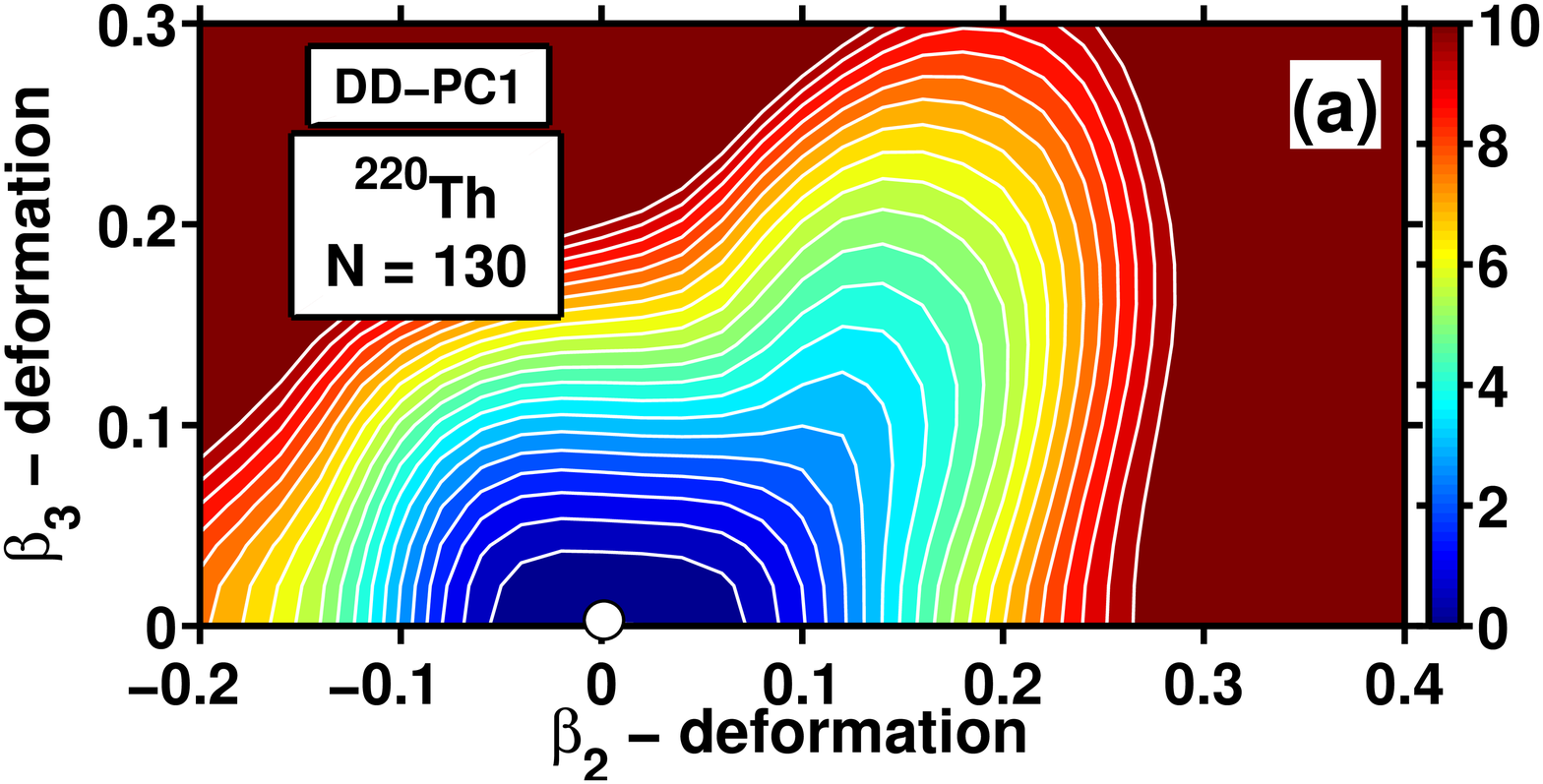}
  \includegraphics[angle=0,width=5.9cm]{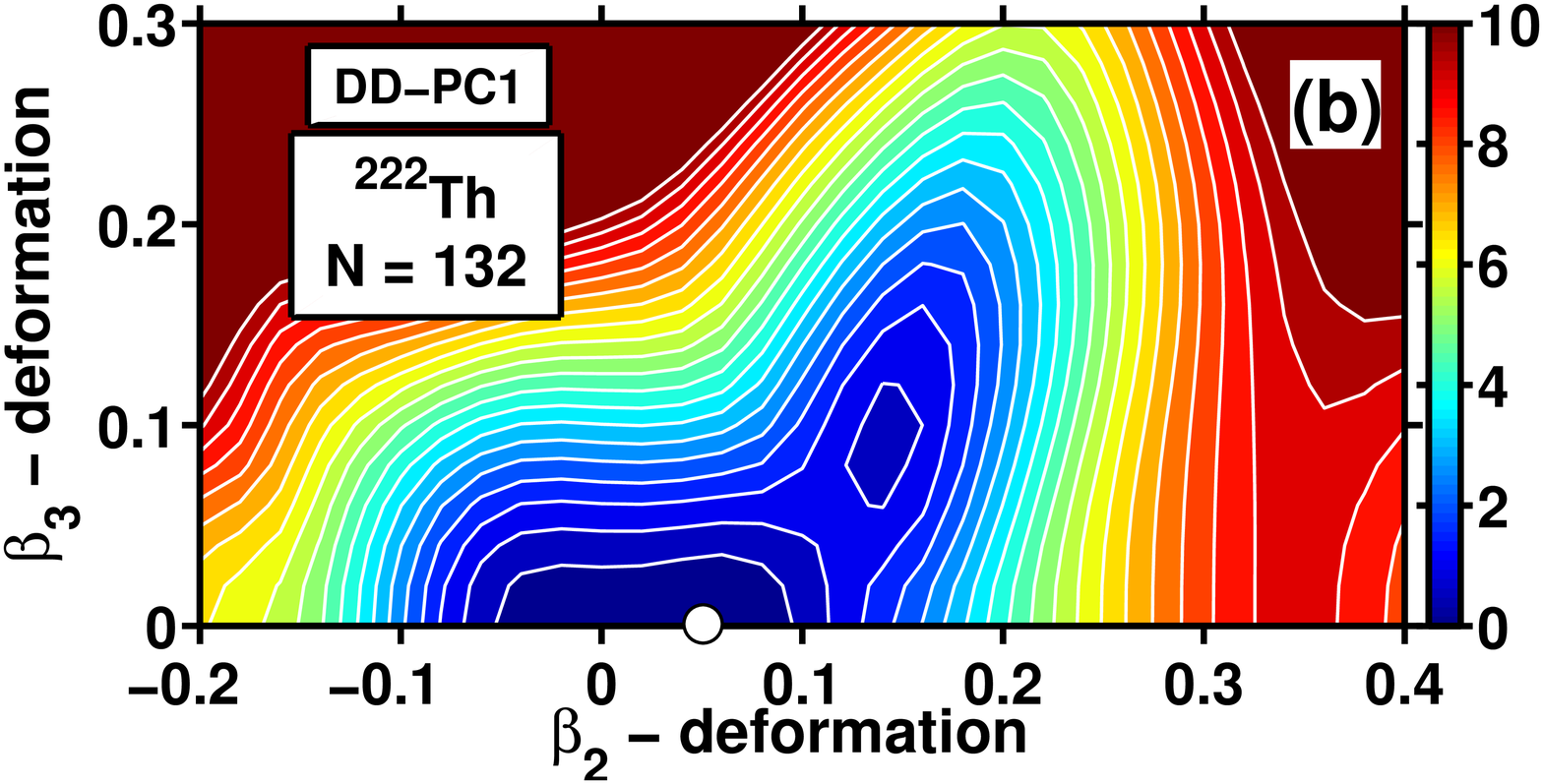}
  \includegraphics[angle=0,width=5.9cm]{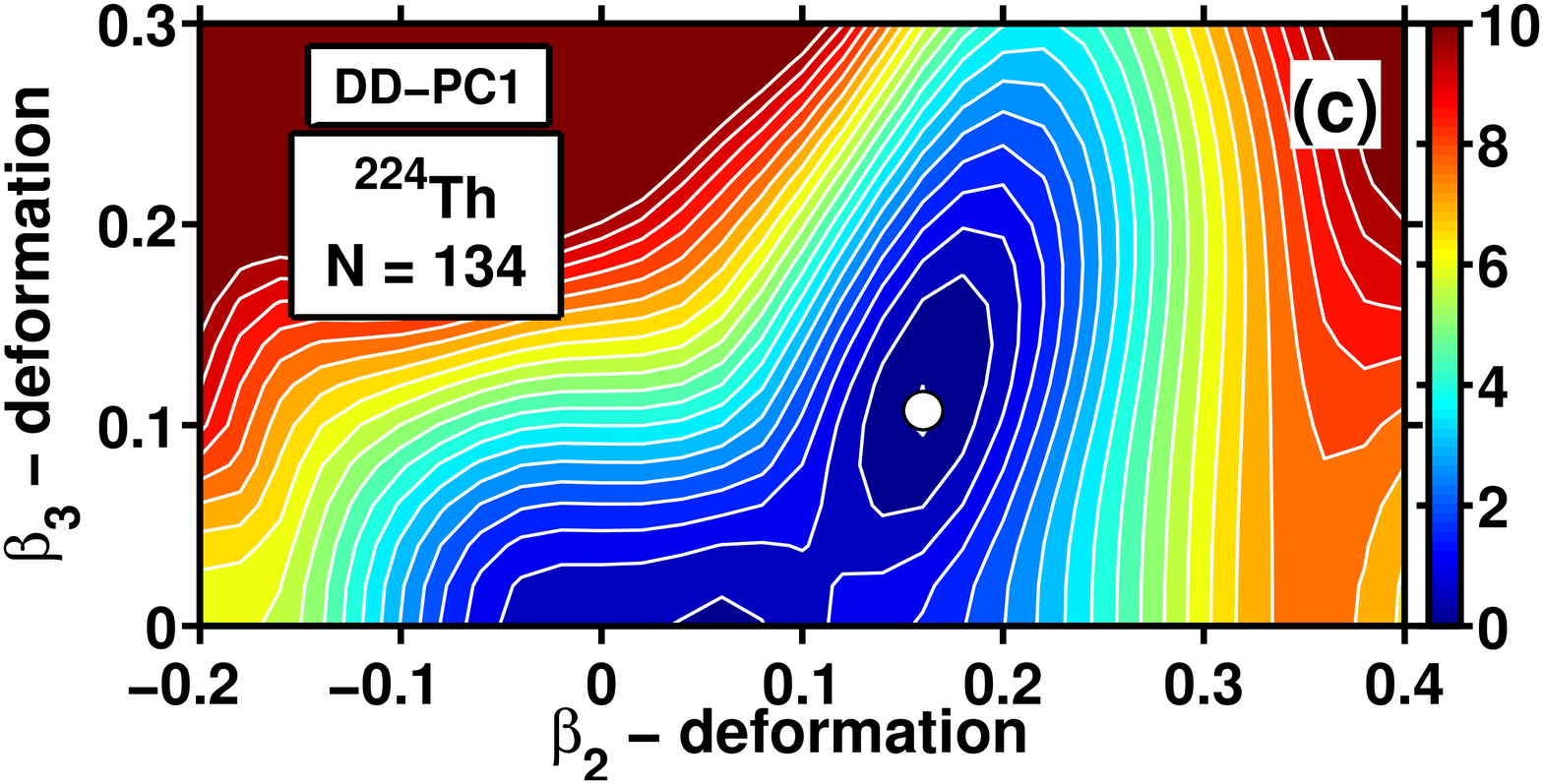}
  \includegraphics[angle=0,width=5.9cm]{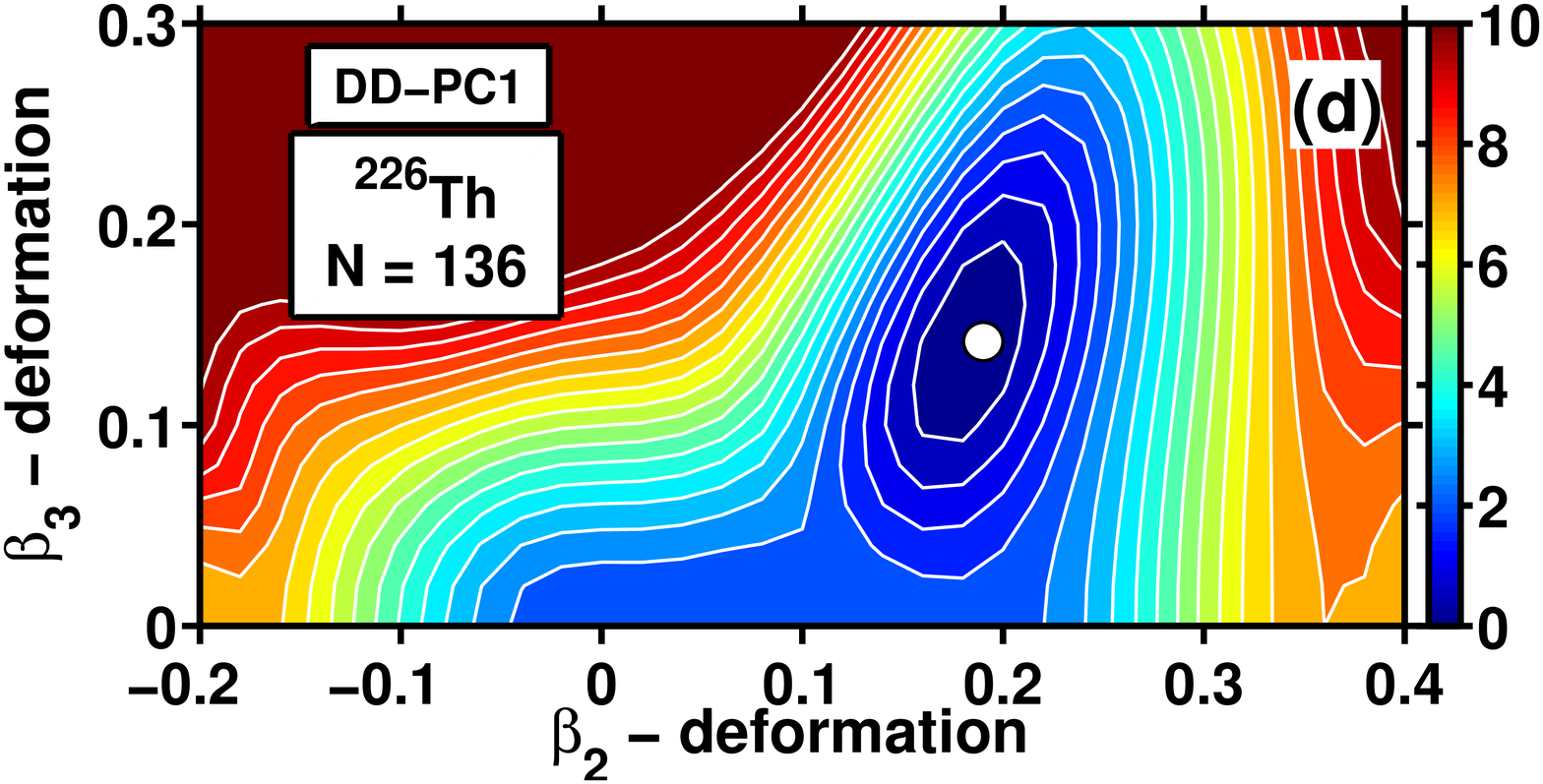}
  \includegraphics[angle=0,width=5.9cm]{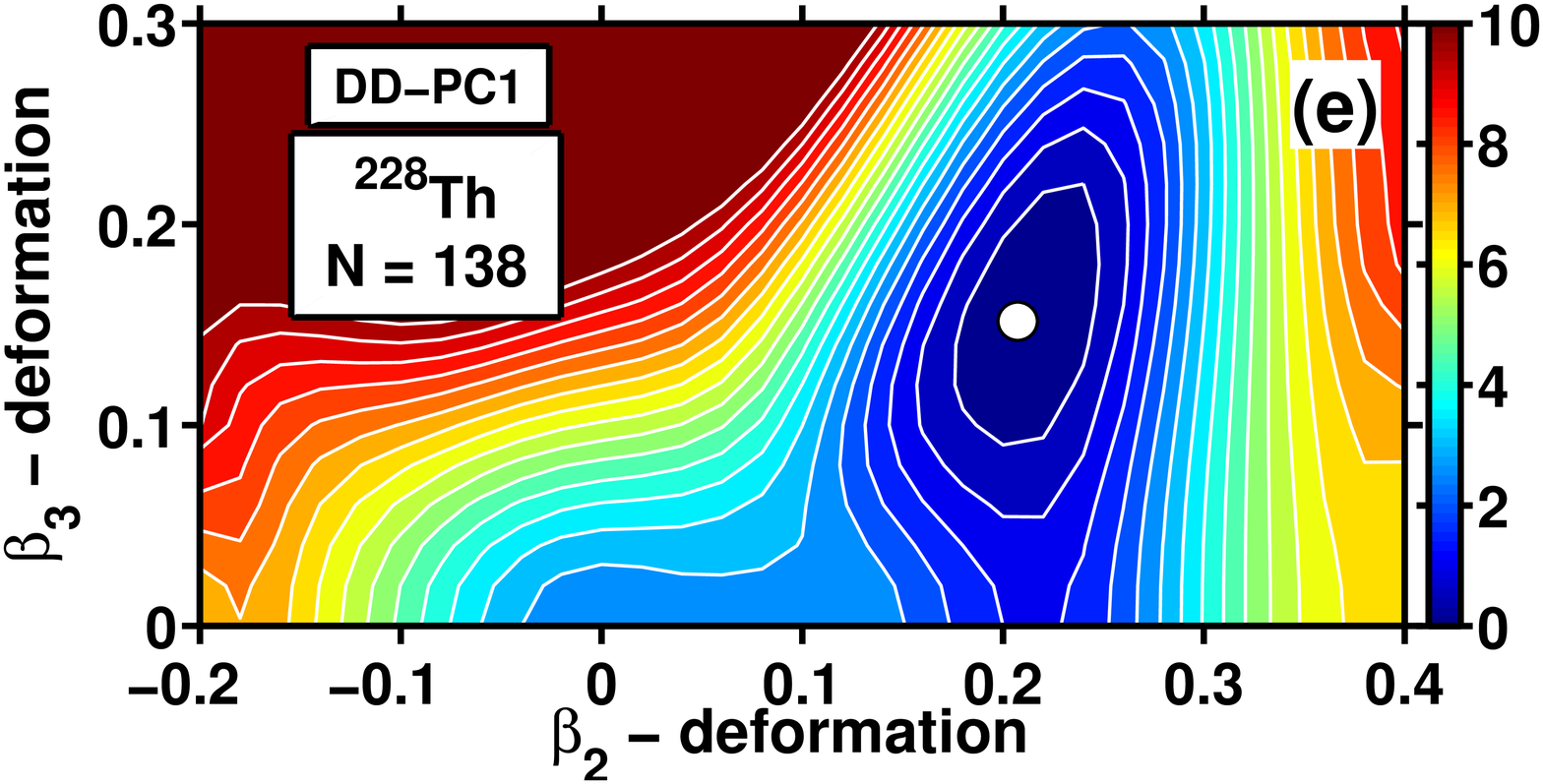}
  \includegraphics[angle=0,width=5.9cm]{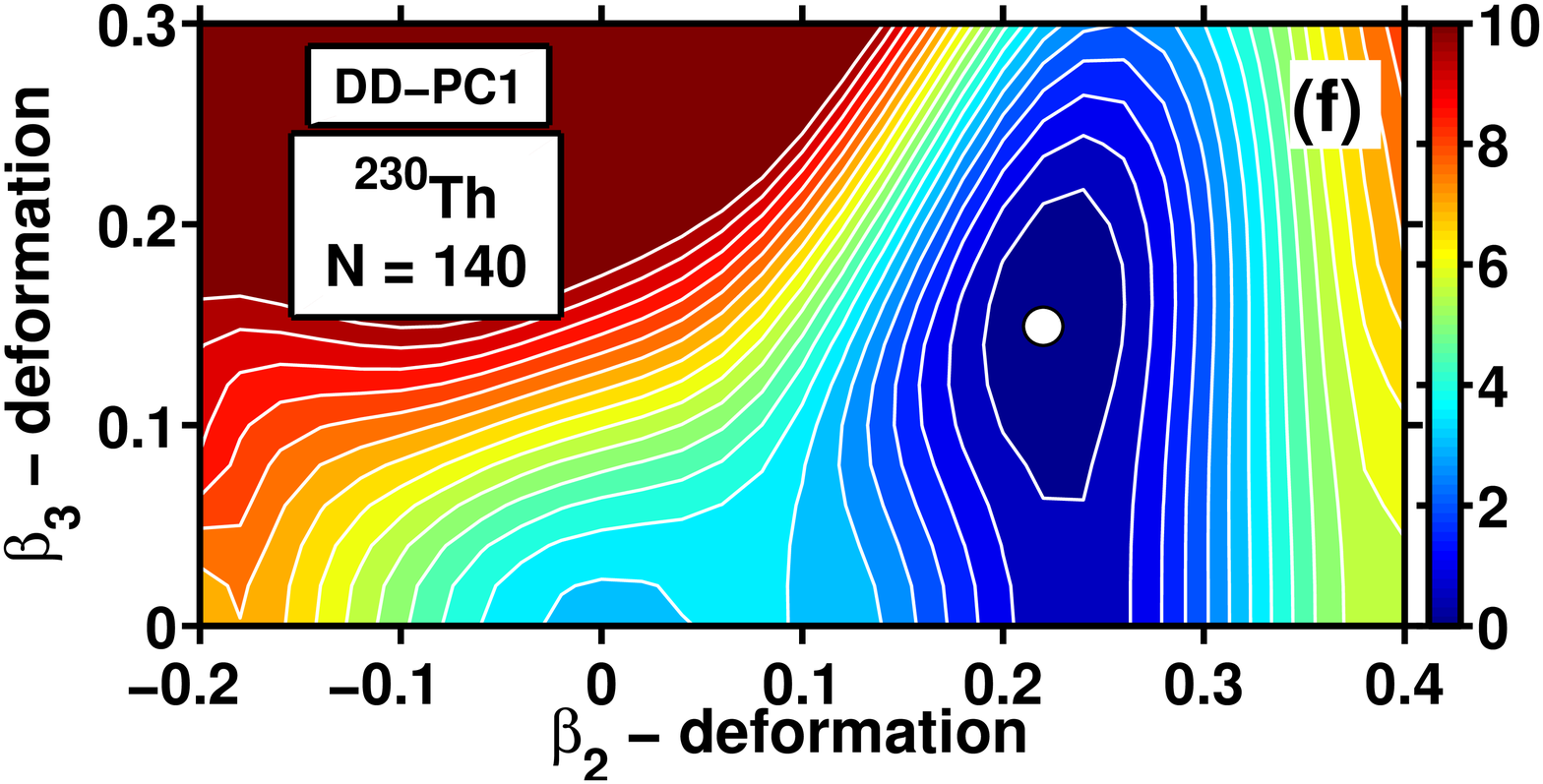}
  \includegraphics[angle=0,width=5.9cm]{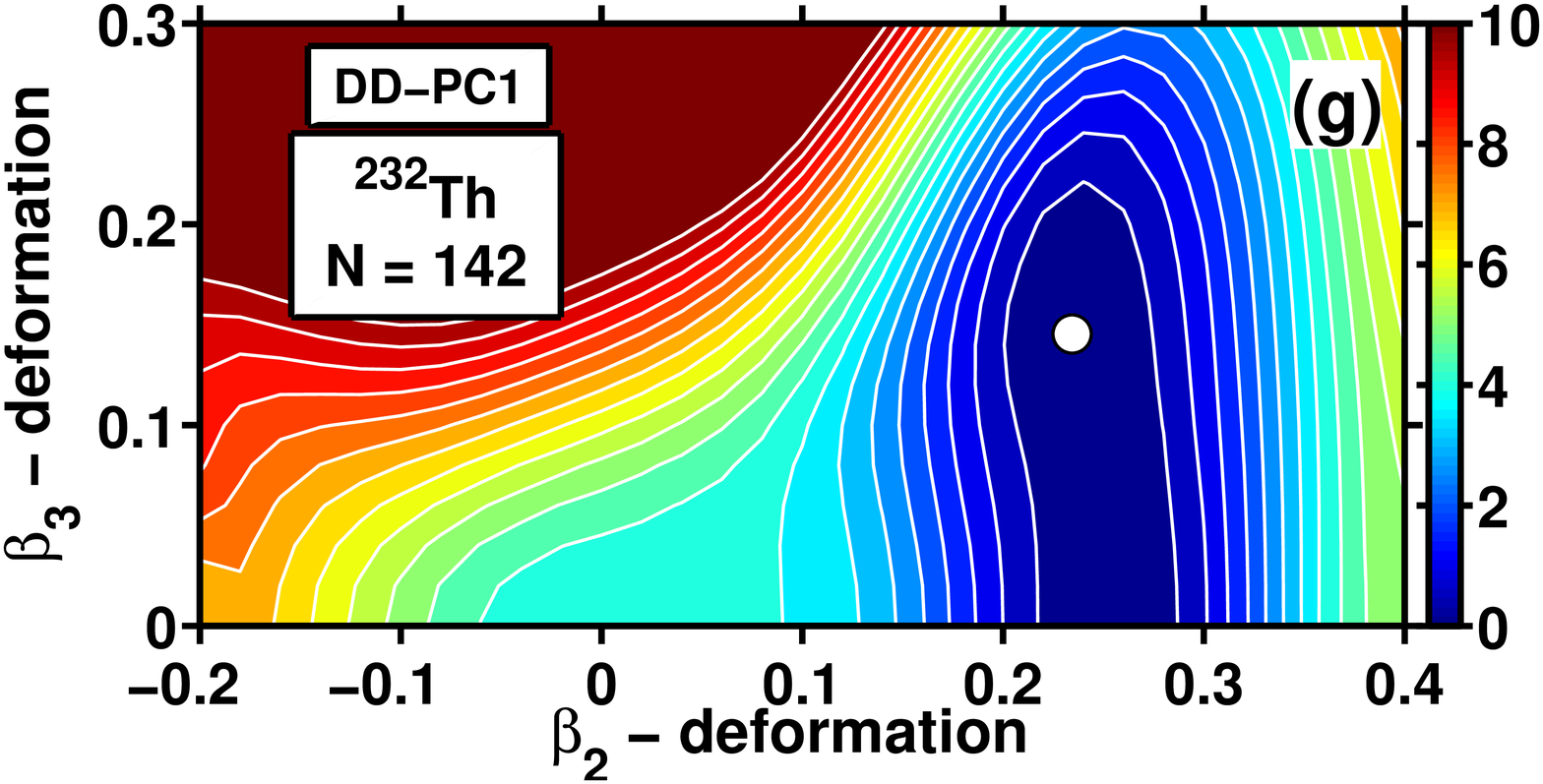}
  \caption{(Color online) The same as Fig.\ \ref{Rn_DD-PC1},
           but for the Th isotopes.}
\label{Th_DD-PC1}
\end{figure*}

  The maximum gain in binding energy due to octupole deformation 
$|\Delta E^{oct}|$ takes place at $N\sim 136$ for the CEDFs PC-PK1,
DD-ME2 and DD-PC1 and at $N=138$ for NL3* (Fig.\ \ref{massdependence}).
For these functionals the maximum $|\Delta E^{oct}|$ values vary from
around 1 MeV for NL3* and PC-PK1 up to 2 MeV for DD-ME2. The DD-ME$\delta$ 
functional does not predict octupole deformation for the nuclei of 
interest which contradicts both to experimental data (see Ref.\  
\cite{BN.96}) and the predictions of other models (see below).

  Experimental data suggest that in the Ra isotopes the maximum effect of octupole
deformation is seen at $N\sim 136$ \cite{BN.96}.
For example, the $N=136$ isotope has the lowest energy of the $1^-$
bandhead of the negative parity band. There are some differences in the
predictions  of the various models for the range of nuclei with octupole
deformation and for the neutron numbers at which the maximum gain in binding
due to octupole deformation takes place. For example, the MM
calculations based on folded Yukawa \cite{MBCOISI.08} (see also
Table \ref{table-global}) and Woods-Saxon potentials \cite{NORDLMR.84}
predict octupole deformation in the $N=130-138$ and $N=134-138$
isotopes, respectively. In these models, the maximum gain in binding
due to octupole deformation takes place at $N=132$ and $N=136$,
respectively. Note that the results obtained with a Woods-Saxon
potential are closer to experimental data \cite{BN.96}. The HFB
calculations with the D1S Gogny force\ \cite{ER.89} and the 
Barcelona-Catania-Paris (BCP) energy density functional 
\cite{RBSV.10} show that there exists non-vanishing
octupole deformation for the Ra isotopes with $N=130-140$ and with $N=130-142$,
respectively. In both cases the maximum gain in binding energy due
to octupole deformation takes place at $N=134$.

\subsubsection{Th isotopes}

  The evolution of the topology of the PESs of the Th isotopes is
presented in Fig.\ \ref{Th_DD-PC1} as a function
of neutron number. It is similar to the
one discussed above for the Ra isotopes. Well pronounced octupole minima
exist in the $N=136$ and 138 Th isotopes.

  The calculations with the DD-PC1 and DD-ME2 functionals suggest
that in whole region under study the Th isotopes are characterized
by the strongest octupole deformation effects (see Fig.\ \ref{massdependence}).
However, this is not the case for PC-PK1, where the octupole
deformation effects are comparable in the Th and U nuclei, and for
NL3*, where they are more pronounced in the U isotopes. In
the Th isotopes the maximum of $|\Delta E^{oct}|$ is located at $N=136$
for DD-PC1, DD-ME2, and PC-PK1 and at $N=138$ for NL3*
(Fig.\ \ref{massdependence}). The predictions of the
DD-ME$\delta$ functional disagree with all other model calculations
and experimental data.

\begin{figure*}
  \includegraphics[angle=0,width=5.9cm]{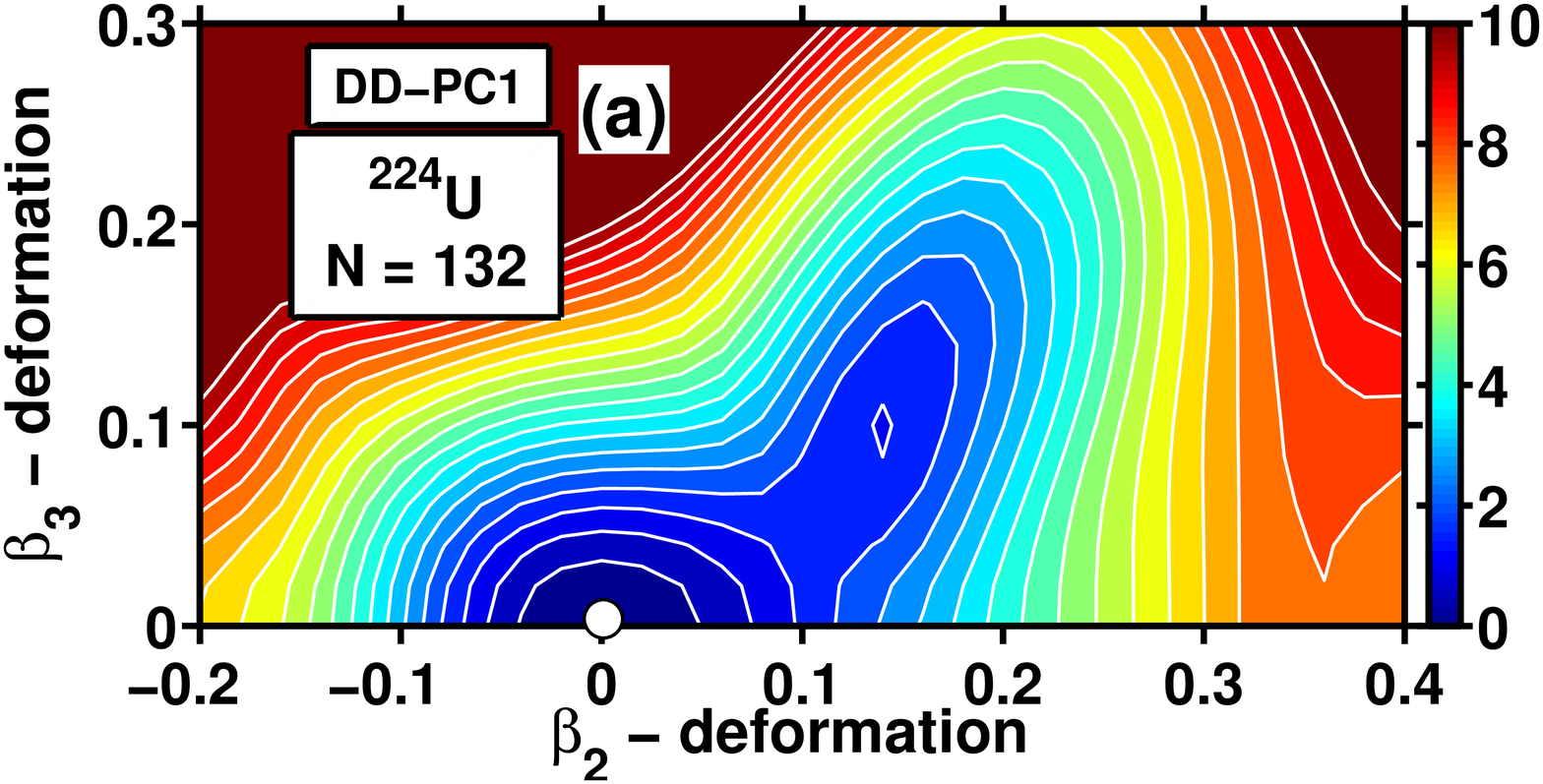}
  \includegraphics[angle=0,width=5.9cm]{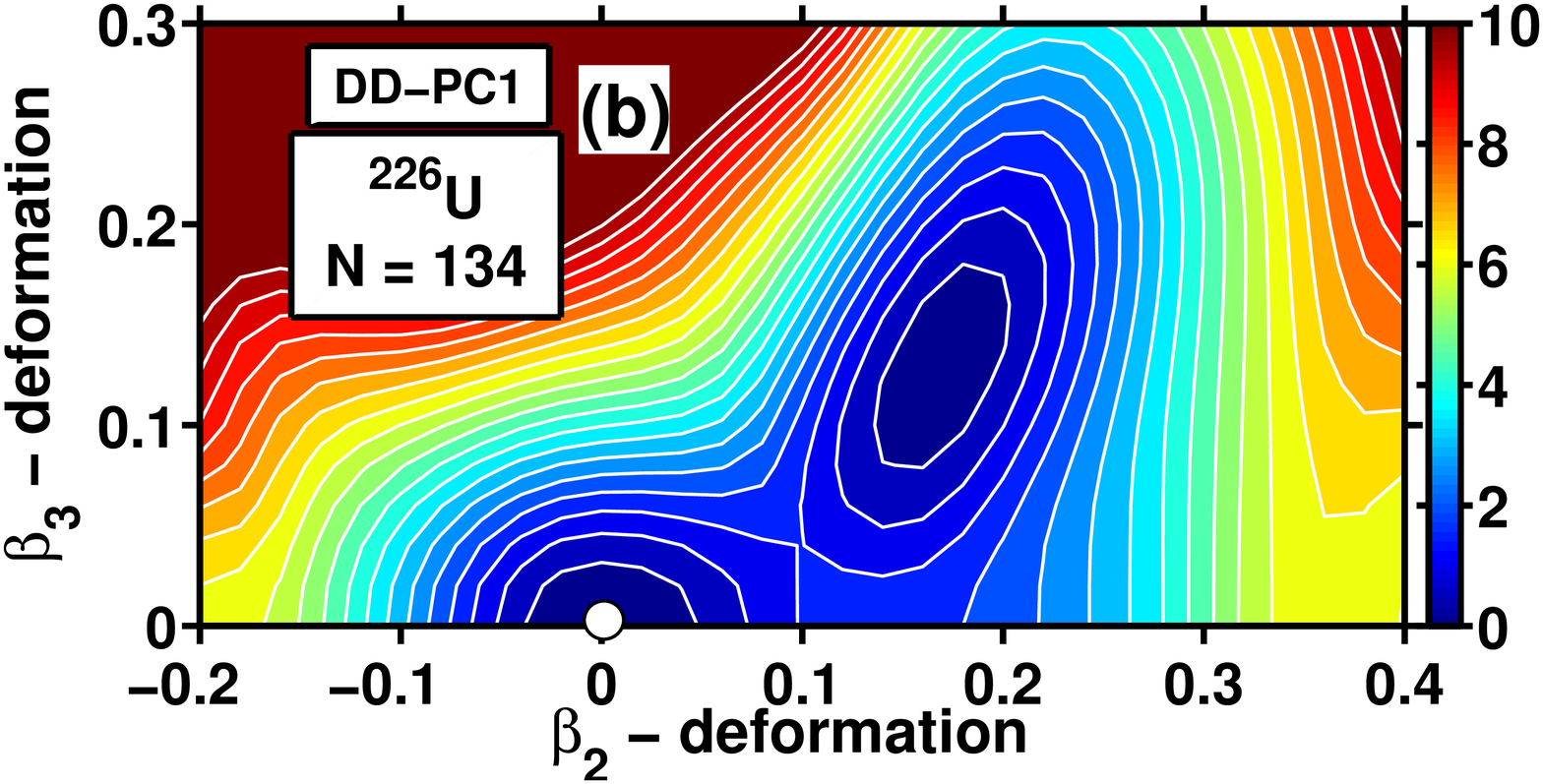}
  \includegraphics[angle=0,width=5.9cm]{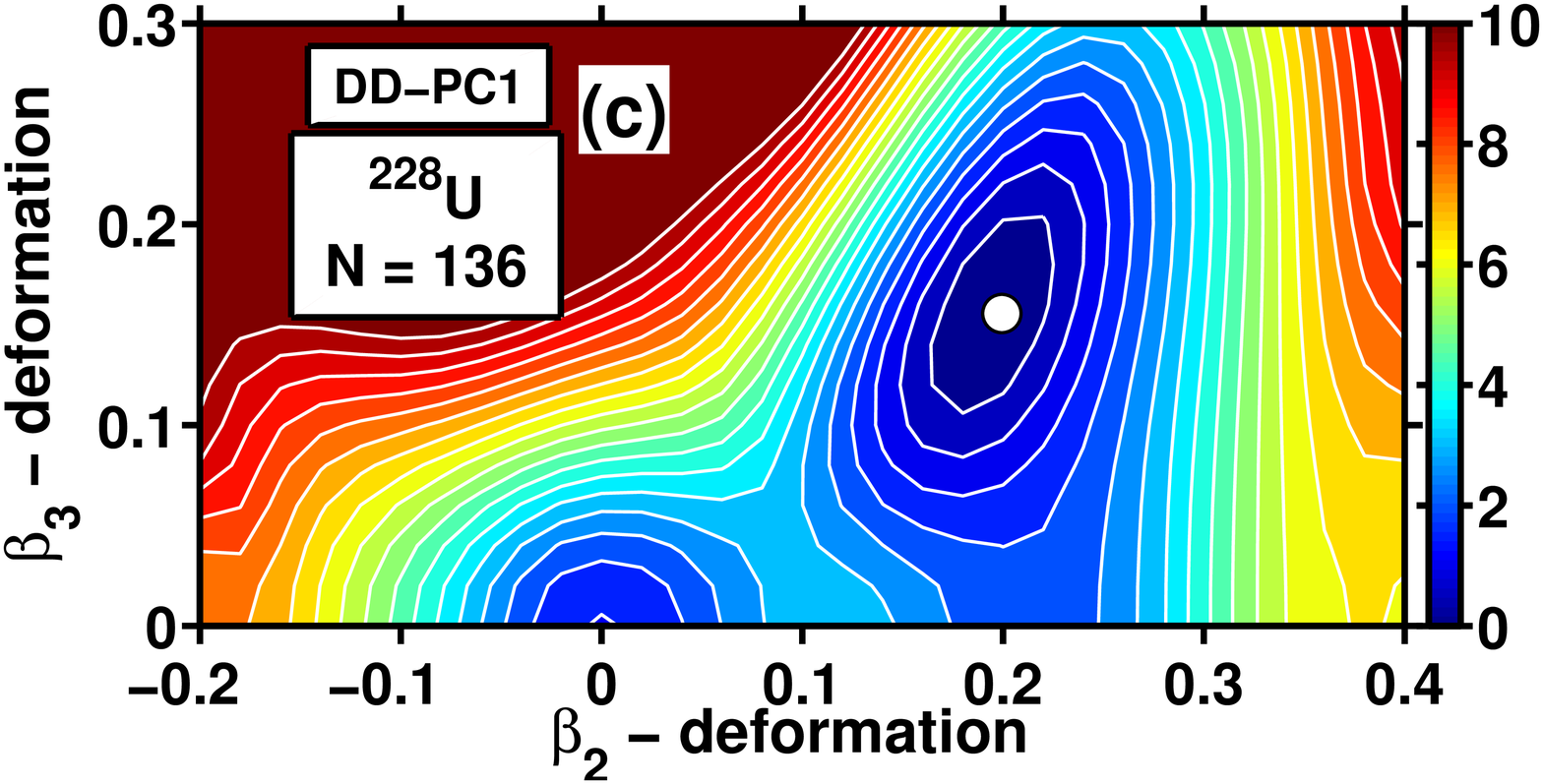}
  \includegraphics[angle=0,width=5.9cm]{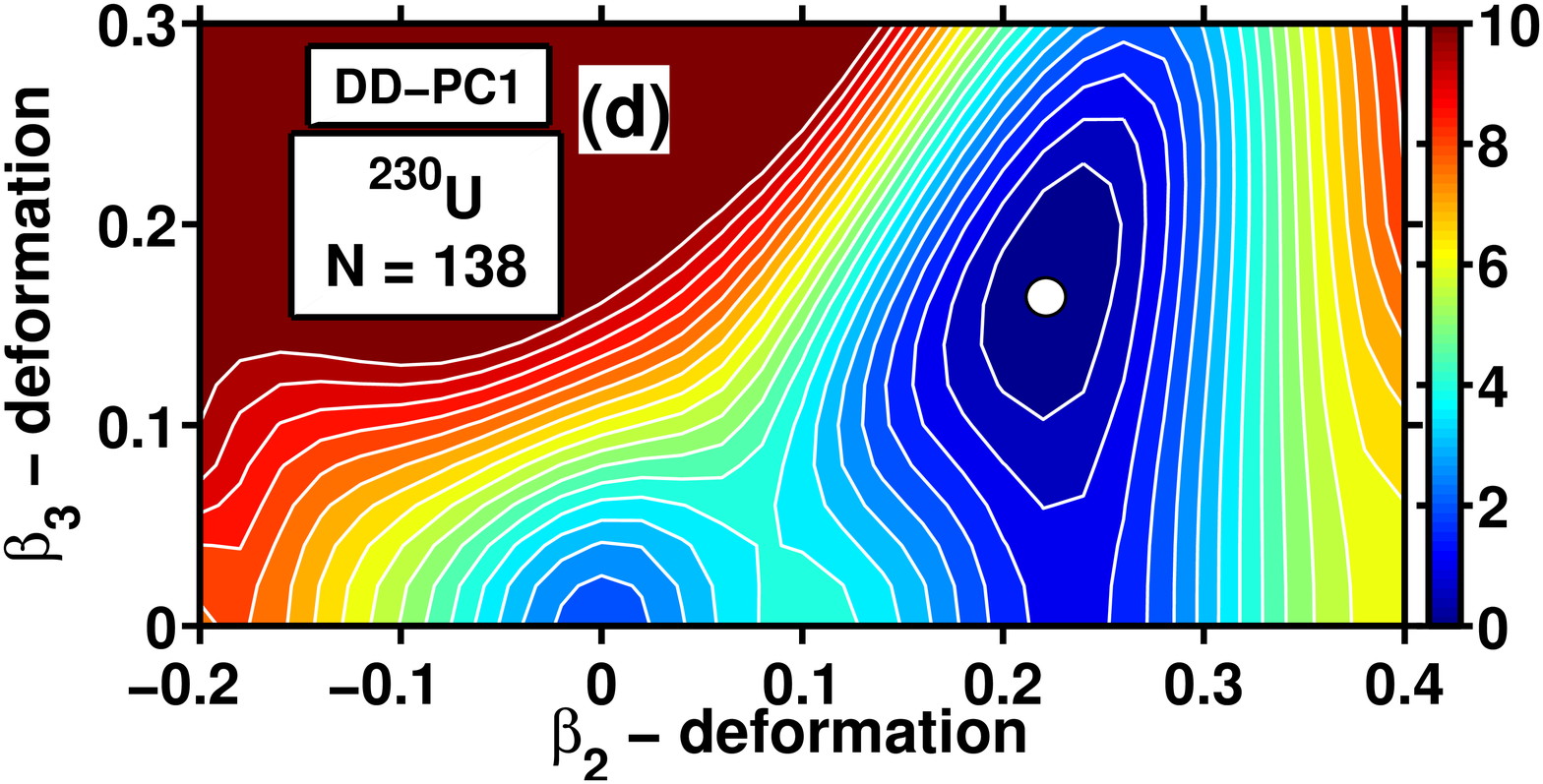}
  \includegraphics[angle=0,width=5.9cm]{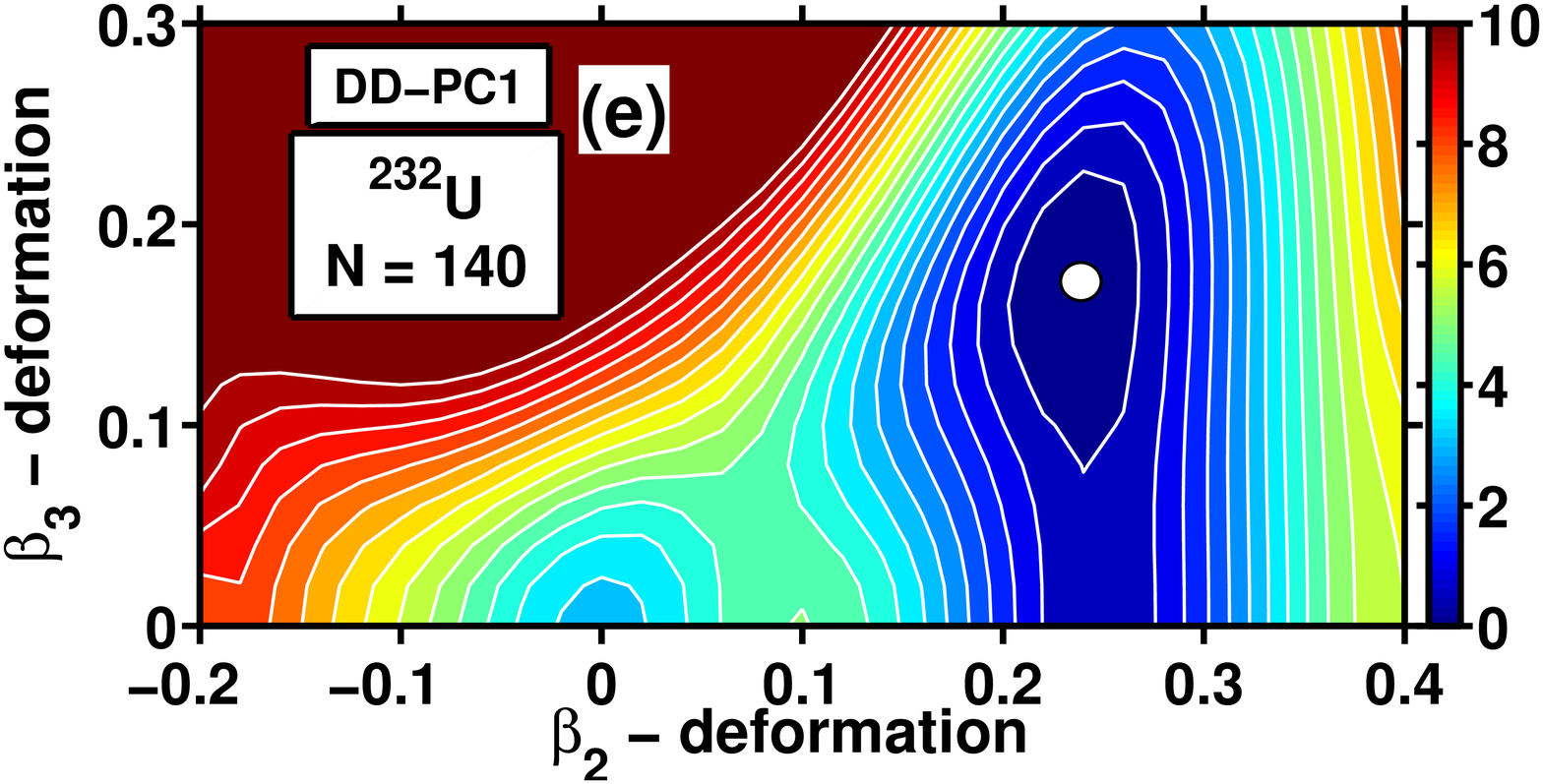}
  \includegraphics[angle=0,width=5.9cm]{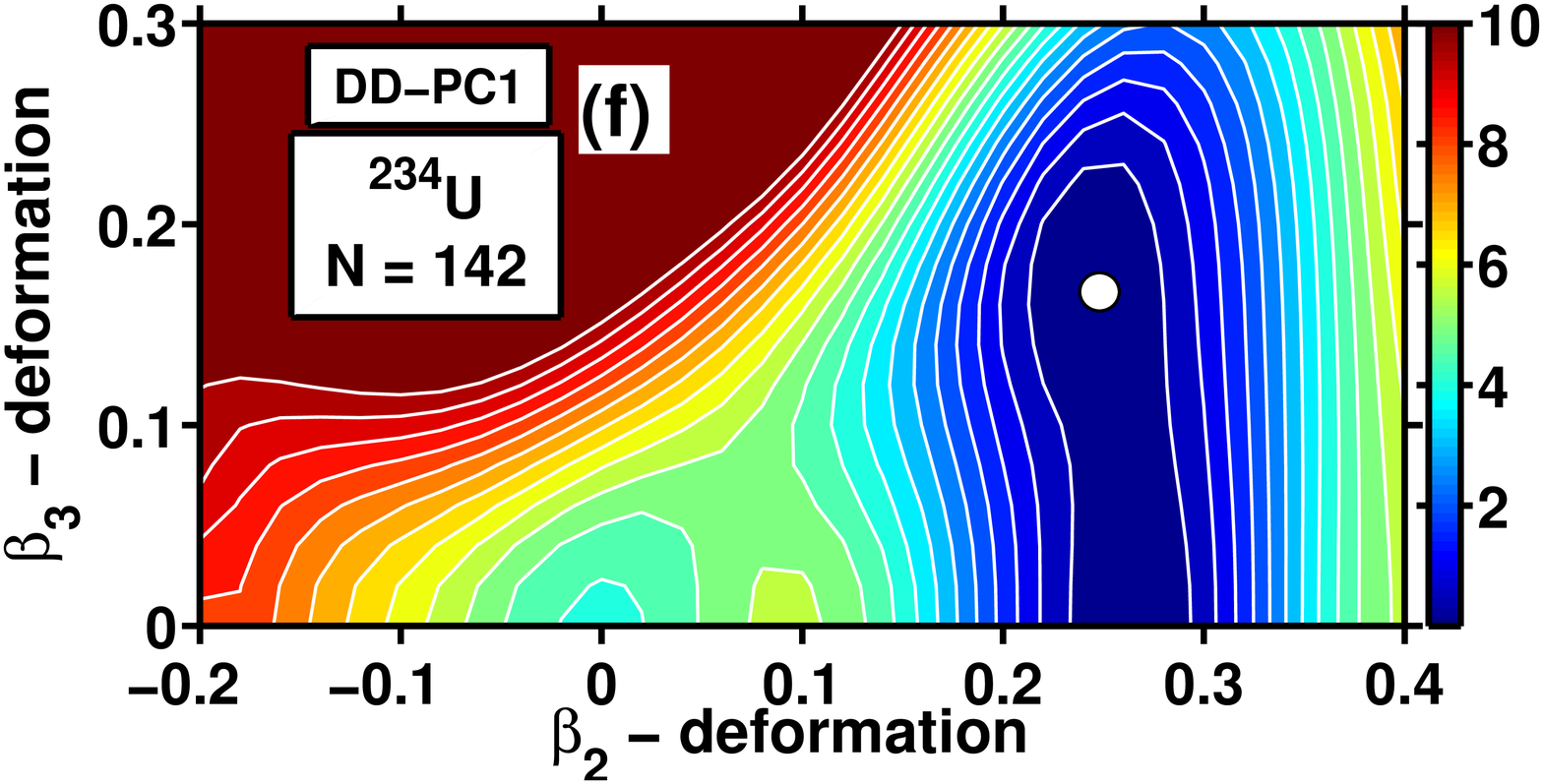}
  \caption{(Color online) The same as Fig.\ \ref{Rn_DD-PC1},
           but for the U isotopes.}
\label{U_DD-PC1}
\end{figure*}

  Experimental data suggest that the maximal effects of octupole
deformation in the Th isotopes are seen at $N\sim 136$ \cite{BN.96}.
For example, the $N=136$ isotope has the lowest energy of the $1^-$
bandhead of the negative parity band. The MM calculations based on folded
Yukawa \cite{MBCOISI.08} and Woods-Saxon \cite{NORDLMR.84} potentials
predict octupole deformation in the $N=130-138$ (see Table
\ref{table-global}) and $N=132-138$ (see Ref.\ \cite{NORDLMR.84}) isotopes,
respectively. In these models, the maximum gain in binding due to
octupole deformation takes place at $N=132$ and $N=134$, respectively.
The results obtained with Woods-Saxon potential are closer to experimental
data \cite{BN.96}. The HF+BCS calculations with the Gogny D1S force \cite{ER.89}
predict octupole deformation in the ground states of the $^{222-228}$Th 
nuclei with $N=132-138$; the maximum gain in binding due to octupole 
deformation is located at $N=132, 134$.

\subsubsection{U isotopes}

The PESs of the U isotopes calculated with DD-PC1 are shown in Fig.\ \ref{U_DD-PC1} 
and Fig.\ \ref{U_DD-PC1_rot} below. The spherical minimum is the lowest in energy for
the  isotope $^{224}{\rm U}$ with $N=132$. The coexistence of spherical and octupole
deformed minima is clearly seen in the $N=134$ and $N=136$ isotopes. However,
at higher neutron number the potential energy landscape is dominated by 
an octupole deformed minimum which becomes extremely soft above $N=140$. At and 
above $N=148$ octupole deformation vanishes and only a quadrupole deformed minimum
is present (see Fig.\ \ref{U_DD-PC1_rot} below).

  The maximum gain in binding energy due to octupole deformation
takes place at $N\sim 136$ with $|\Delta E^{oct}| \geq 1.0$
MeV for $N=134, 136$ and 138 for DD-ME2 and PC-PK1 and for
$N=136$ and 138 for NL3* and DD-PC1 (Fig.\ \ref{massdependence}). Again, the results 
for DD-ME$\delta$ are in contradiction with all other functionals and 
with experiment. An alternative parity rotational band, indicative of static 
octupole deformation, is observed in the nucleus $^{226}$U with $N=134$ in 
Ref.\ \cite{U236.98}. The experimental data on $^{228}$U is restricted to the 
ground and $2^+$ states \cite{Eval-data}; so no decision about the presence 
of static octupole deformation is possible in this nucleus. Octupole vibrational 
bands with $1^-$ bandheads, located at extremely low excitation energies of 
366.65 keV and 562.19 keV, are observed in $^{230}$U and $^{232}$U, respectively 
\cite{Eval-data}. They are indicative of extreme octupole softness of the 
potential energy surfaces.

  The MM calculations based on different phenomenological potentials and on 
different liquid
drop formulas predict octupole deformed U nuclei at $N=128-134$ (see Fig.\ 2 in
Ref.\ \cite{WLX.12}). In these calculations the maximum gain in binding due to
octupole deformation takes place at lower $N$ as compared with CDFT. 
For example, it is located at $N=132$ (Table \ref{table-global})
in the MM calculations of Ref.\ \cite{MBCOISI.08}. The HFB calculations with 
several versions of the Gogny force show that the U isotopes with $N=130-138$ 
have non-zero octupole deformation \cite{RR.12}.

\subsubsection{Pu isotopes}

\begin{figure*}
  \includegraphics[angle=0,width=5.9cm]{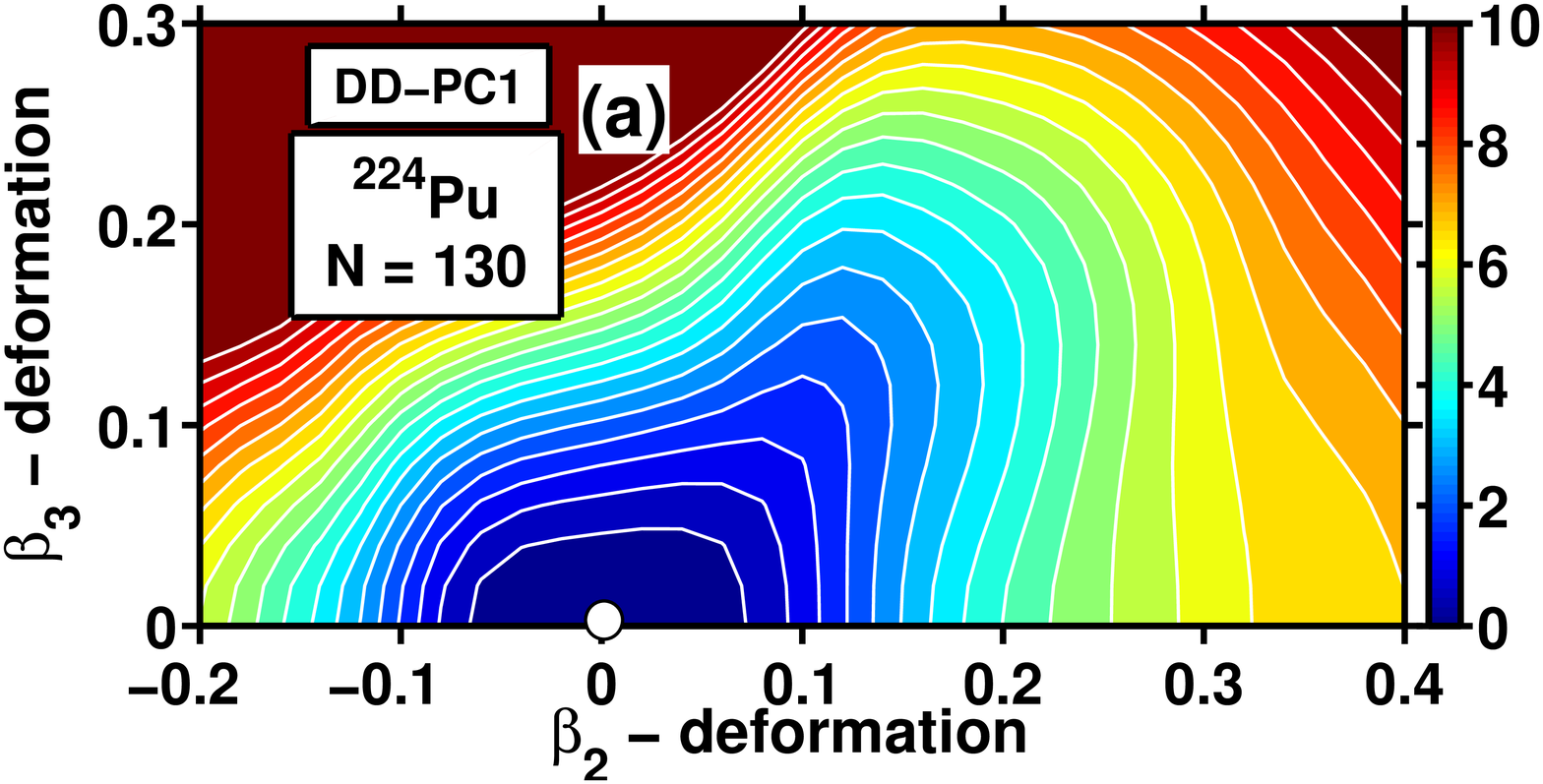}
  \includegraphics[angle=0,width=5.9cm]{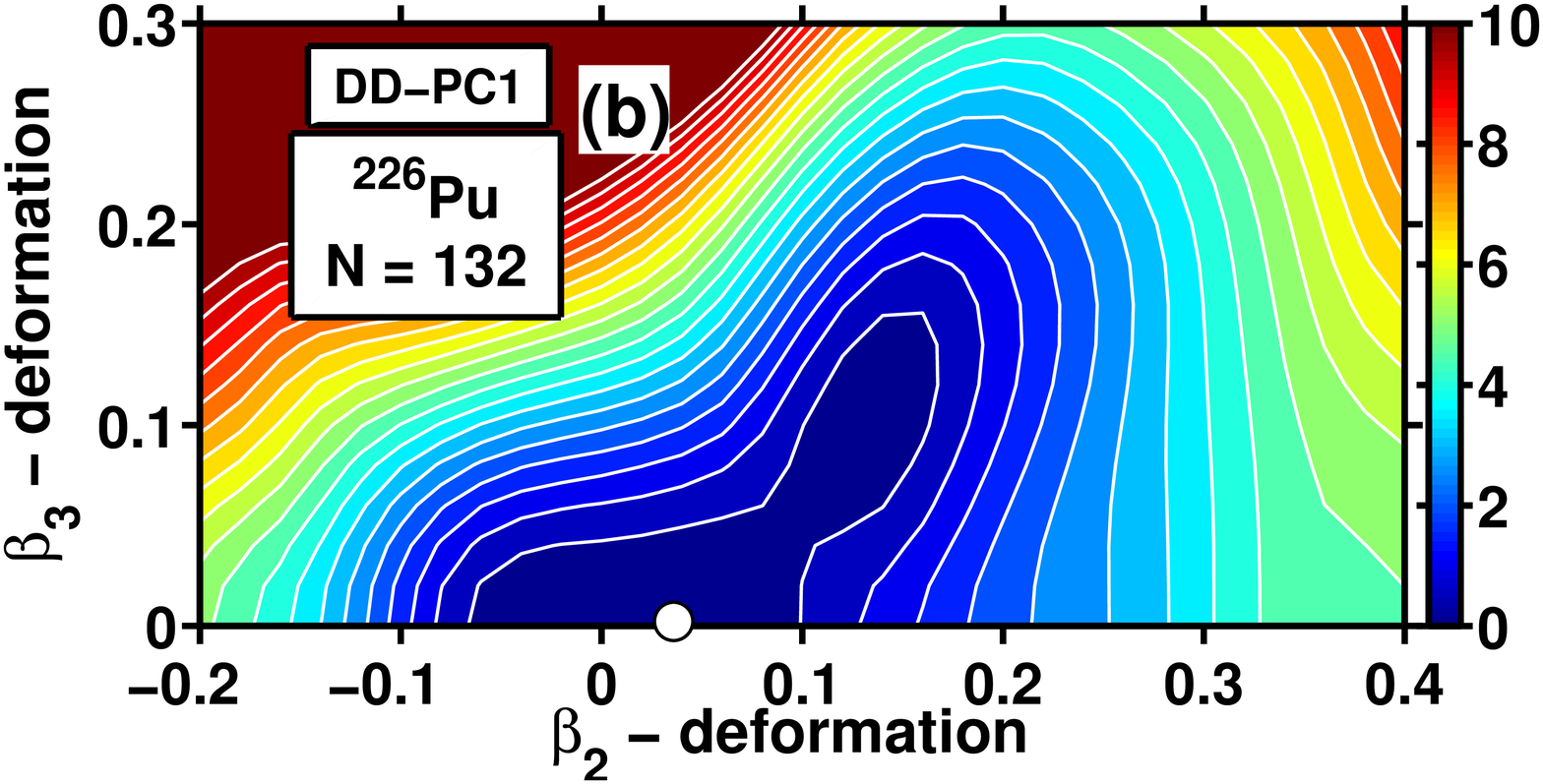}
  \includegraphics[angle=0,width=5.9cm]{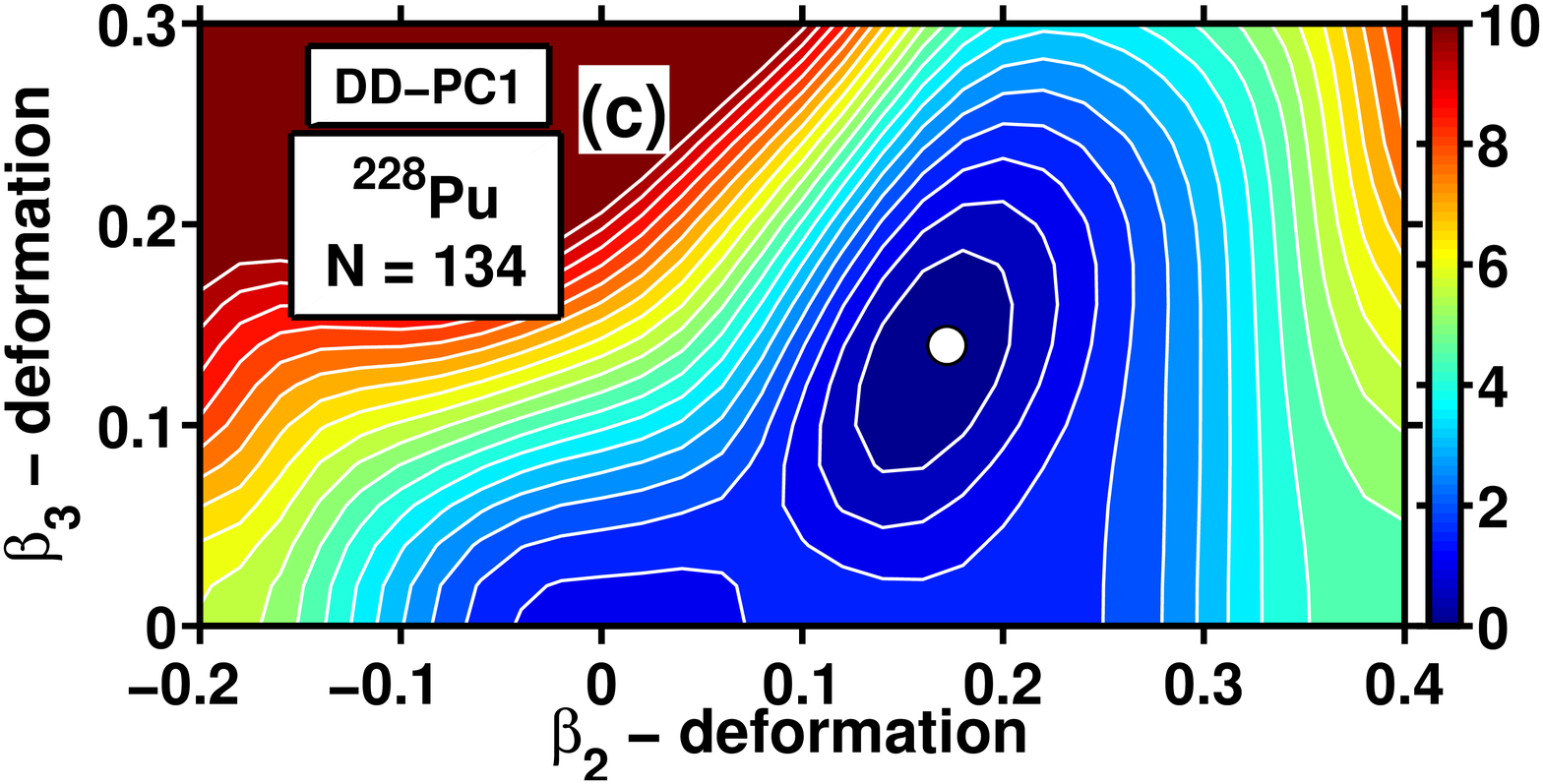}
  \includegraphics[angle=0,width=5.9cm]{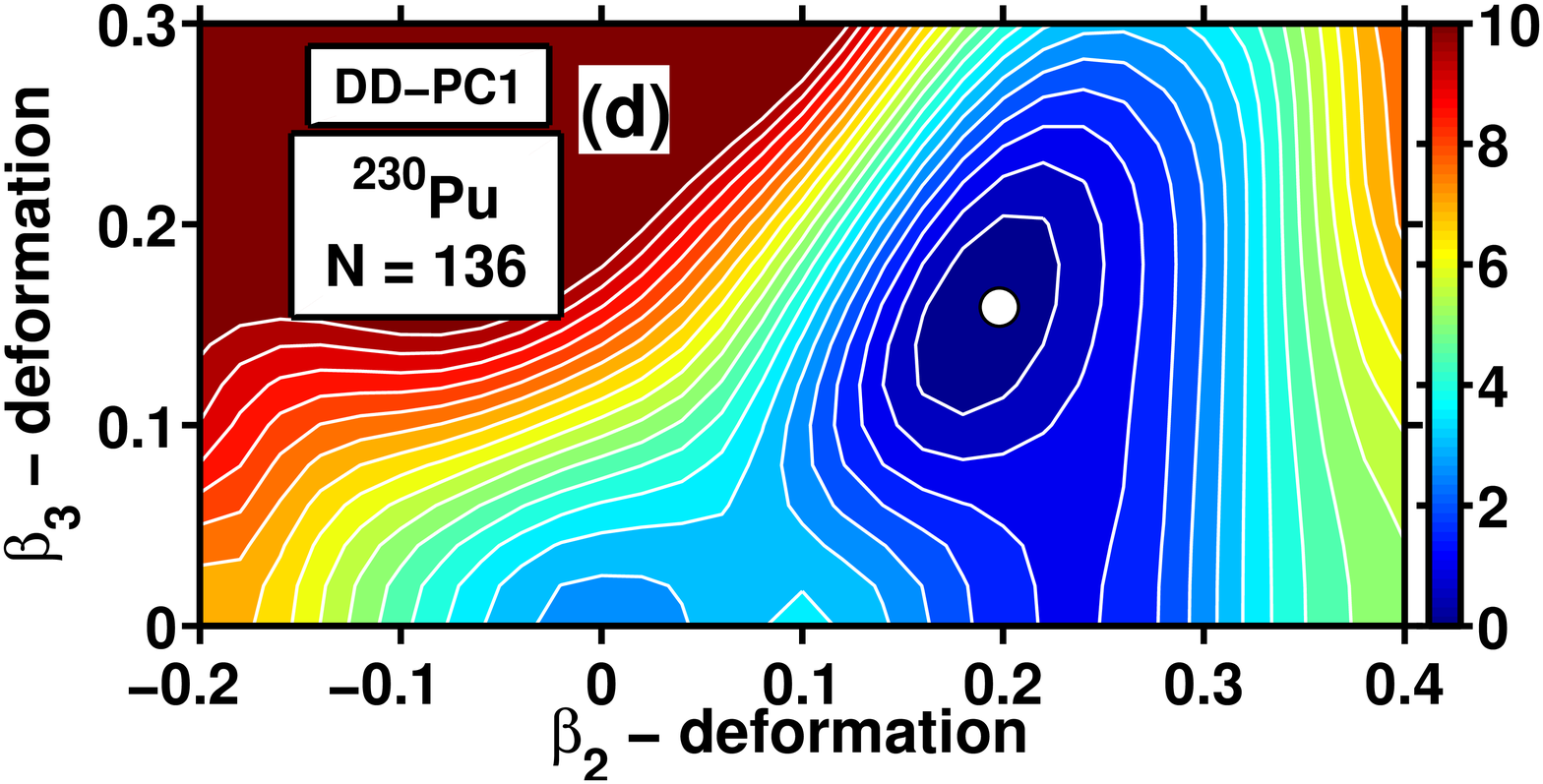}
  \includegraphics[angle=0,width=5.9cm]{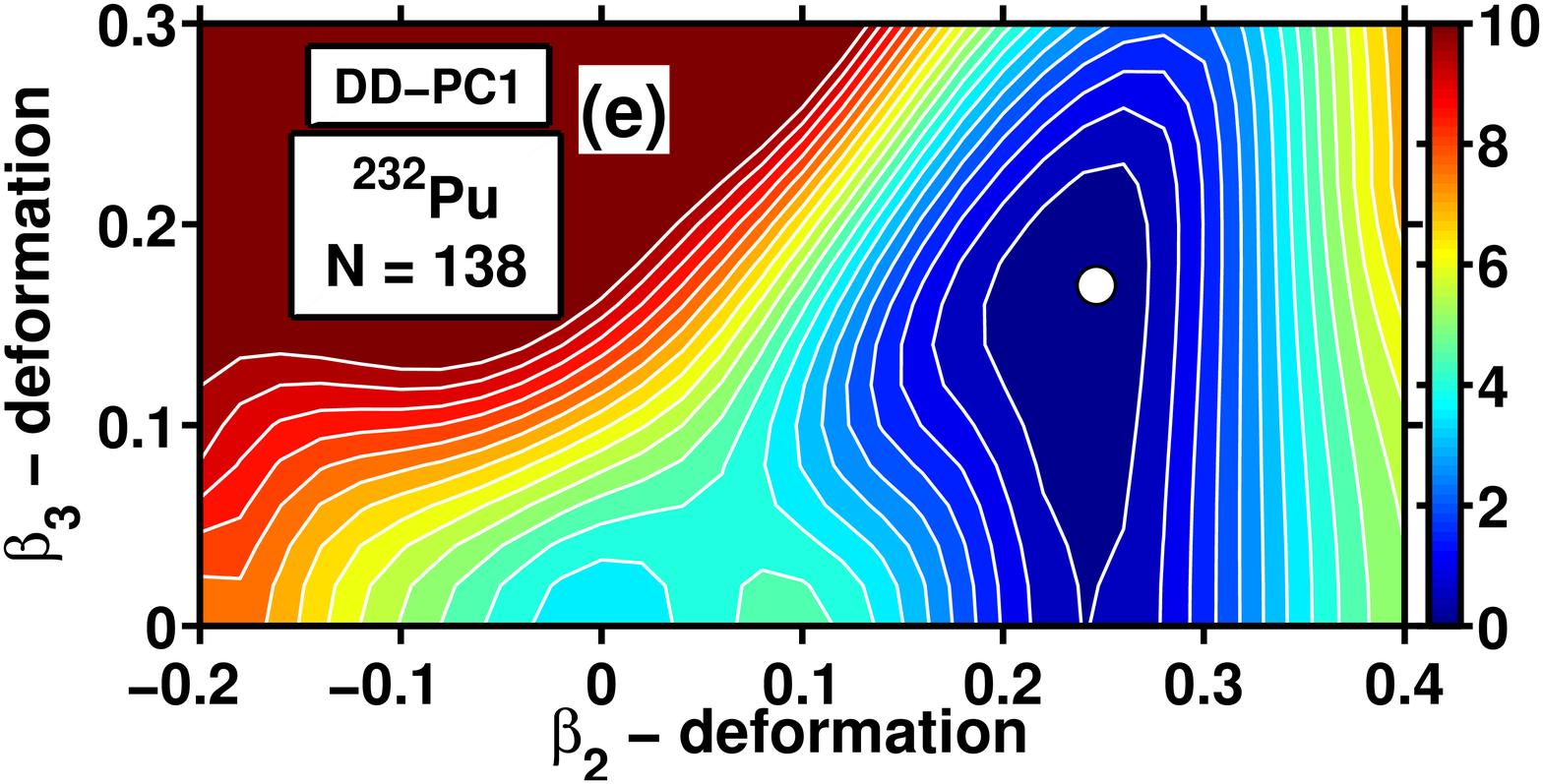}
  \includegraphics[angle=0,width=5.9cm]{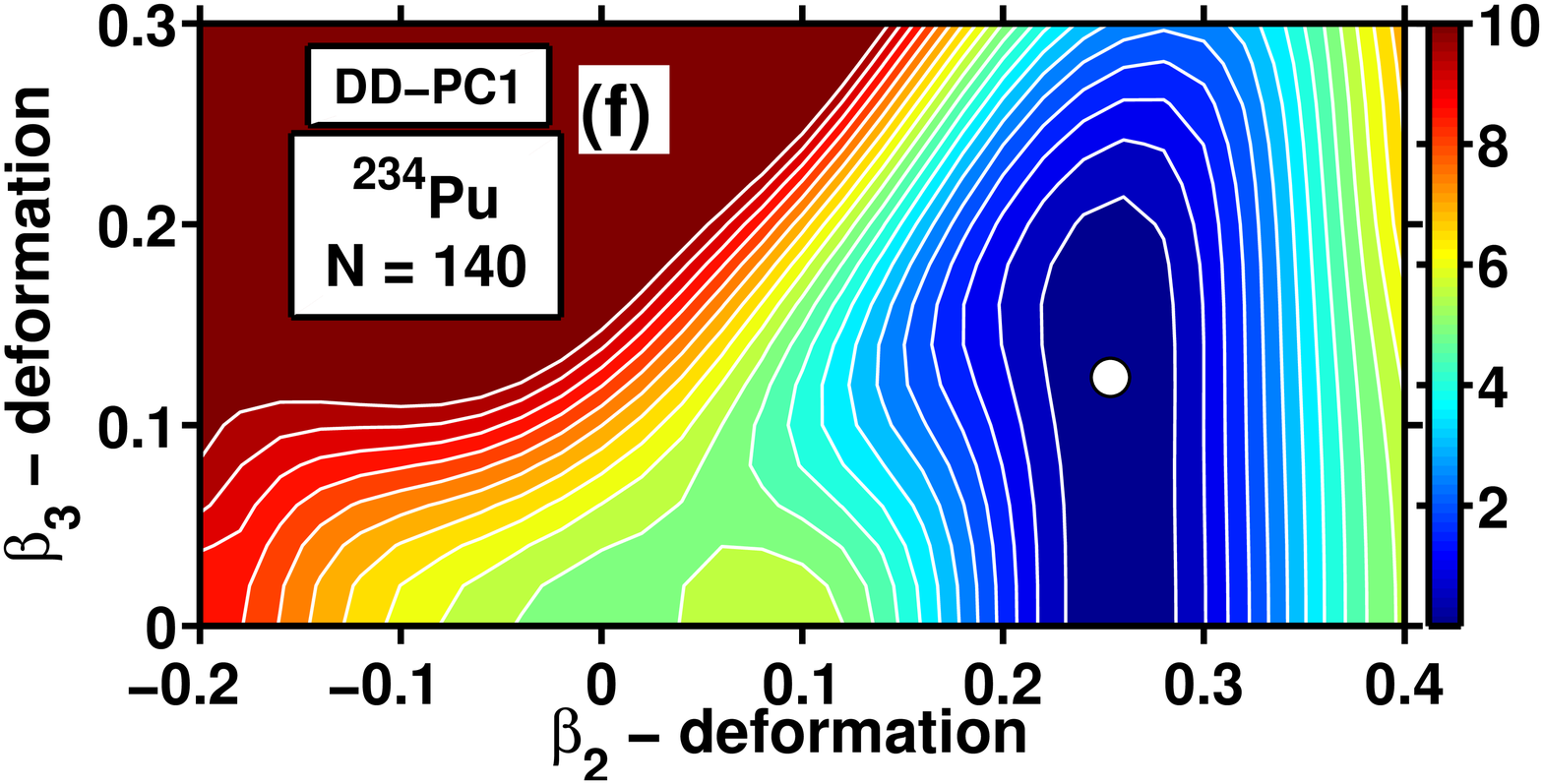}
  \includegraphics[angle=0,width=5.9cm]{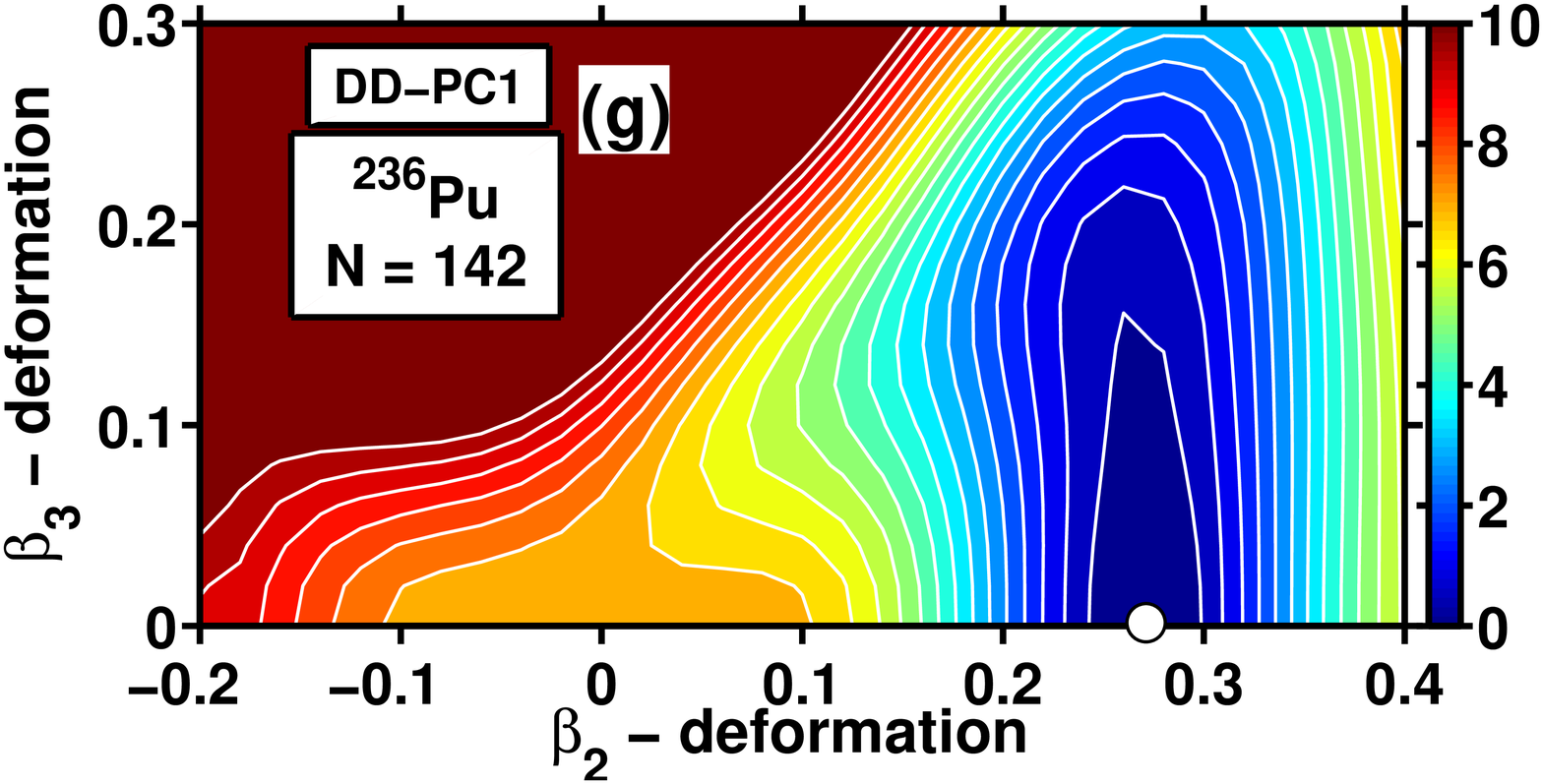}
  \caption{(Color online) The same as Fig.\ \ref{Rn_DD-PC1},
           but for the Pu isotopes.}
\label{Pu_DD-PC1}
\end{figure*}

\begin{figure*}
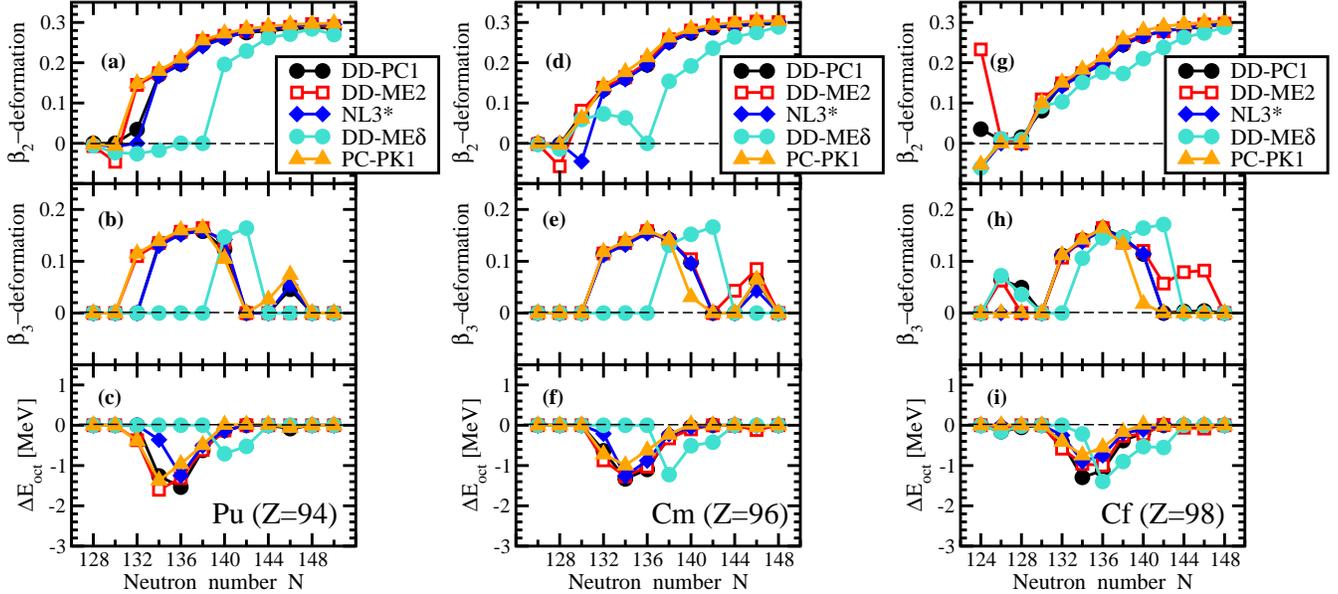

  \includegraphics[angle=0,width=5.8cm]{fig-7-a-new.eps}
  \includegraphics[angle=0,width=5.8cm]{fig-7-b-new.eps}
  \includegraphics[angle=0,width=5.8cm]{fig-7-c-new.eps}
  \caption{(Color online) The same as Fig.\ \ref{massdependence},
           but for the Pu, Cm and Cf isotopes.}
\label{massdependence-Pu-etc}
\end{figure*}

  The PESs of the Pu isotopes
calculated with DD-PC1 are shown in Fig.\ \ref{Pu_DD-PC1} and Fig.\
\ref{Pu_DD-PC1_NL3s} below. The spherical minimum is the lowest in energy for
the isotope $^{224}{\rm Pu}$ with $N=130$. The coexistence of a spherical (which is the
lowest  in energy) and an octupole deformed minima is clearly seen in the
$N=132$ isotope. A well pronounced octupole deformed minimum exists in
the $N=134$ and 136 isotopes. However, at higher neutron number the PESs become
extremely soft in octupole deformation so that the position of the minimum
(finite octupole deformation at $N=140$ and $146$ or vanishing octupole deformation 
at $N=142, 144$ and 148) depends on fine details of the underlying single-particle 
structure. Note that the energy gain due to octupole deformation is very small for 
$N=146$, namely, less than 100 keV (Table \ref{table-global}). The maximum gain in
binding energy due to octupole deformation is found at $N=134$ for PC-PK1 and 
DD-ME2  and at $N=136$ for NL3* and DD-PC1 (Fig.\ \ref{massdependence-Pu-etc}). 
Only in these isotopes the gain due to octupole deformation $|\Delta E^{oct}|$ is 
close to or exceeds $1.0$ MeV. The results for DD-ME$\delta$ are again in 
contradiction with all other functionals.

The predictions of CDFT differ from those of the non-relativistic models.
The HFB calculations with Gogny forces of Ref.\ \cite{RR.12} indicate
that the isotopes $^{224-232}$Pu with $N=130-138$ have finite octupole deformations
in the ground states; the maximum gain in binding due to octupole deformation
takes place at $N=132$. The MM calculations of Ref.\ \cite{MBCOISI.08}
predict octupole deformation only for the isotopes $^{222-228}$Pu ($N=128-134$)
with a maximum gain in binding due to octupole deformation at $N=130$ (see
Table \ref{table-global}).

At present, it is impossible to discriminate between the predictions of
these models because of the limitations of experimental data. Only the $0^+$ ground
state has been observed in the nuclei $^{228-234}$Pu with $N=134-140$ (which does
not allow to define the presence or absence of octupole deformation) and
no experimental data are available for lighter Pu isotopes \cite{Eval-data}.
On the other hand, octupole vibrational bands have been observed \cite{Eval-data} in
the isotopes $^{236-230}$Pu ($N=142-146$) with bandheads located at low
excitation energies of 698, 605 and 597 keV, respectively.
This suggests a substantial octupole softness of the potential energy surfaces of
these nuclei. Indeed, there are some indications of the stabilization of
octupole deformation at high spin in $^{240}$Pu \cite{Pu240,F.08} (see also
Sec.\ \ref{rotation}).

\subsubsection{Cm isotopes.}

\begin{figure*}
  \includegraphics[angle=0,width=5.9cm]{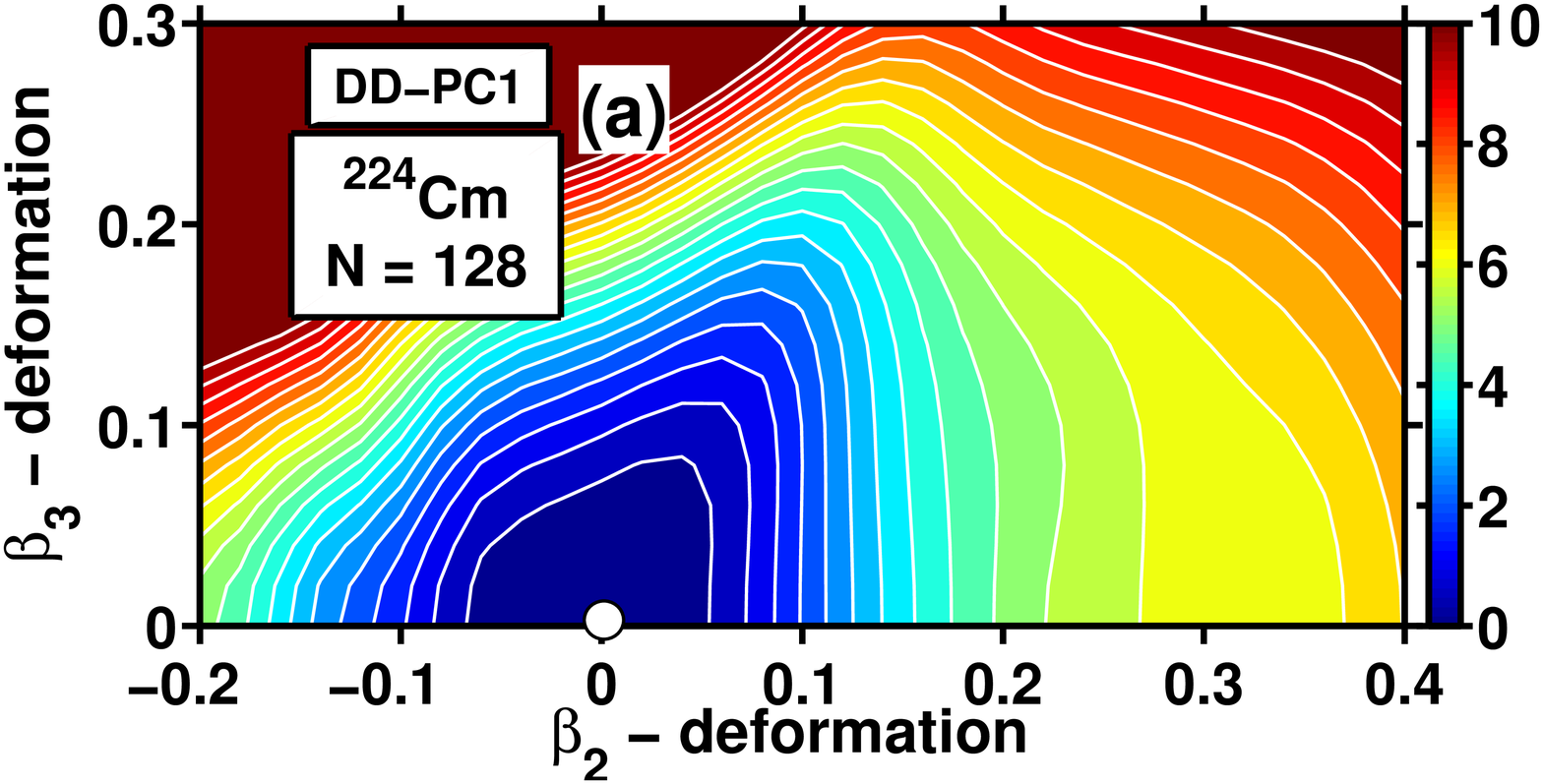}
  \includegraphics[angle=0,width=5.9cm]{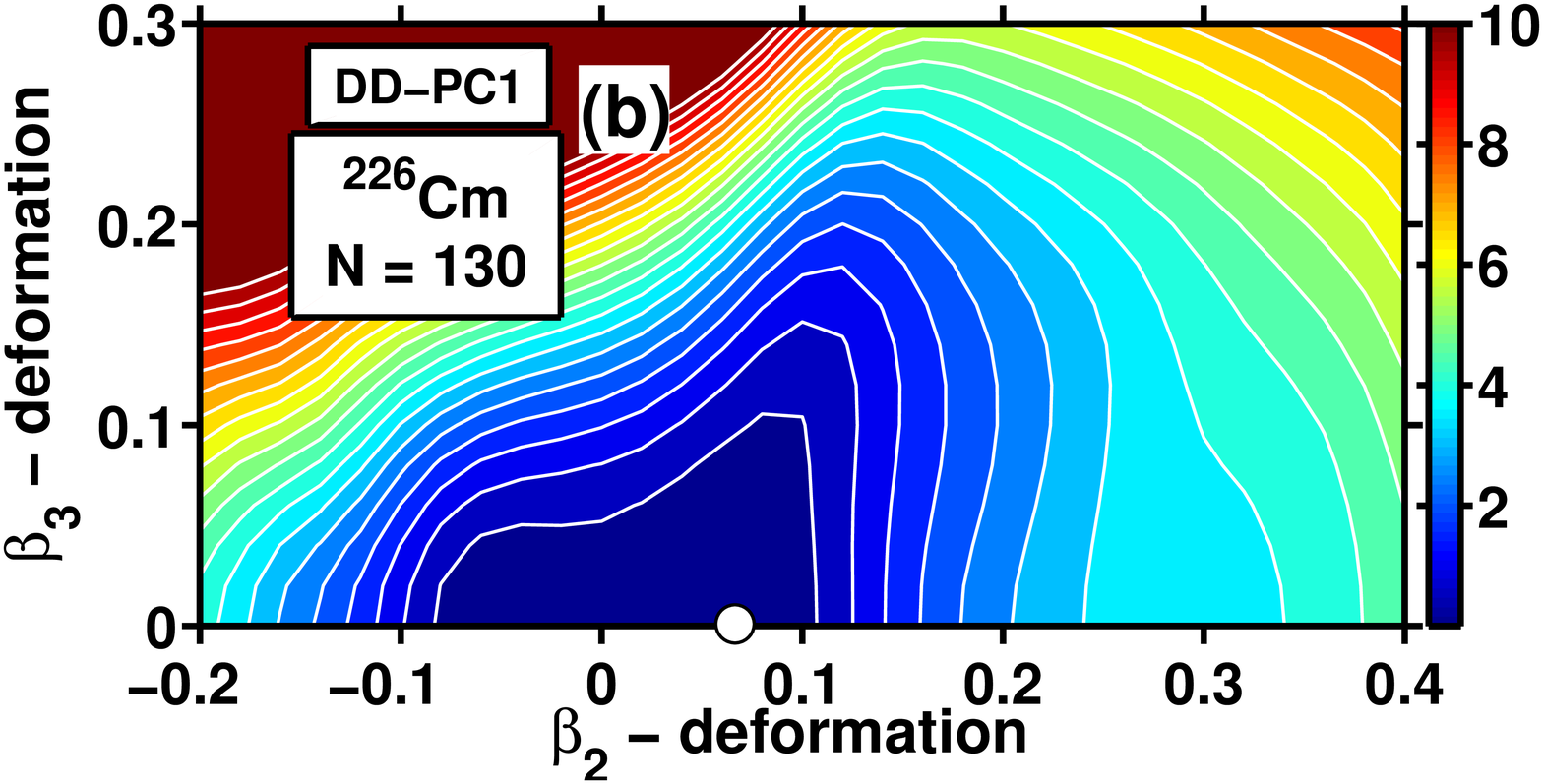}
  \includegraphics[angle=0,width=5.9cm]{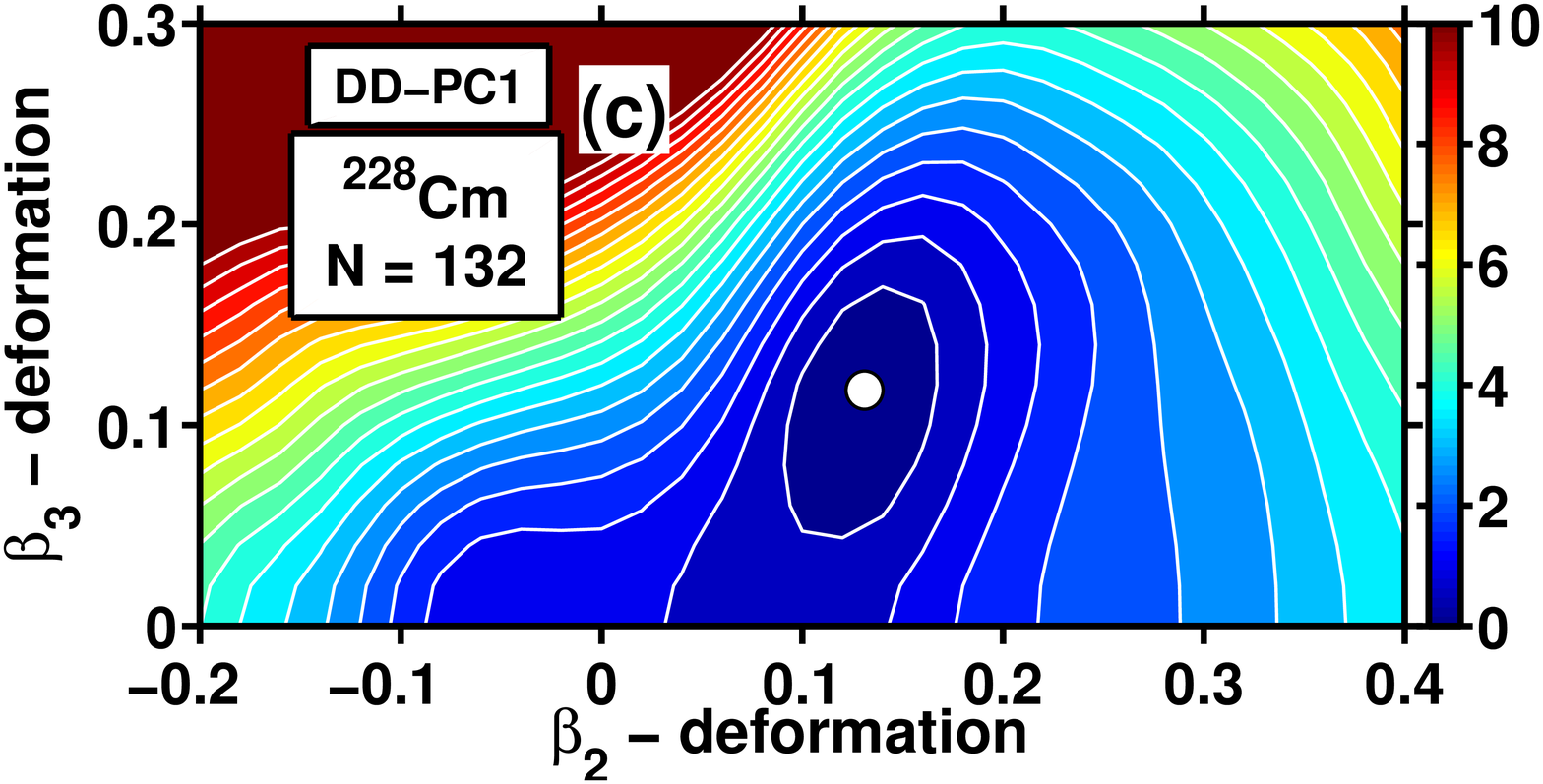}
  \includegraphics[angle=0,width=5.9cm]{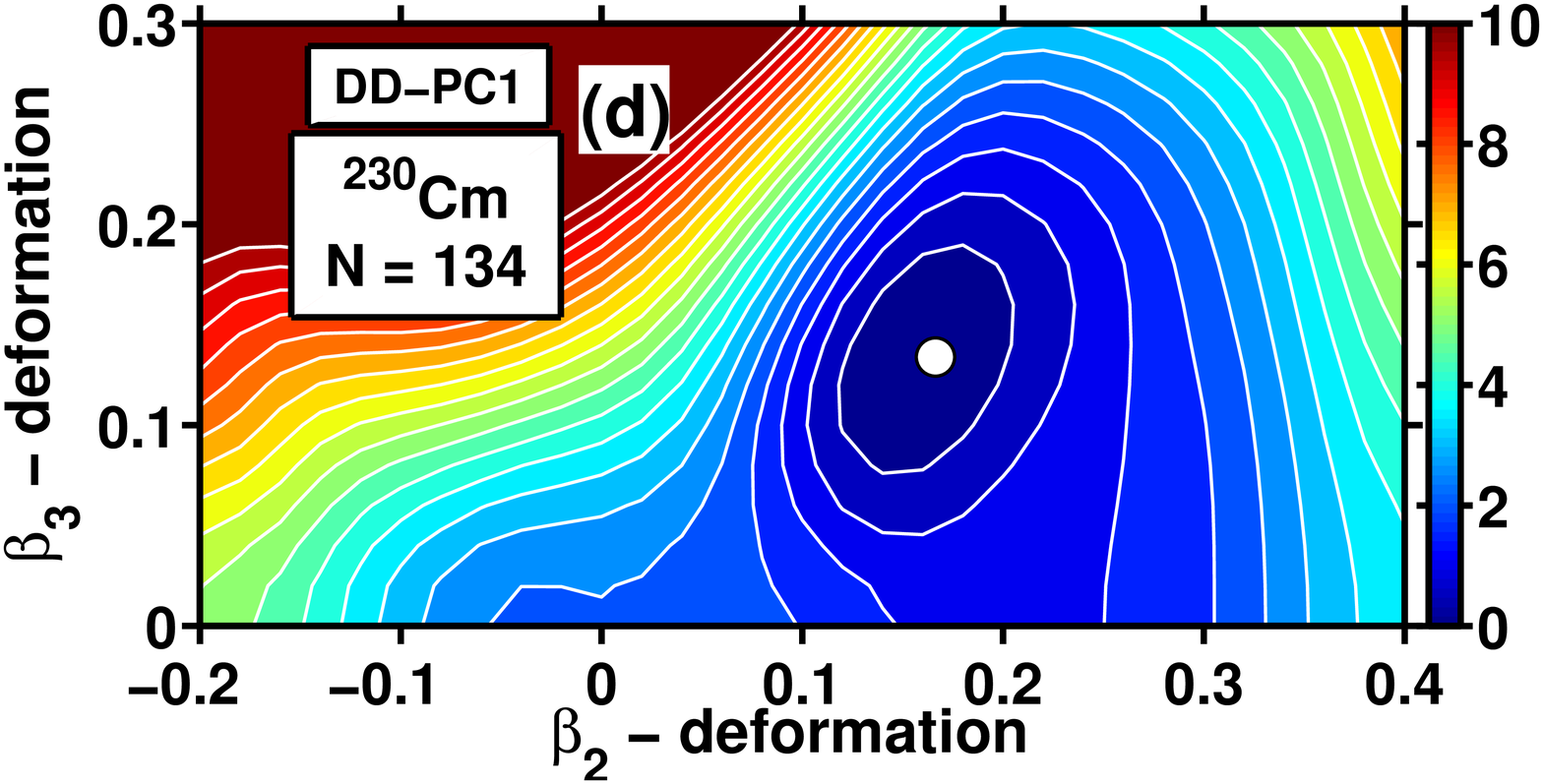}
  \includegraphics[angle=0,width=5.9cm]{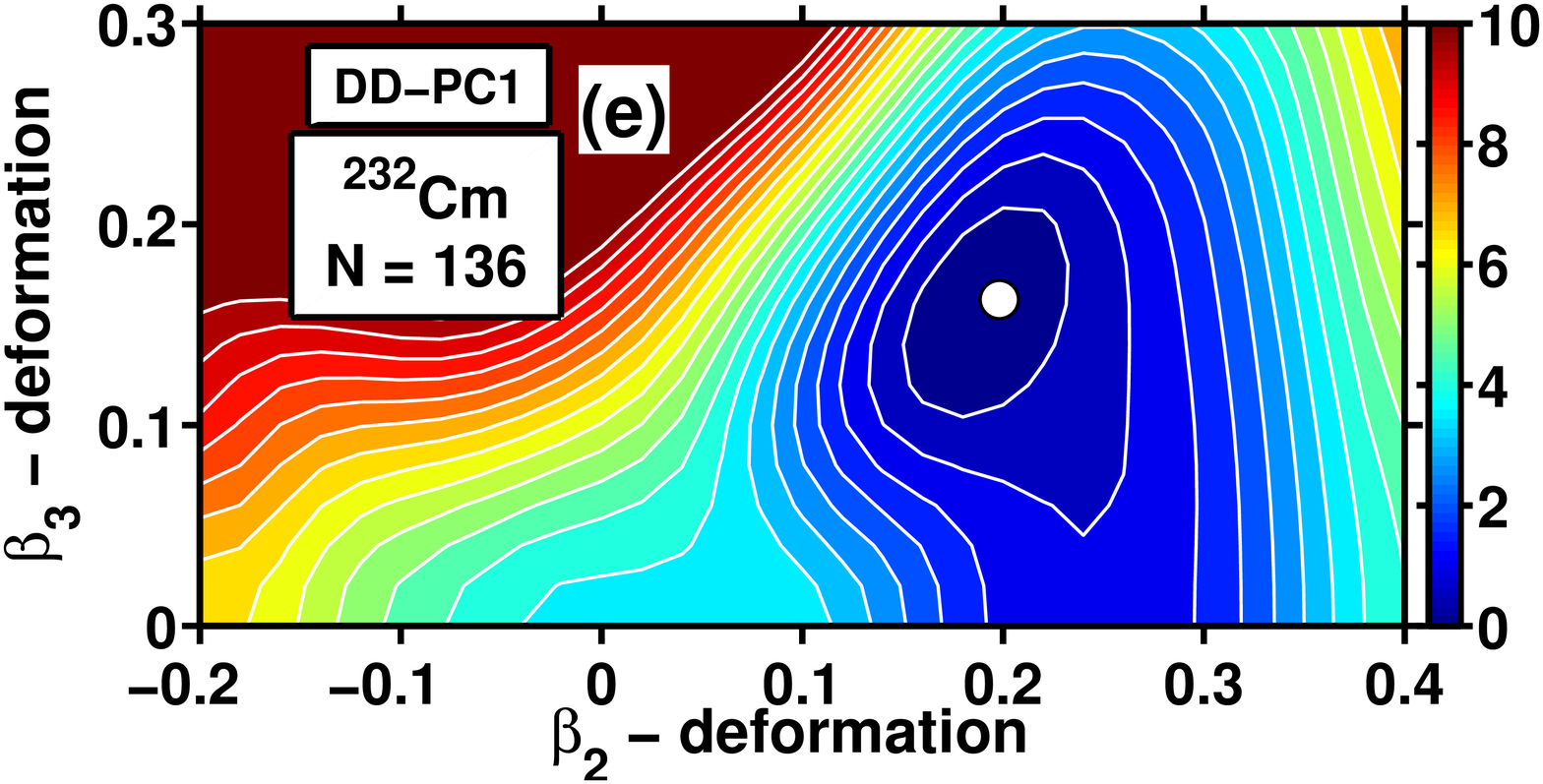}
  \includegraphics[angle=0,width=5.9cm]{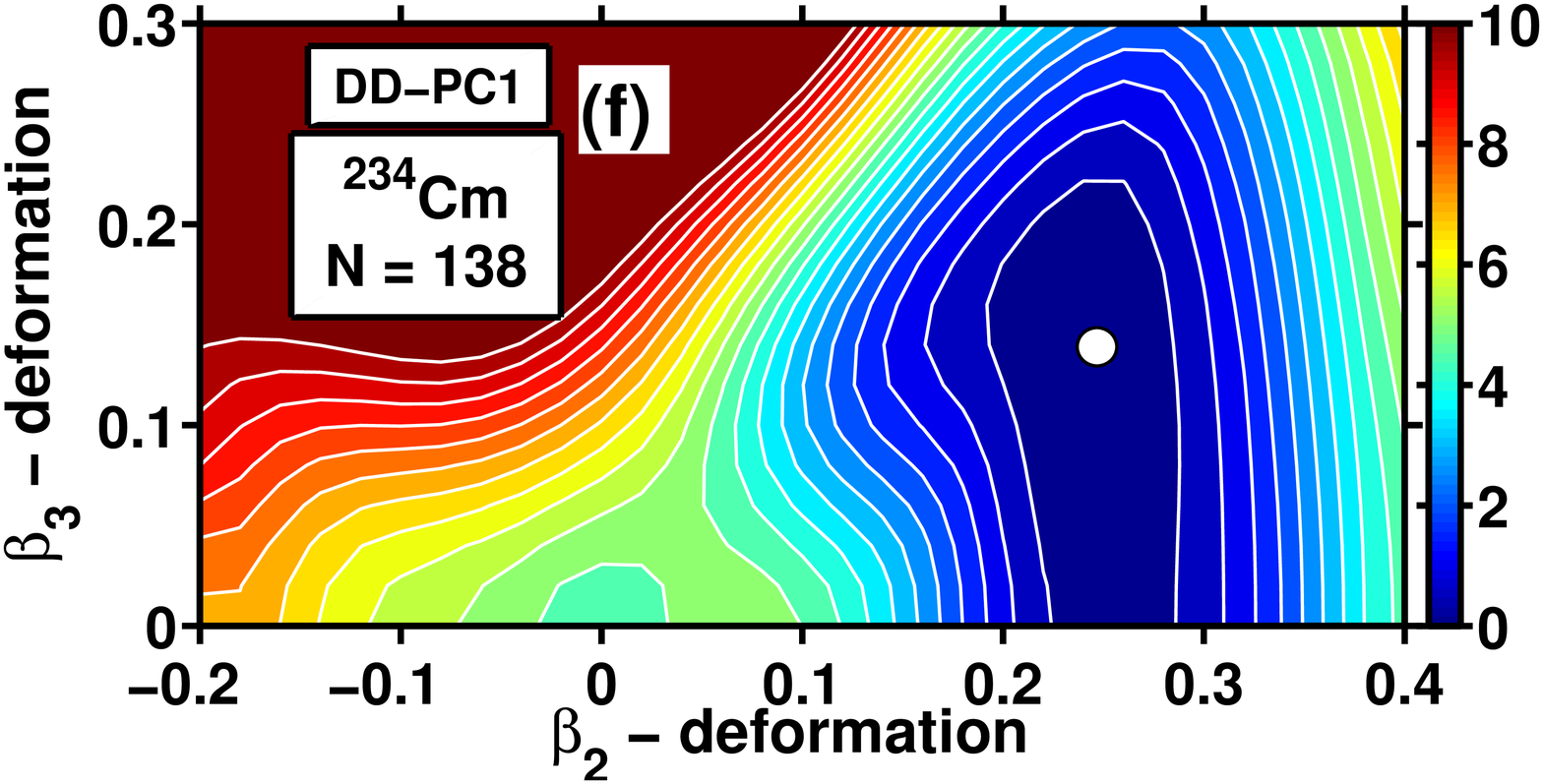}
  \includegraphics[angle=0,width=5.9cm]{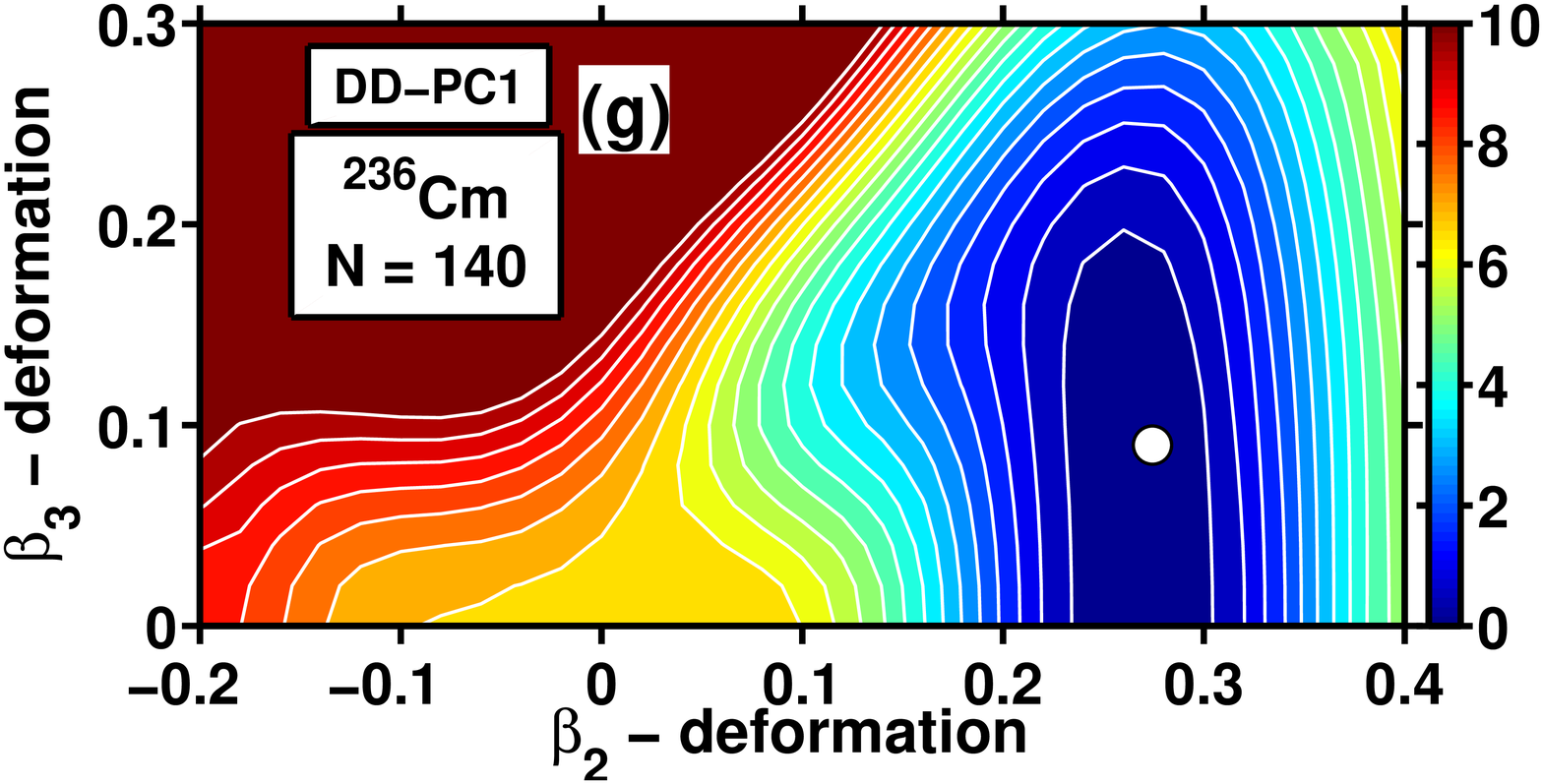}
  \caption{(Color online) The same as Fig.\ \ref{Rn_DD-PC1},
           but for the Cm isotopes.}
\label{Cm_DD-PC1}
\end{figure*}

PESs of Cm isotopes are presented in Fig.\ \ref{Cm_DD-PC1}. Spherical and
weakly deformed minima are seen in the isotopes $^{224,226}$Cm with $N=128-130$.
The increase of neutron number leads to the development of an octupole deformed
minimum which becomes especially pronounced in the nuclei $^{230,232}$Cm with $N=134,136$. 
With a further increase of neutron number the potential energy surfaces become extremely soft in octupole
direction. The maximum of the gain in
binding energy due to octupole deformation ($|\Delta E^{oct}| \sim 1.0$ MeV)
takes place at $N\sim 134$ for the NL3*, PC-PK1, DD-ME2 and DD-PC1 CEDFs
(Fig.\ \ref{massdependence-Pu-etc}). The results for DD-ME$\delta$ are in
contradiction with all other functionals.

 The HFB calculations with Gogny forces of Ref.\ \cite{RR.12} indicate
that the isotopes $^{226-230}$Cm with $N=130-134$ have non-zero octupole 
deformation in the ground states; the maximum gain in binding due to 
octupole deformation
of around several hundred keV takes place at $N=130$. The MM calculations of
Ref.\ \cite{MBCOISI.08}  predict octupole deformation only for the isotopes
$^{224-228}$Cm ($N=128-132$) with a maximum gain in binding due to octupole deformation
($|\Delta E^{oct}|\sim 0.8$ MeV) at $N=130$ (see Table \ref{table-global}).
Since experimental data do not exist for the Cm isotopes with $N\leq 136$ and
only ground states are observed in the nuclei $^{234,236}$Cm with $N=138, 140$ 
\cite{Eval-data}, there is no way to discriminate between the predictions of
the models.

\subsubsection{Cf isotopes}
\label{Cf-isotopes}

The PESs of the Cf isotopes are displayed in Fig.\ \ref{Cf_DD-PC1}. An unusual 
feature of the $^{224,226}$Cf isotopes with $N=126, 128$ is the presence of the 
minimum in the PES with almost zero quadrupole deformation and octupole 
deformation $\beta_3\sim 0.1$. However, the PESs are soft in octupole 
direction in $^{224}$Cf and both in quadrupole and octupole directions in $^{226}$Cf. 
An octupole deformed minimum starts to develop in $N=132$
$^{230}$Cf and becomes especially pronounced in $^{232,234}$Cf. A further increase of
neutron number leads to PES which are extremely soft in
octupole direction. The maximum of the gain in binding energy due to octupole
deformation ($|\Delta E^{oct}| \sim 1.0$ MeV) takes place either at $N=134$ or
at $N=136$
for NL3*, PC-PK1, DD-ME2 and DD-PC1 (Fig.\ \ref{massdependence-Pu-etc}).
The results for DD-ME$\delta$ are in contradiction with all other functionals.

The HFB calculations with Gogny forces of Ref.\ \cite{RR.12} indicate
that isotopes $^{228-232}$Cf with $N=130-134$ have non-zero octupole deformation
in the ground states. Dependent on the functional the maximum gain in
binding due to octupole deformation of around 0.5 MeV takes place either
at $N=130$ or at $N=132$. The MM calculations of Ref.\ \cite{MBCOISI.08}
predict octupole deformation only for the isotopes $^{224-228}$Cf ($N=126-130$)
with a maximum gain in binding due to octupole deformation
($|\Delta E^{oct}|\sim 0.6$ MeV) at $N=128$ (Table \ref{table-global}).
Since experimental data do not exist for the Cf isotopes with $N\leq 138$ and
only the ground state is observed in the $N=140$ $^{238}$Cf nucleus
\cite{Eval-data}, there is no way to discriminate between the predictions of
the models.

\subsubsection{Fm isotopes}
\label{Fm-isotopes}

\begin{figure*}
  \includegraphics[angle=0,width=5.9cm]{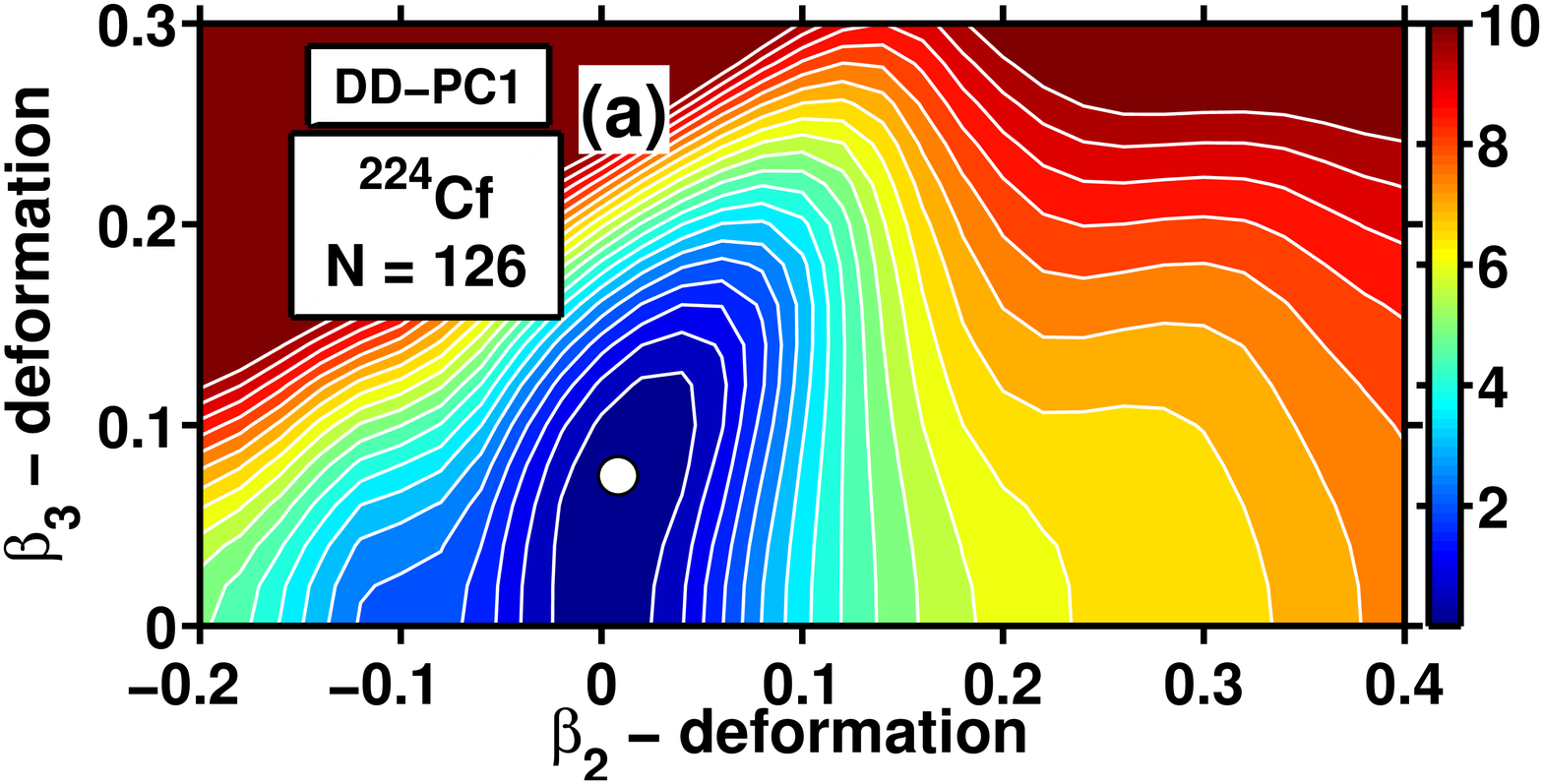}
  \includegraphics[angle=0,width=5.9cm]{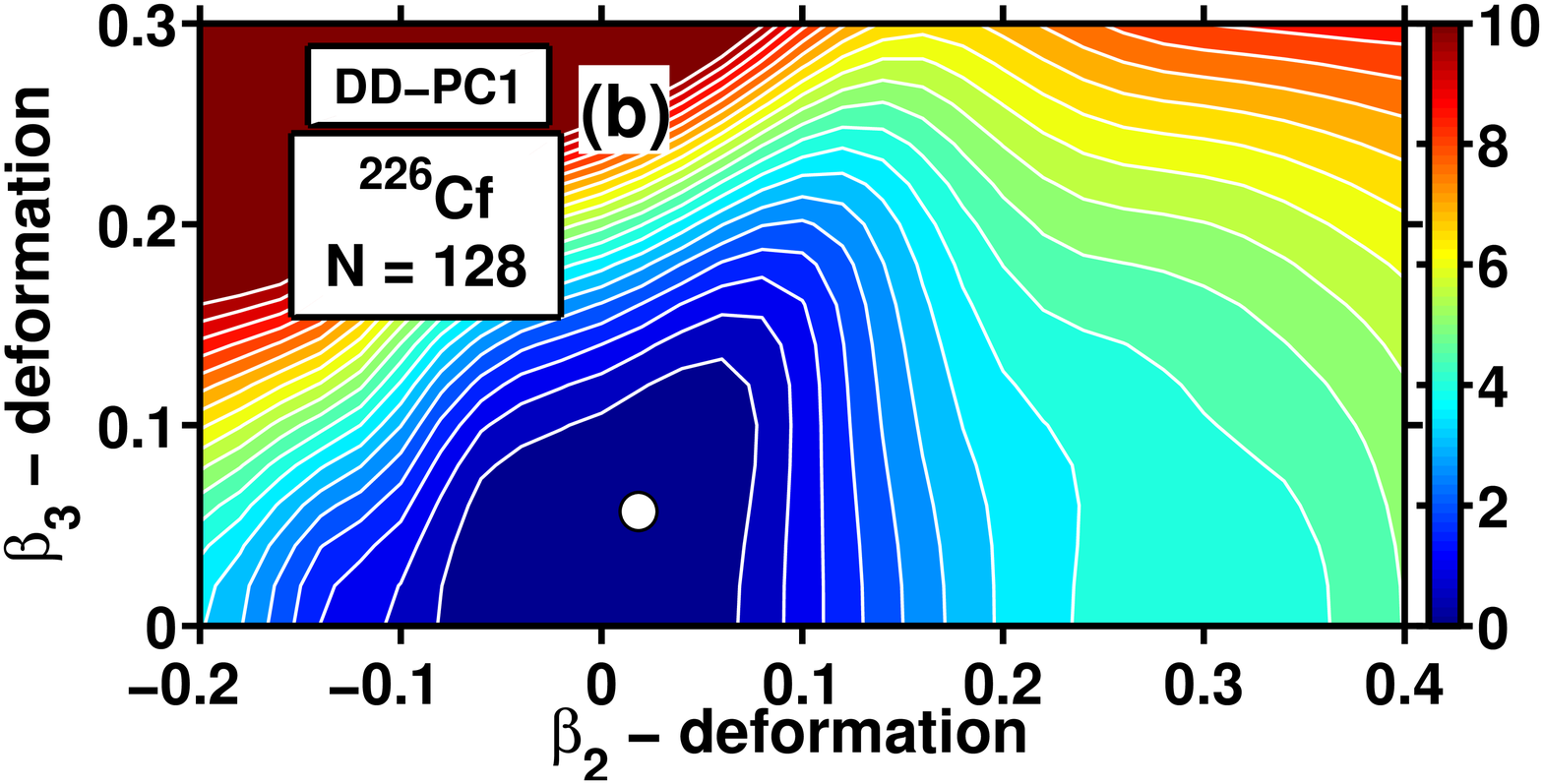}
  \includegraphics[angle=0,width=5.9cm]{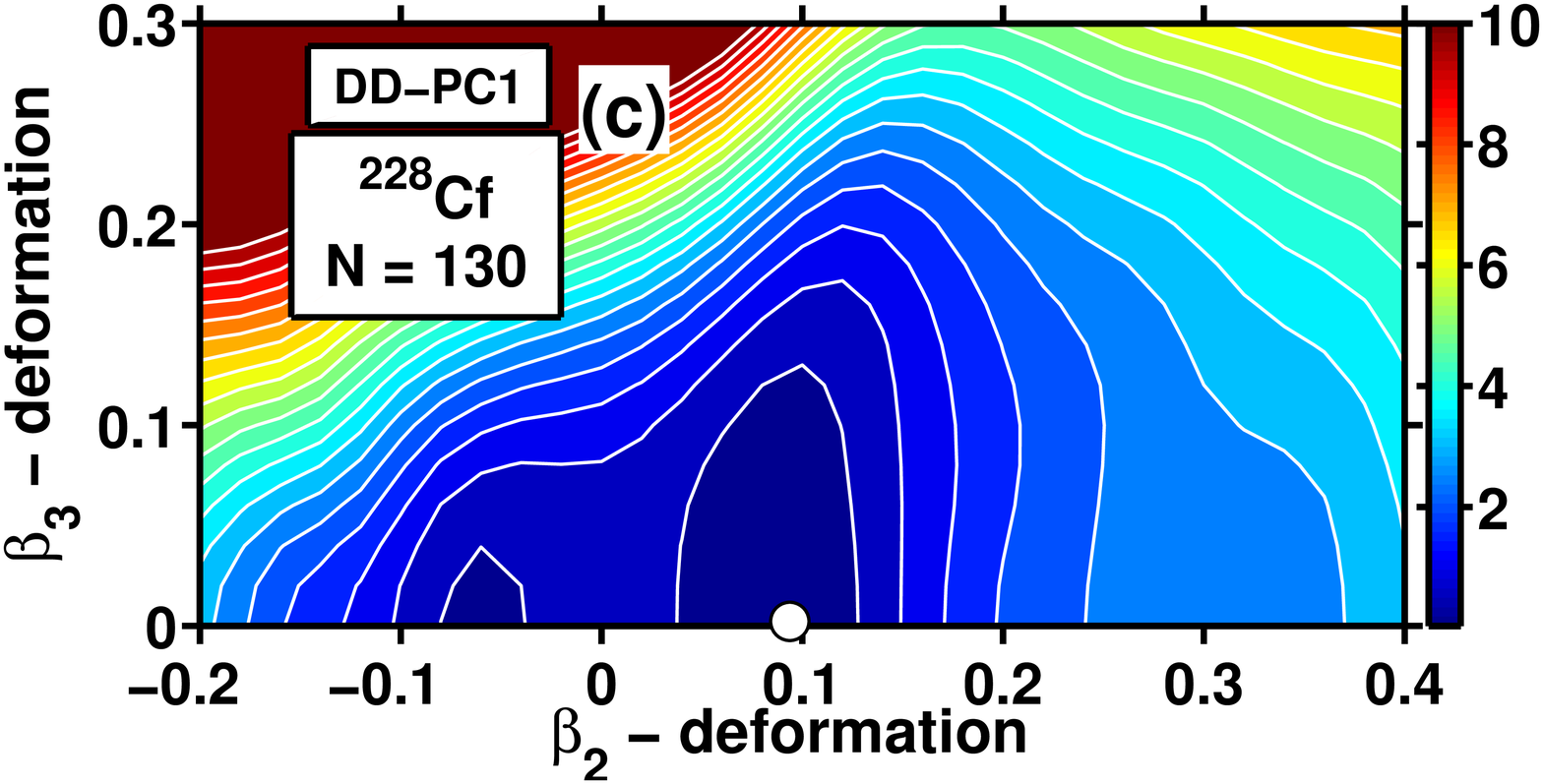}
  \includegraphics[angle=0,width=5.9cm]{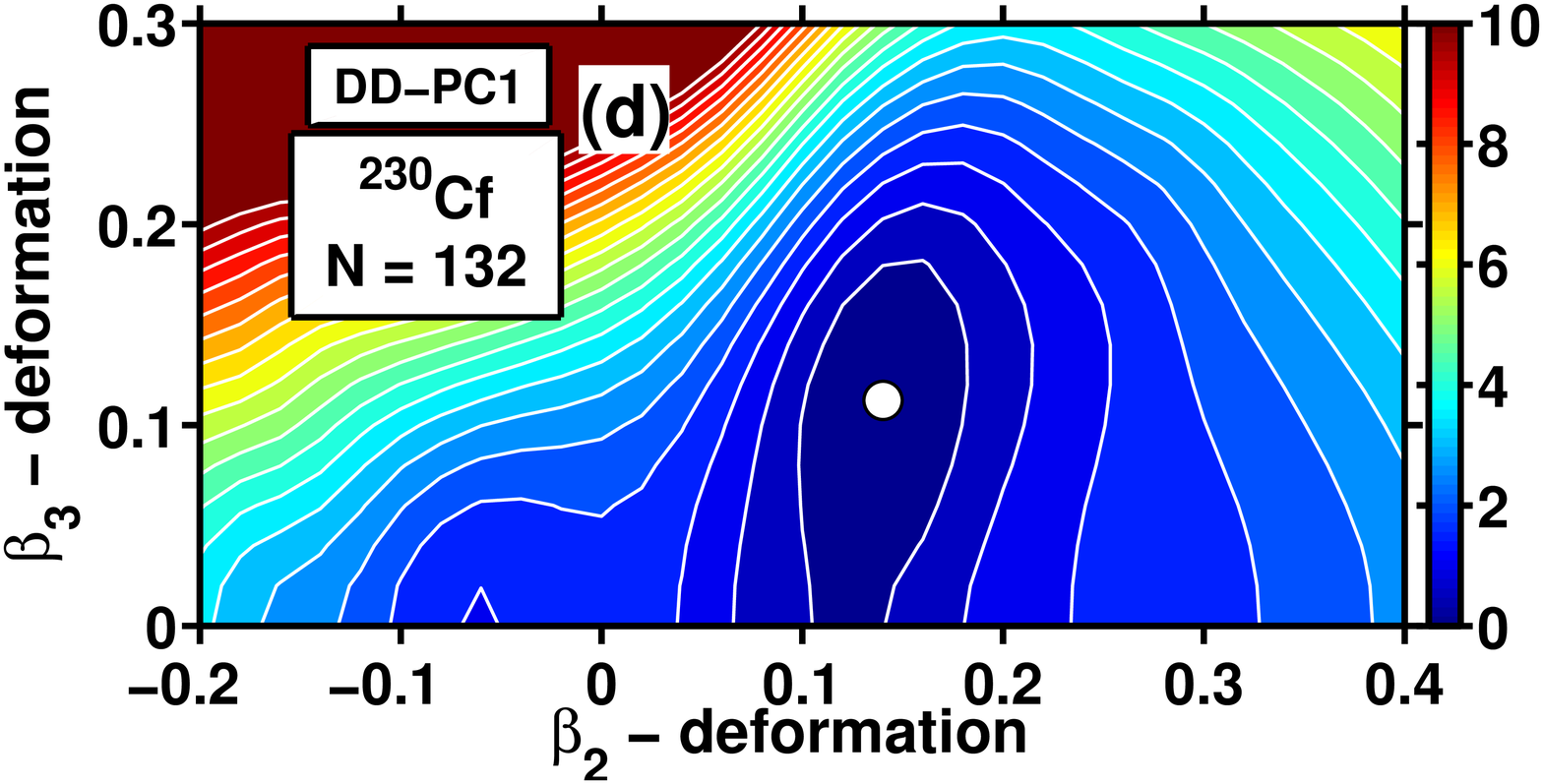}
  \includegraphics[angle=0,width=5.9cm]{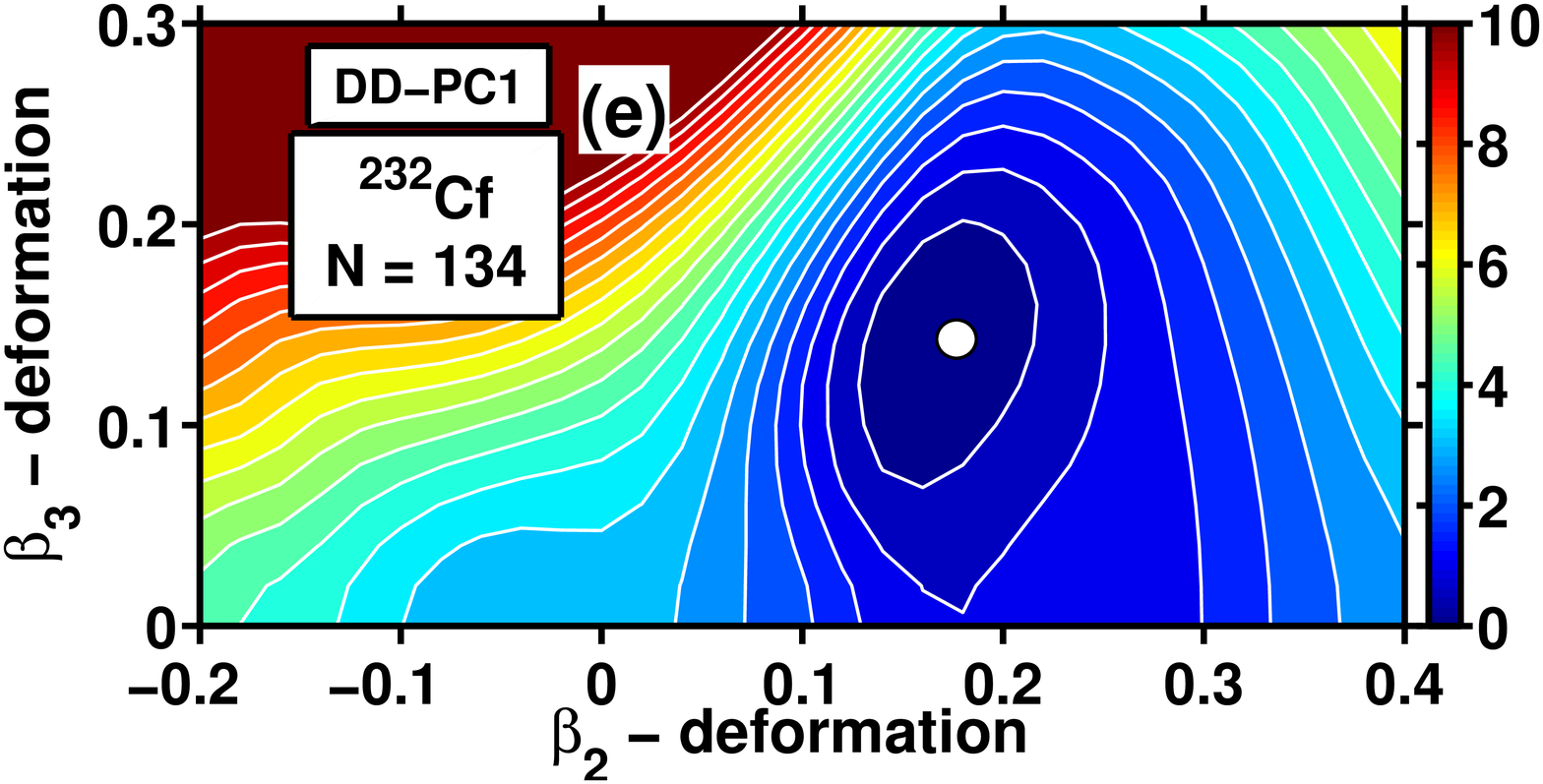}
  \includegraphics[angle=0,width=5.9cm]{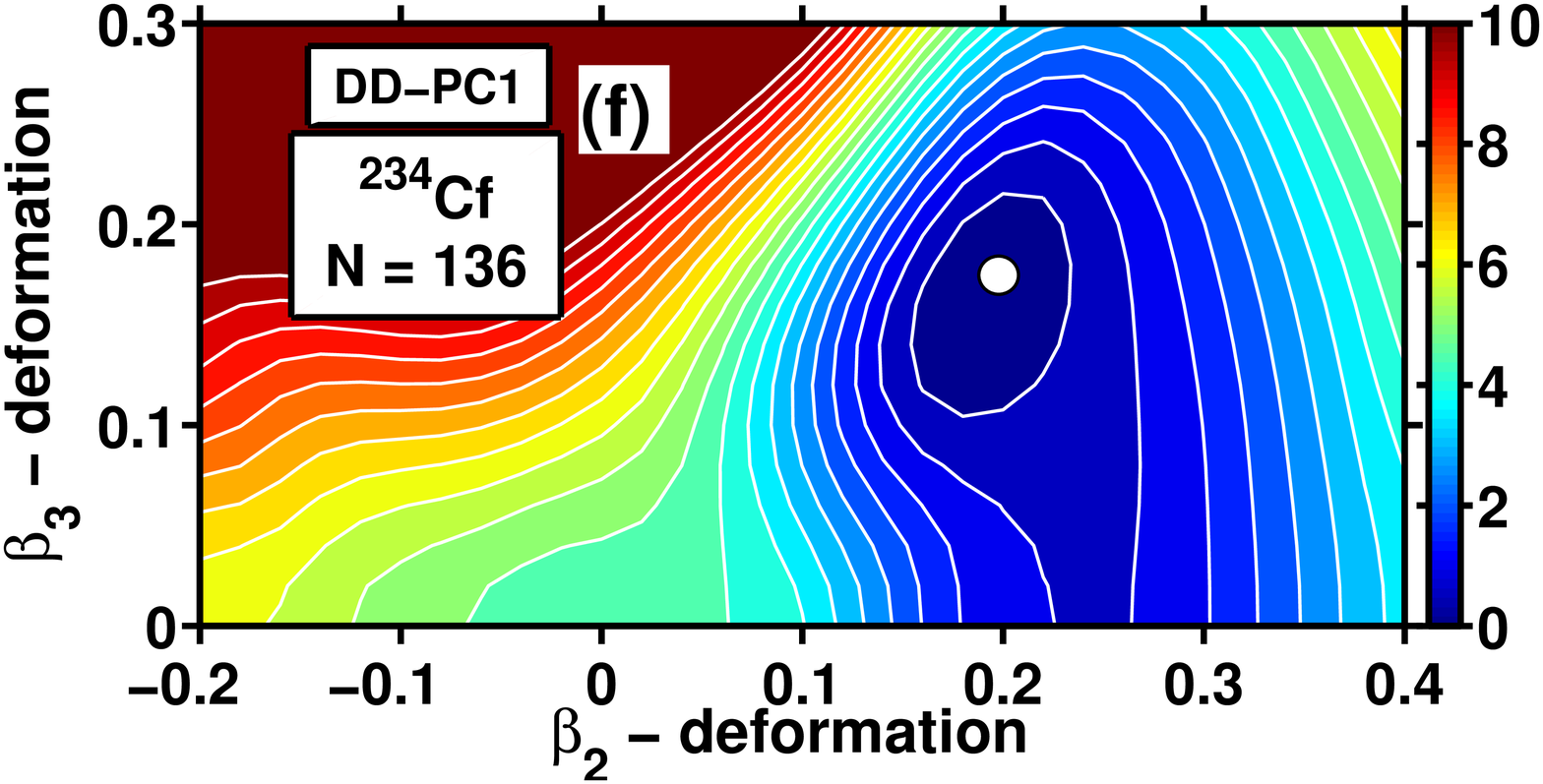}
  \includegraphics[angle=0,width=5.9cm]{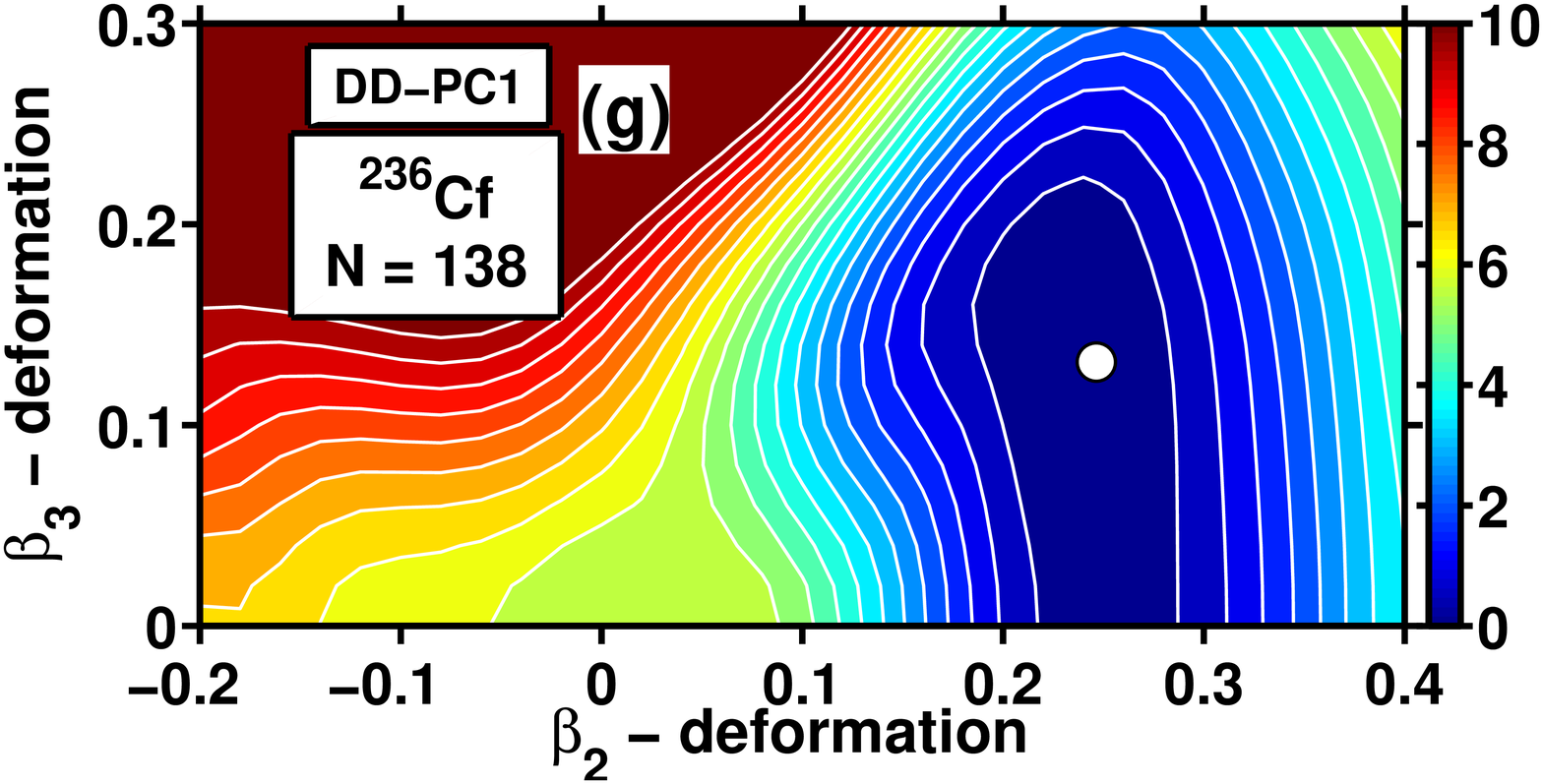}
  \includegraphics[angle=0,width=5.9cm]{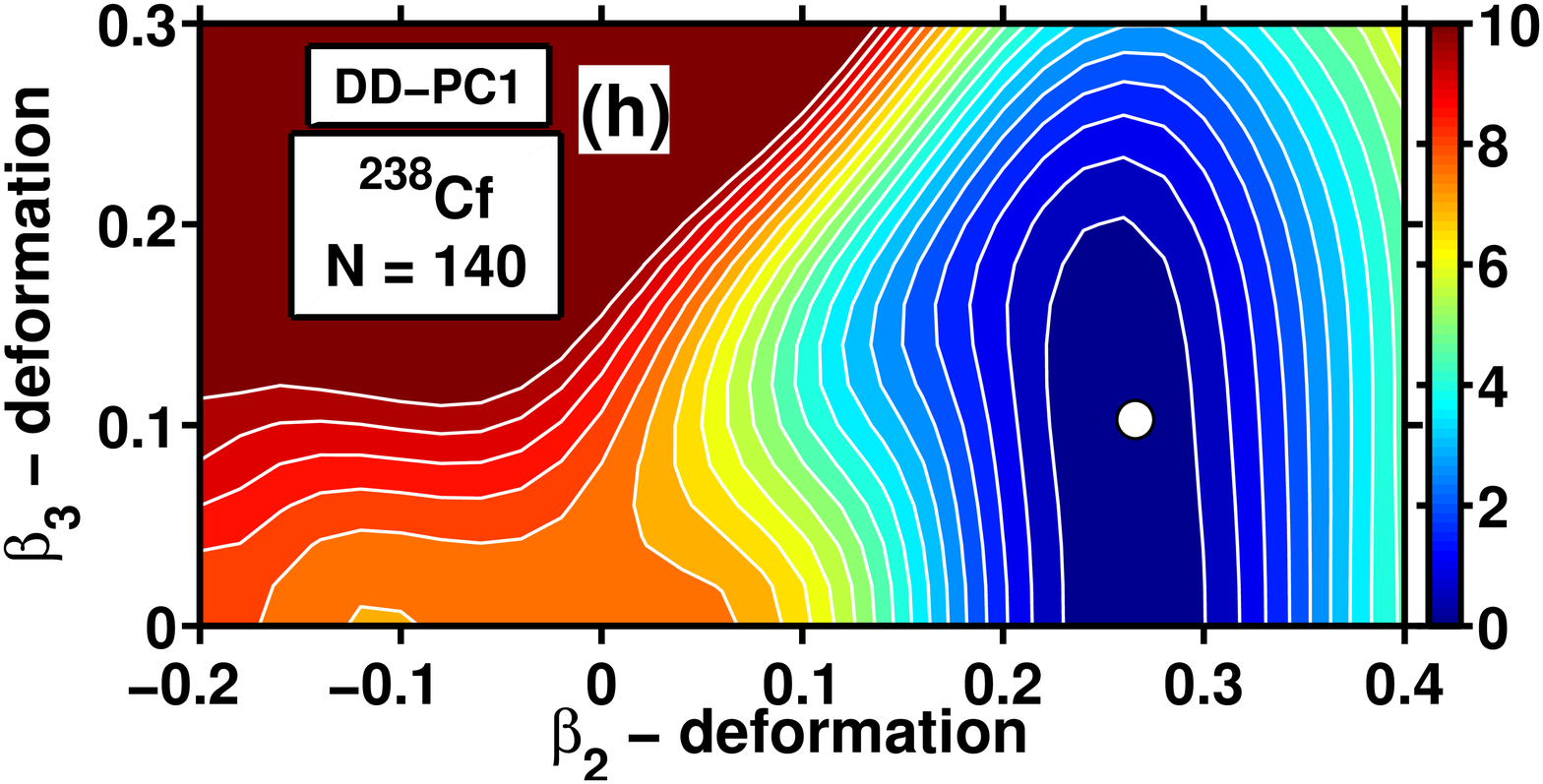}
  \caption{(Color online) The same as Fig.\ \ref{Rn_DD-PC1}, but for
           the Cf isotopes.}
\label{Cf_DD-PC1}
\end{figure*}

  For a given neutron number, the PESs of the Fm isotopes are very similar
to the ones of the Cf isotopes (compare Figs.\ \ref{Fm_DD-PC1} and
\ref{Cf_DD-PC1}). Thus, the discussion of the evolution of PESs as a
function of the neutron number presented in Sec.\ \ref{Fm-isotopes} is
also applicable to the Fm isotopes. The results of the calculations
with NL3*, PC-PK1, DD-ME2 and DD-PC1 show a similar evolution of
$\Delta E^{oct}$ as a function of neutron number (Fig.\
\ref{massdependence-Fm}). The maximum in $|\Delta E^{oct}|$ is seen
either at $N=134$ (DD-PC1 and PC-PK1) or at $N=136$ (NL3* and DD-ME2);
its value is located in the $0.5-1.0$ MeV range. The results
for DD-ME$\delta$ differ substantially from all other functionals.

  The MM approach predicts octupole deformation only in the $N=126, 128$
nuclei (Table \ref{table-global}) with rather modest gains in binding due
to octupole deformation. The maximum of $|\Delta E^{oct}|$ is located at
$N=128$. The HFB calculations with Gogny forces show the presence of
octupole deformation in the isotopes $^{230-232}$Fm with  $N=130-132$ \cite{RR.12}.
No experimental data are available for the Fm isotopes with $N\leq 140$
\cite{Eval-data}. Thus, the predictions of different models cannot
be discriminated.

\subsubsection{No, Rf and Sg isotopes}

Our RHB calculations predict octupole deformed No, Rf and Sg isotopes   
(see Table \ref{table-global}). However, most of these
nuclei are octupole soft with marginal gains in binding due to octupole
deformation. In addition, there is a substantial difference between
the NL3* and DD-PC1 functionals. For example, only the $N=136$ $^{238}$No
nucleus is predicted to be octupole deformed in the calculations with
DD-PC1. On the contrary, the $N=134-140$ $^{236-242}$No nuclei are octupole
deformed in NL3*. No octupole deformed Rf nuclei are predicted in
DD-PC1, while the $N=138-142$ $^{242-246}$Rf nuclei have nonzero
octupole deformation in the ground states (Table \ref{table-global}).
Note that the No, Rf and Sg nuclei with nonzero octupole deformation
are located at or near the two-proton drip line (see Fig.\ \ref{fig-global}
below) so their experimental observation could be very difficult. At
present, no experimental data are available for these proton-rich nuclei
\cite{Eval-data}.

  The MM calculations of Ref.\ \cite{MBCOISI.08} do not predict
octupole deformation for the No, Rf and Sg isotopes. The HFB calculations
with Gogny forces of Ref.\ \cite{RR.12} predict nonzero $\beta_3$ values
only in the $^{232-234}$No nuclei.

\subsection{General observations}
\label{actinides-general}

\begin{figure*}
  \includegraphics[angle=0,width=5.9cm]{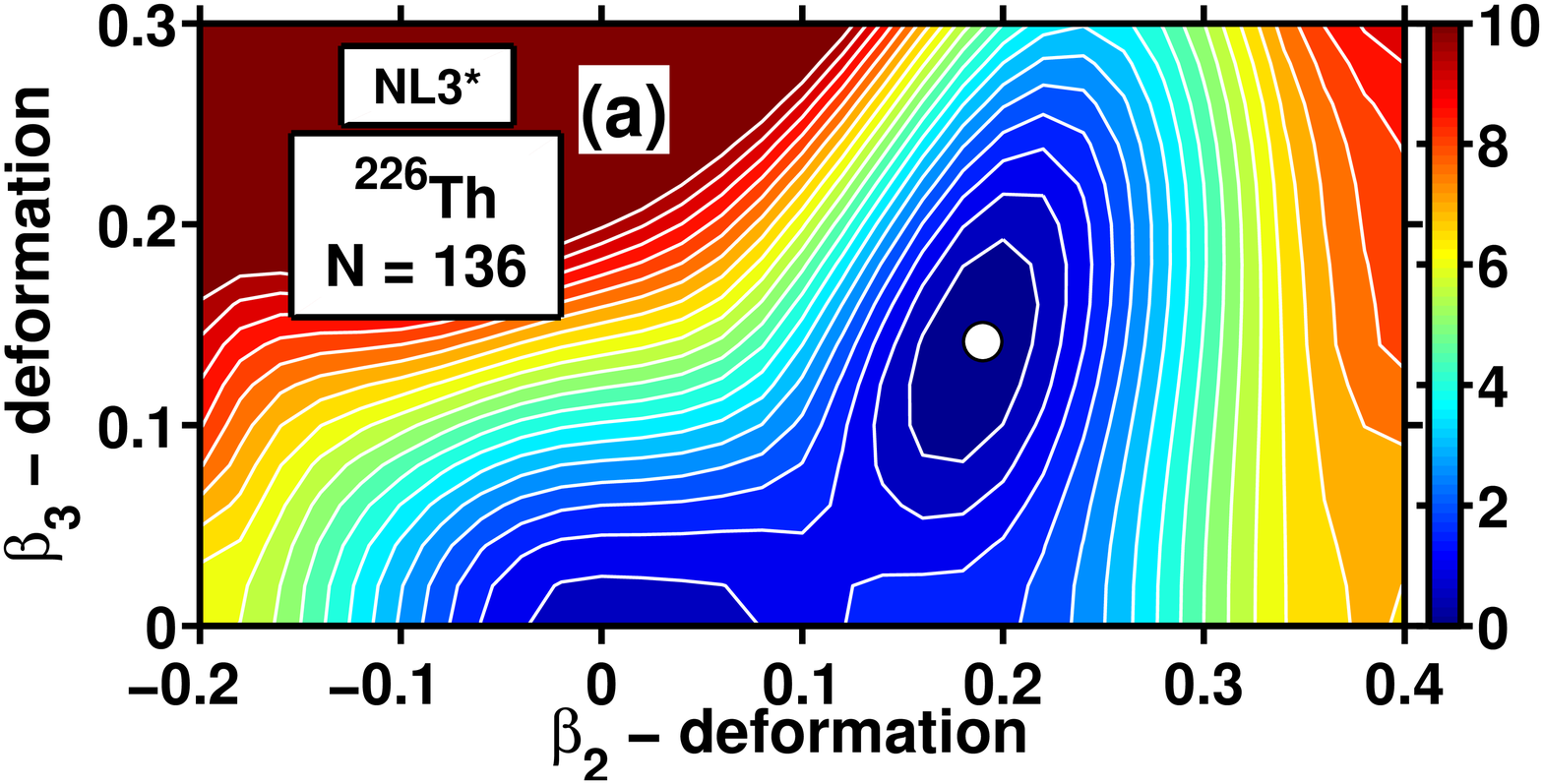}
  \includegraphics[angle=0,width=5.9cm]{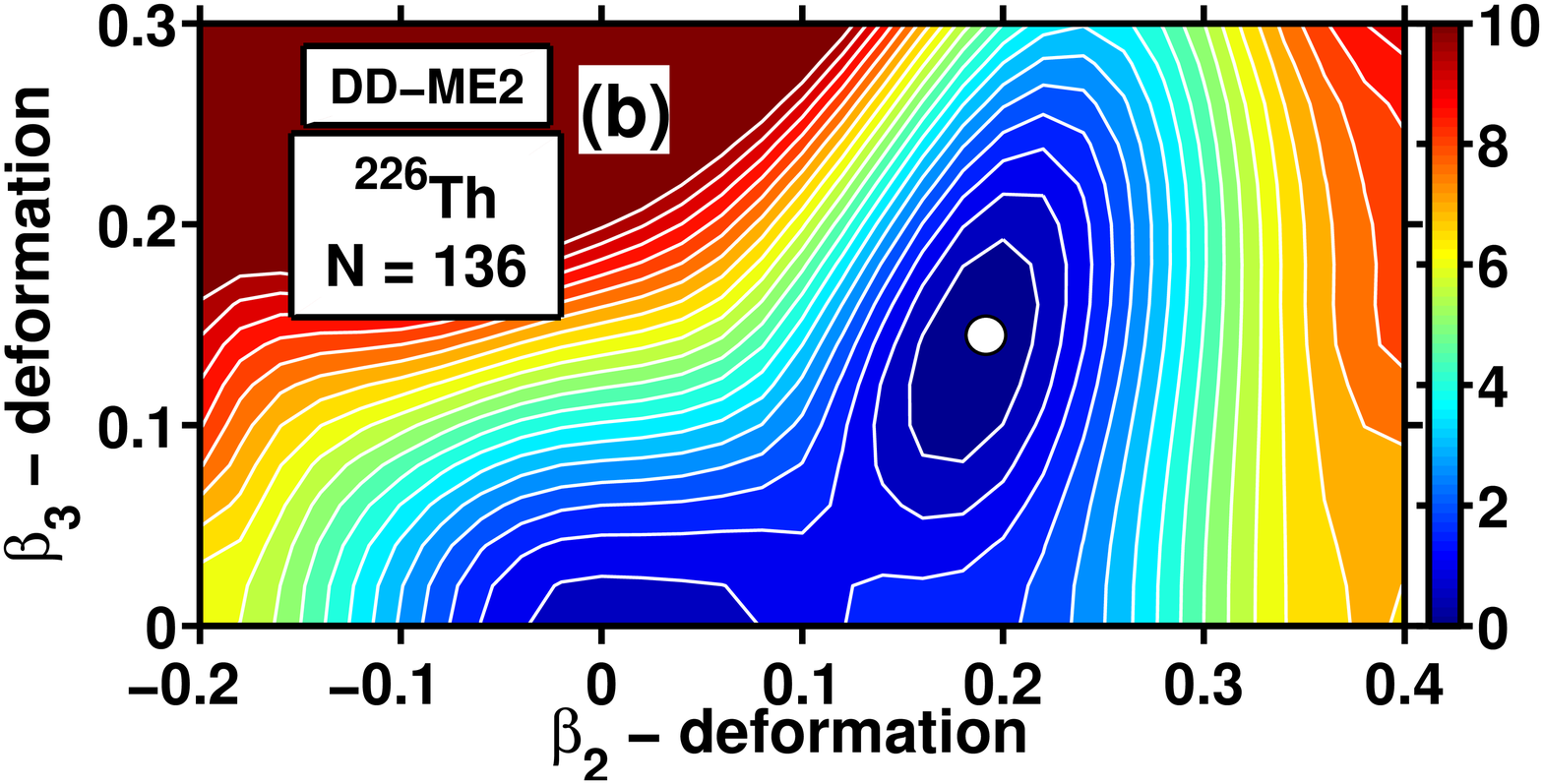}
  \includegraphics[angle=0,width=5.9cm]{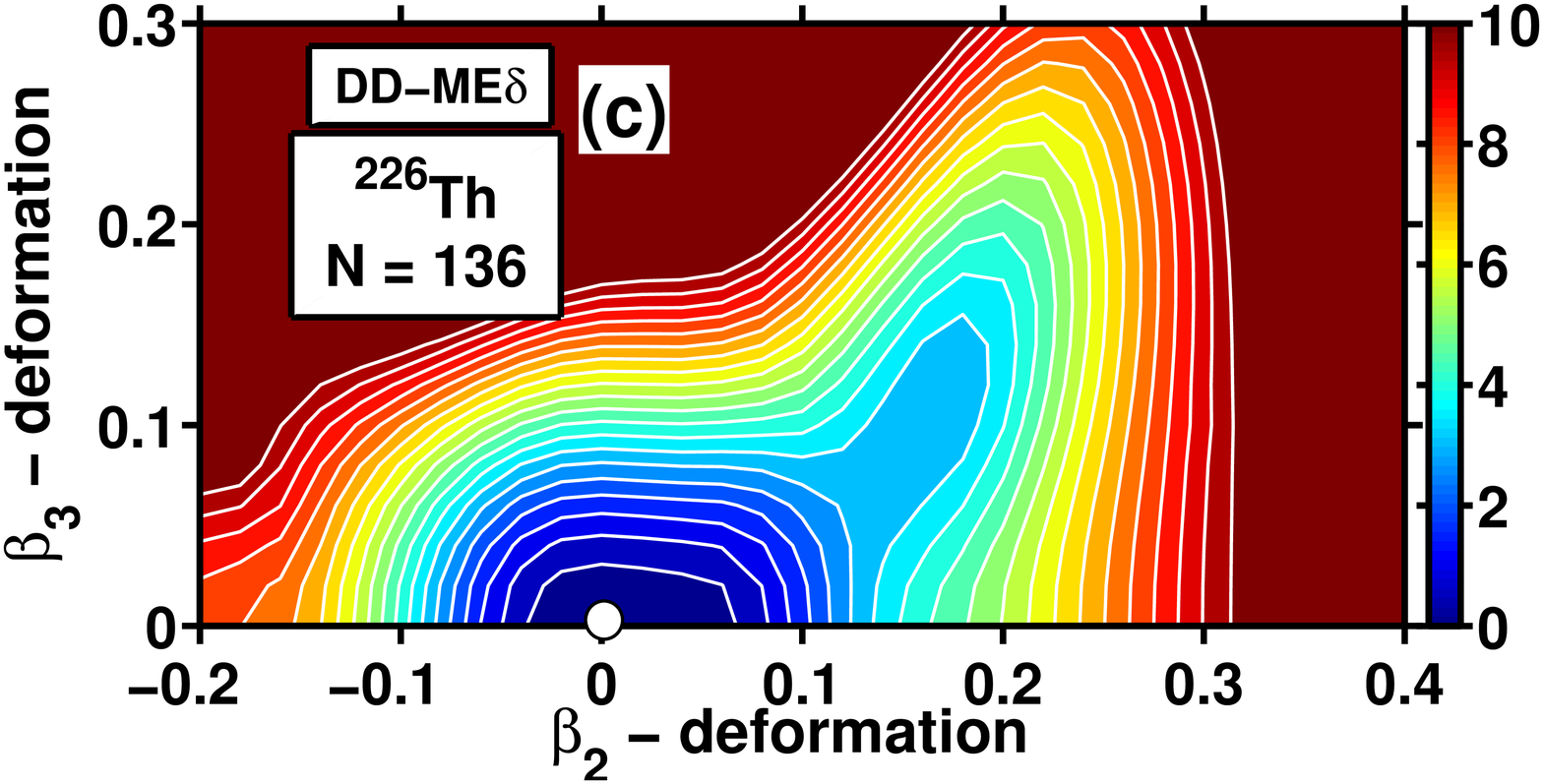}
  \includegraphics[angle=0,width=5.9cm]{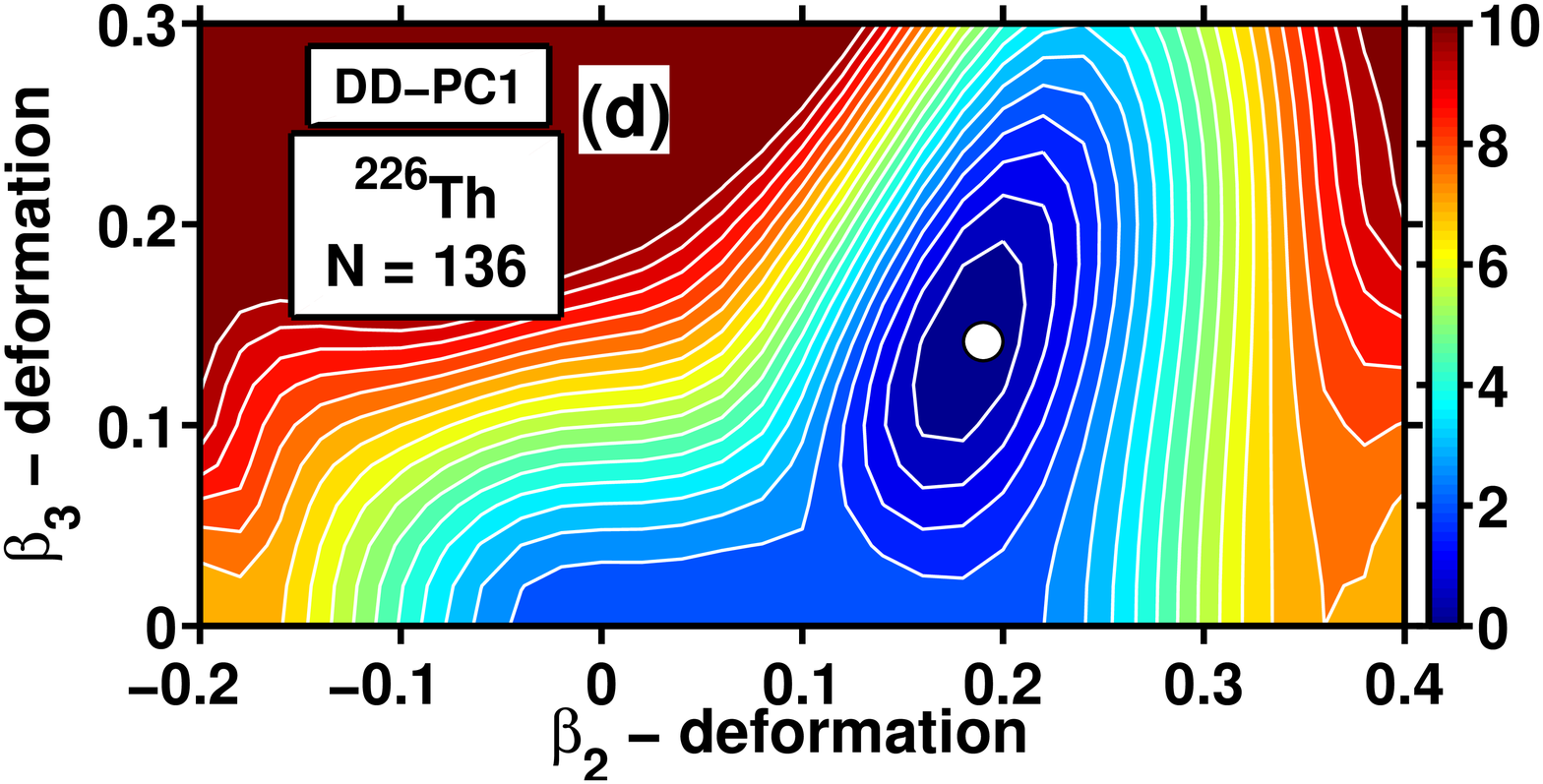}
  \includegraphics[angle=0,width=5.9cm]{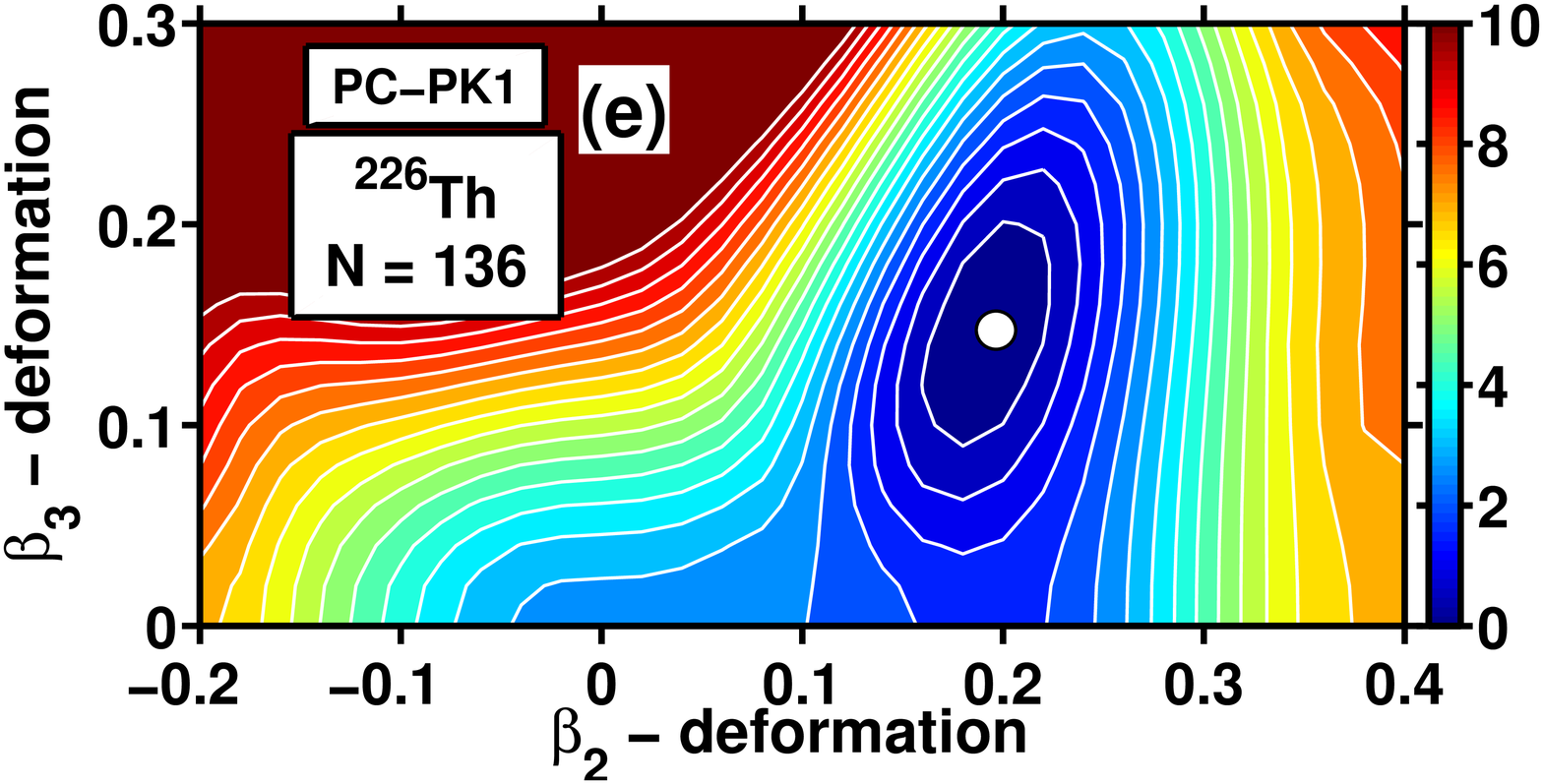}
\caption{(Color online) The same as Fig.\ \ref{Rn_DD-PC1}, but for
           the $^{226}$Th obtained with indicated CEDFs}
\label{Th226-5-cedf}
\end{figure*}

  The analysis of the results of the RHB calculations and the comparison between
different models reveal the following general features:

\begin{itemize}

\item
  The results of the calculations are rather similar for the CEDFs DD-ME2, DD-PC1 and
PC-PK1 (see Figs.\ \ref{massdependence}, \ref{massdependence-Pu-etc} and
\ref{massdependence-Fm}). The results with NL3* only slightly deviate from
the ones for these functionals. Not only quadrupole and octupole equilibrium
deformations as well as the gains in binding due to octupole
deformation displayed in Figs.\ \ref{massdependence}, \ref{massdependence-Pu-etc}
and \ref{massdependence-Fm} show these features, but also the topologies
of the PESs, which affect the results of beyond mean field calculations, are similar
in these four functionals (see Fig.\ \ref{Th226-5-cedf}).
The predictions obtained with the CEDF DD-ME$\delta$ contradict to all available
model calculations and to experimental data. Therefore, the results obtained with this functional
will not be discussed in this subsection.

\begin{figure*}
  \includegraphics[angle=0,width=5.9cm]{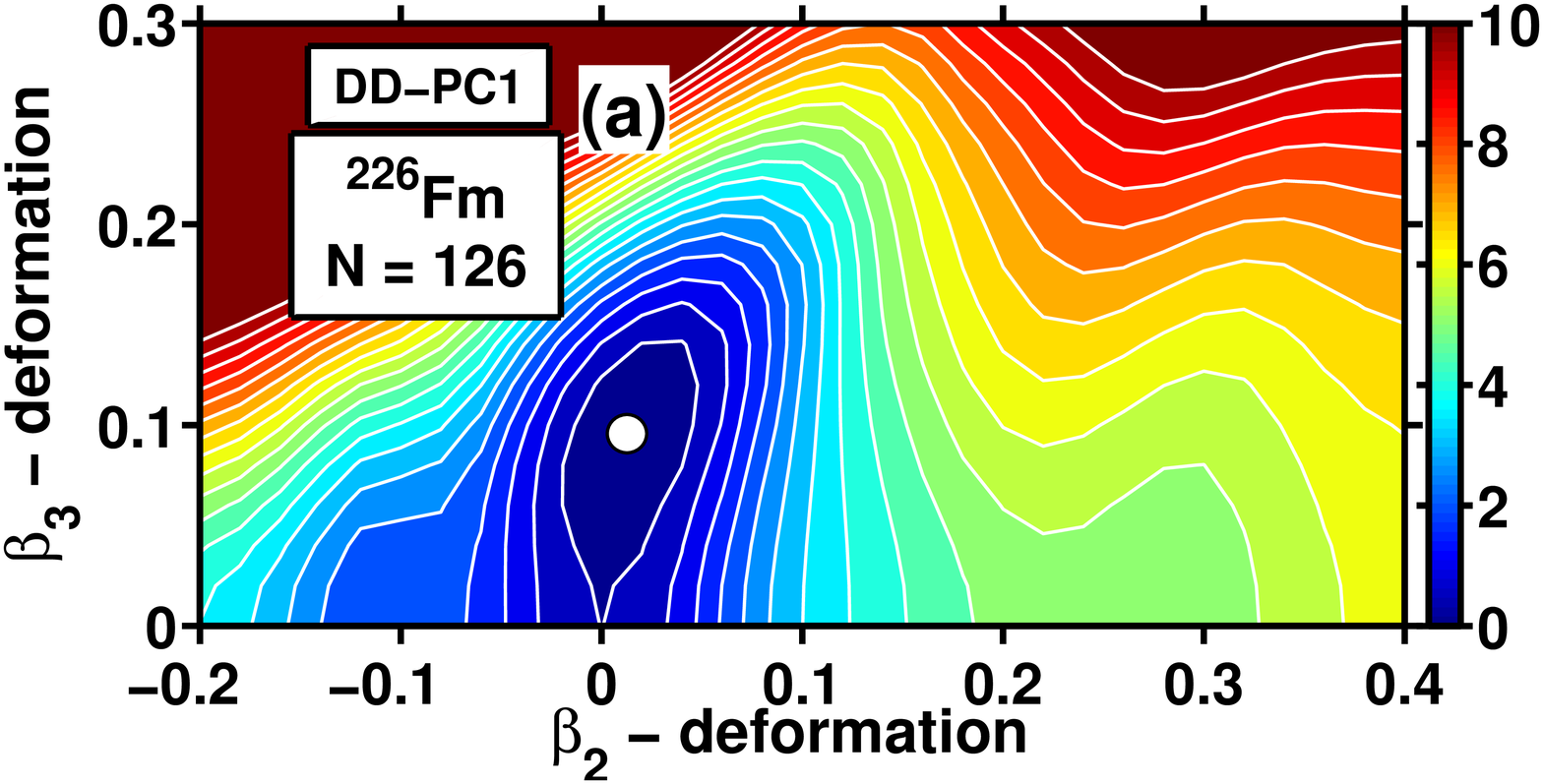}
  \includegraphics[angle=0,width=5.9cm]{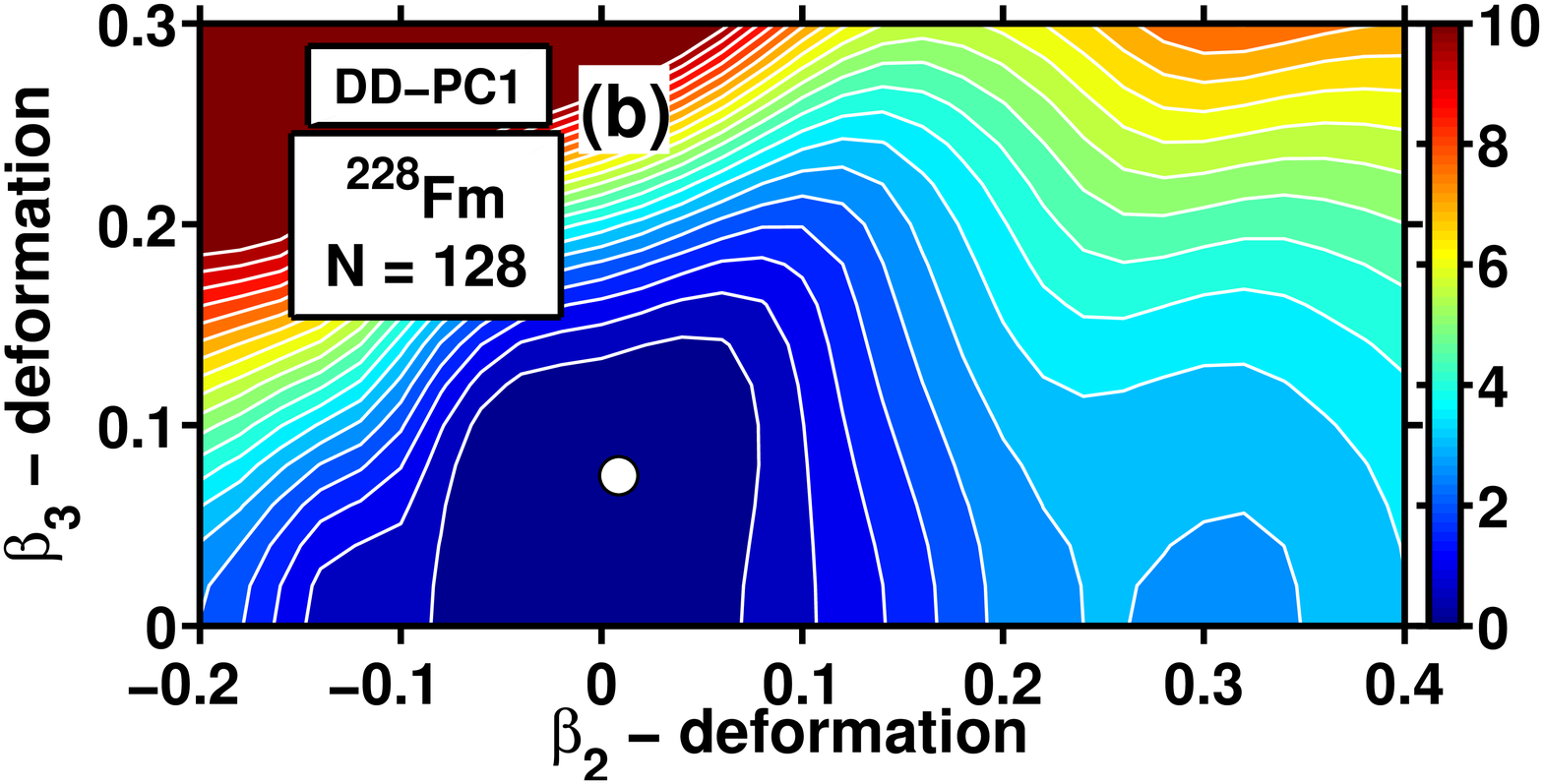}
  \includegraphics[angle=0,width=5.9cm]{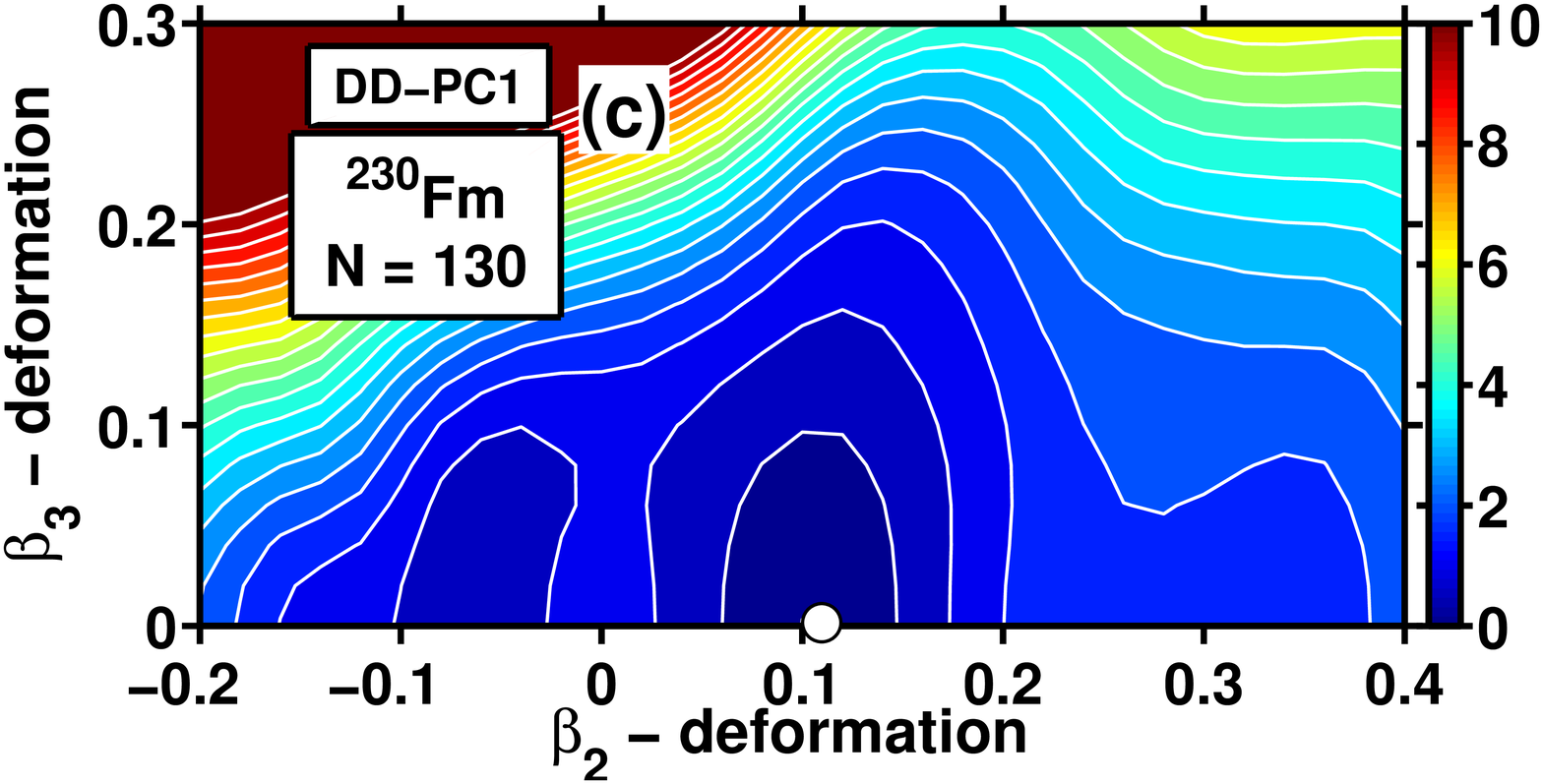}
  \includegraphics[angle=0,width=5.9cm]{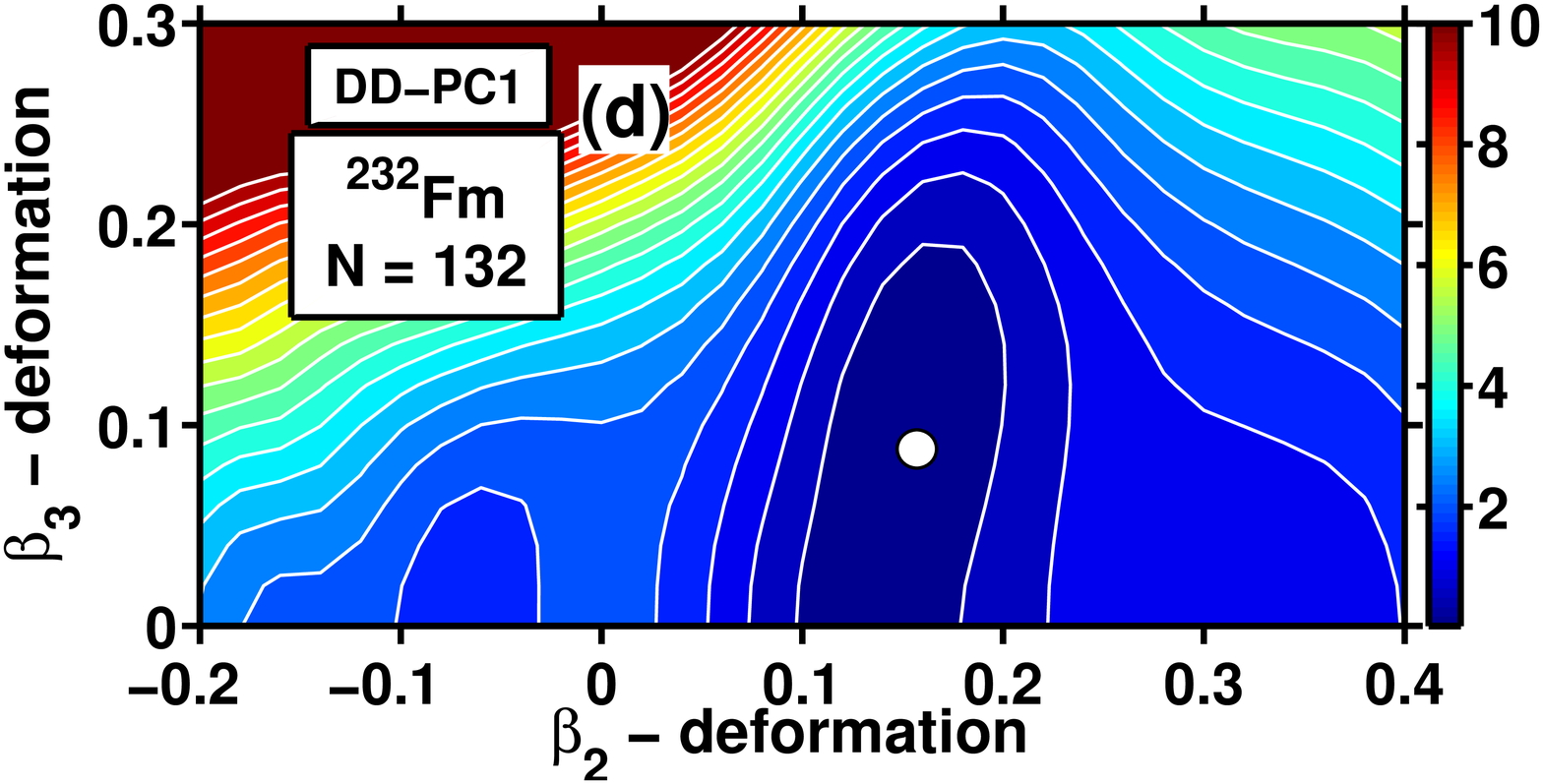}
  \includegraphics[angle=0,width=5.9cm]{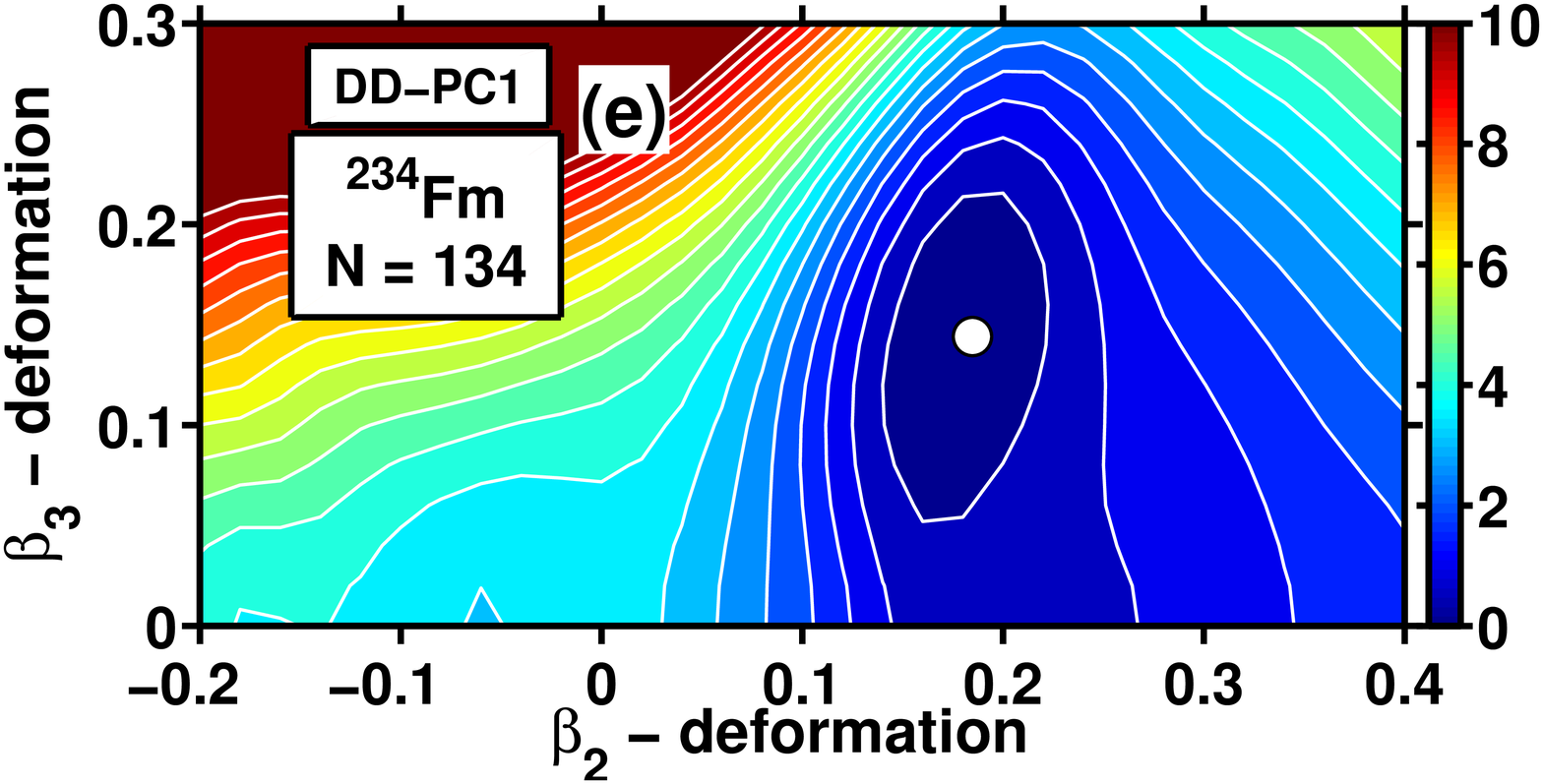}
  \includegraphics[angle=0,width=5.9cm]{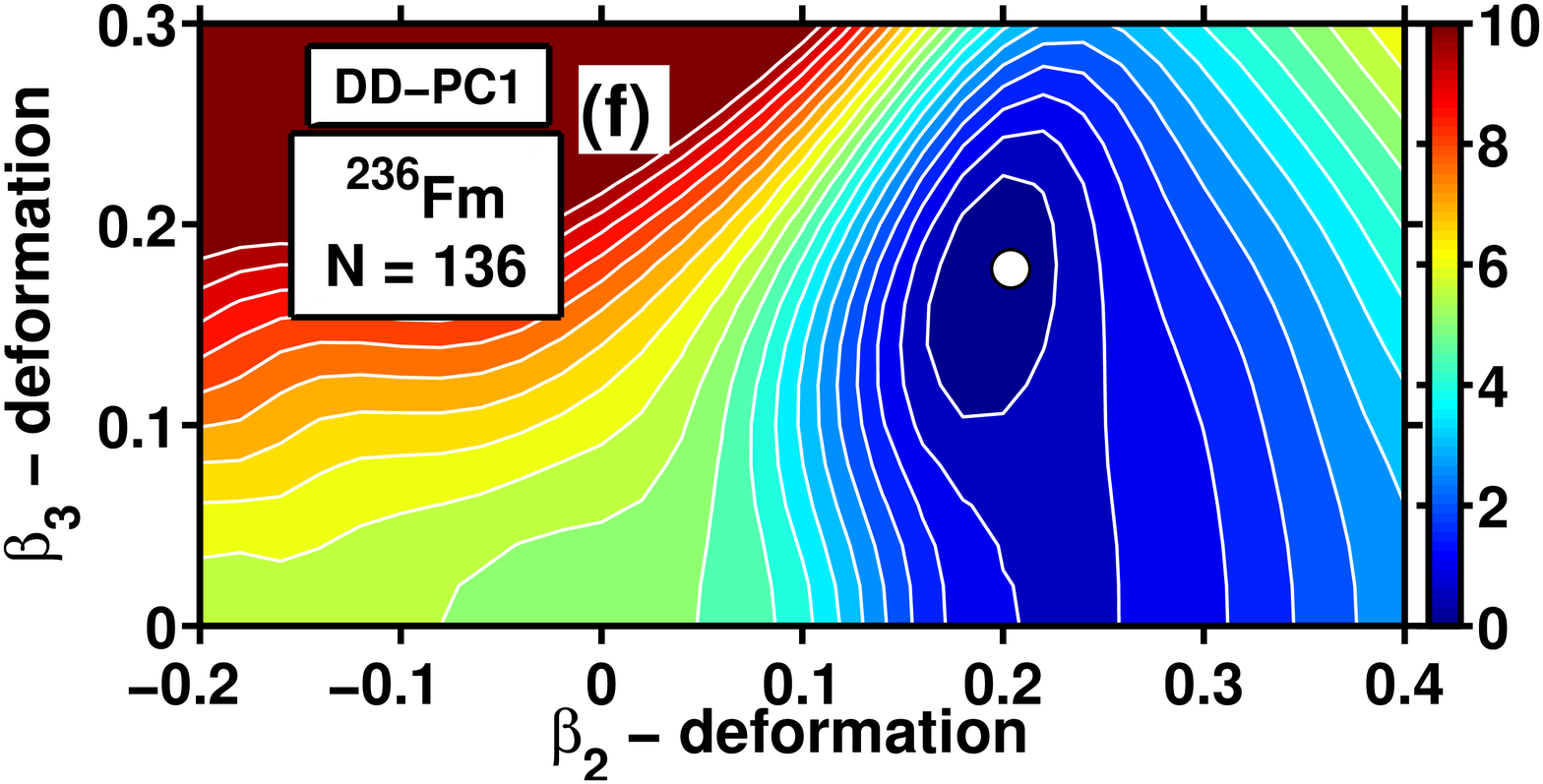}
  \includegraphics[angle=0,width=5.9cm]{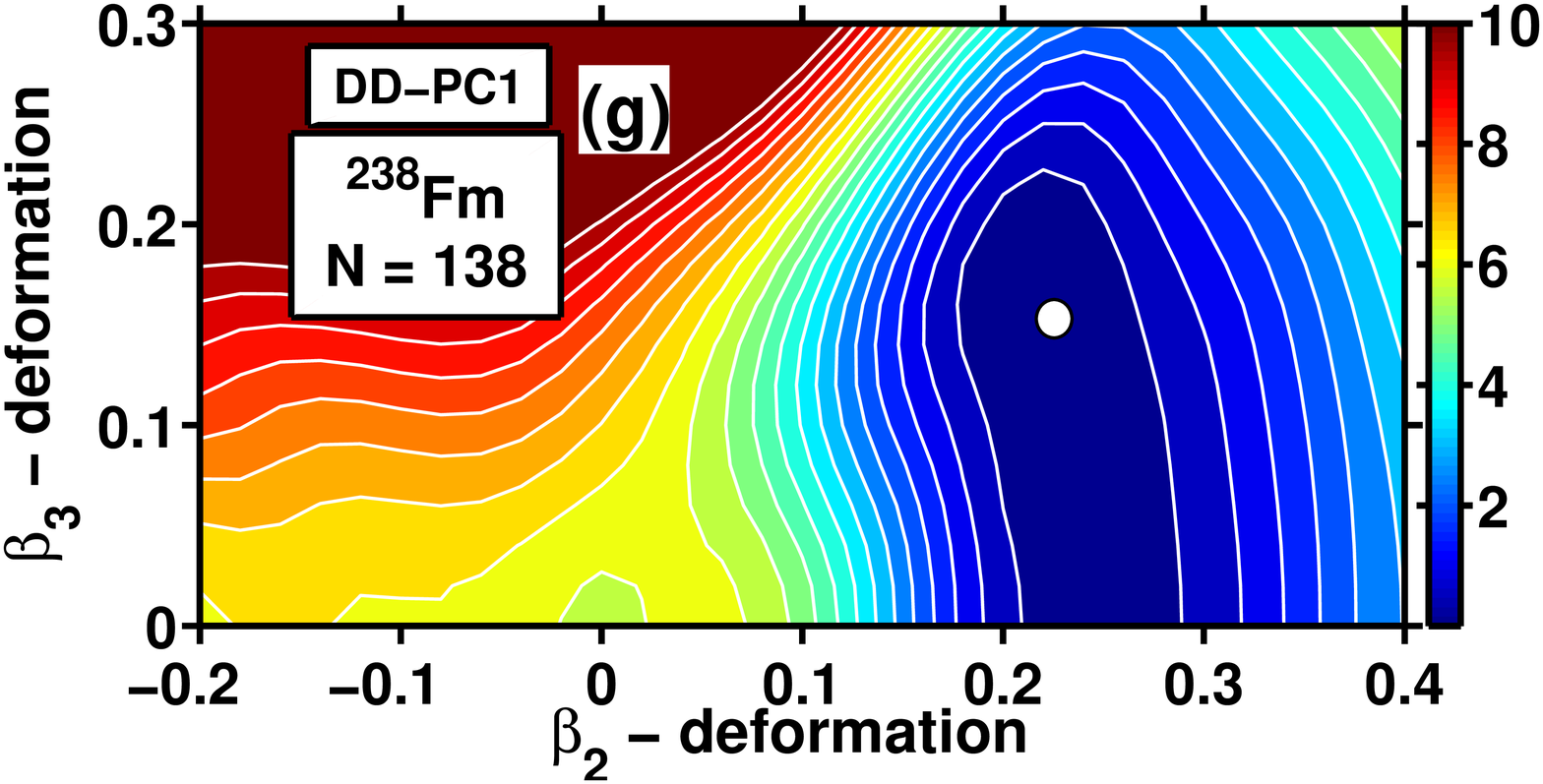}
  \includegraphics[angle=0,width=5.9cm]{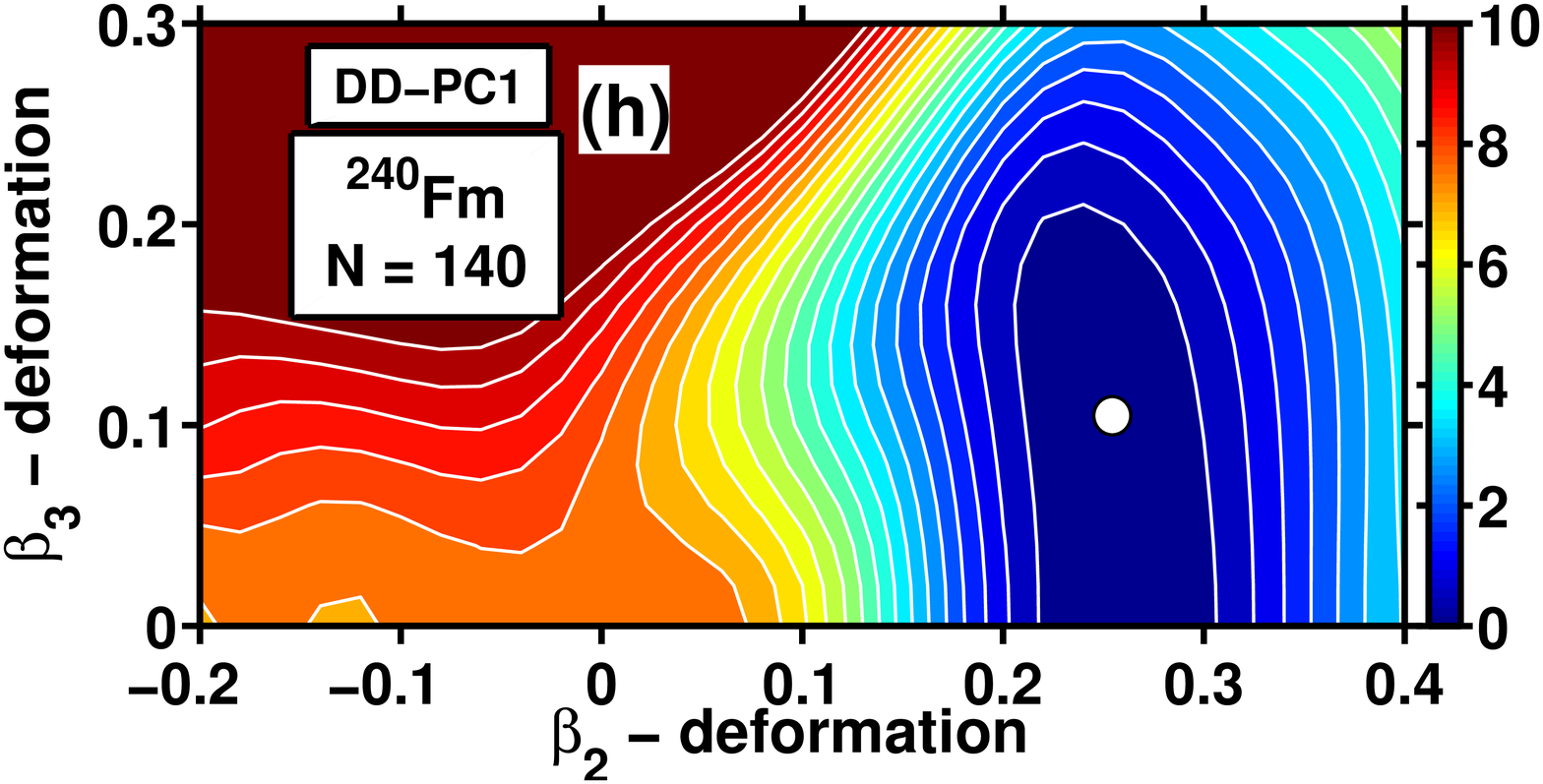}
  \caption{(Color online) The same as Fig.\ \ref{Rn_DD-PC1},
           but for the Fm isotopes.}
\label{Fm_DD-PC1}
\end{figure*}

\begin{figure}
  \includegraphics[angle=0,width=5.8cm]{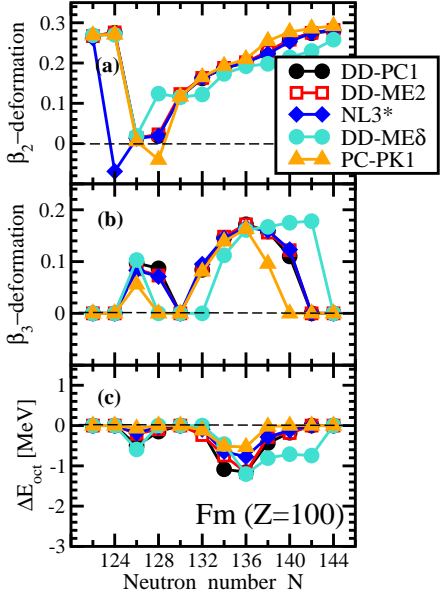}
  \caption{(Color online) The same as Fig.\ \ref{massdependence},
           but for the Fm isotopes.}
\label{massdependence-Fm}
\end{figure}

\item
 In the RHB calculations with DD-ME2, DD-PC1, NL3* and PC-PK1 a gradual 
increase of 
quadrupole deformation is typically seen in the $N=130-140$ range (see
Figs.\ \ref{massdependence}, \ref{massdependence-Pu-etc} and \ref{massdependence-Fm}).
At higher $N$ value, the $\beta_2$ deformation is nearly constant.

\item
  The center of the island of octupole deformation, defined in terms of maximum
gain in binding due to octupole deformation, is located at $Z=90-92$ and
$N=136$ in the RHB calculations with the CEDFs DD-PC1, DD-ME2 and PC-PK1. This agrees with
the experiment \cite{BN.96} and with the results obtained within the HFB framework
based on Gogny forces \cite{RR.12} and the MM calculations based on
Woods-Saxon potentials \cite{NORDLMR.84}. The MM calculations of Ref.\ \cite{MBCOISI.08}
with folded Yukawa potentials favor somewhat lower neutron numbers $N=132-134$.
In these calculations the neutron number $N$ of the
nucleus with maximum gain in binding due to octupole deformation in a given
isotope chain decreases with increasing proton number $Z$ 
down to $N=128$ in the Cf and Fm isotopes. A similar but less pronounced trend
is seen in the HFB calculations with the Gogny forces. On the contrary, 
for relativistic functionals this decrease is only 2 neutrons in going from the Th
and U isotopes to the Cf and Fm isotopes (see Figs.\ \ref{massdependence},
\ref{massdependence-Pu-etc} and \ref{massdependence-Fm}). Thus, the CDFT
predictions favor the experimental observation of these octupole deformed
nuclei as compared with non-relativistic models since the island of octupole
deformed nuclei is located closer to the beta-stability line in the RHB
calculations. In particular, octupole deformed Pu, Cm, and Cf nuclei could
be within the reach of future dedicated experiments.
\end{itemize}

\section{Octupole deformation in the Ba-Ce-Nd-Sm region}
\label{lanth}

 Octupole deformation is predicted also in the ground states of
the Ba, Ce, Nd and Sm isotopes. The results for equilibrium quadrupole
and octupole deformations and the gains in binding due to octupole
deformation are summarized in Fig.\ \ref{massdependence-Ba-etc}.

  Several features are typical for this mass region. First, the
gain in binding due to octupole deformation is substantially smaller
($|\Delta E^{oct}|$ is typically around 0.5 MeV) than in the actinides.
Thus, the stabilization of octupole deformation at the ground state
is less likely in this region as compared with actinides.

\begin{figure*}
  \includegraphics[angle=0,width=5.9cm]{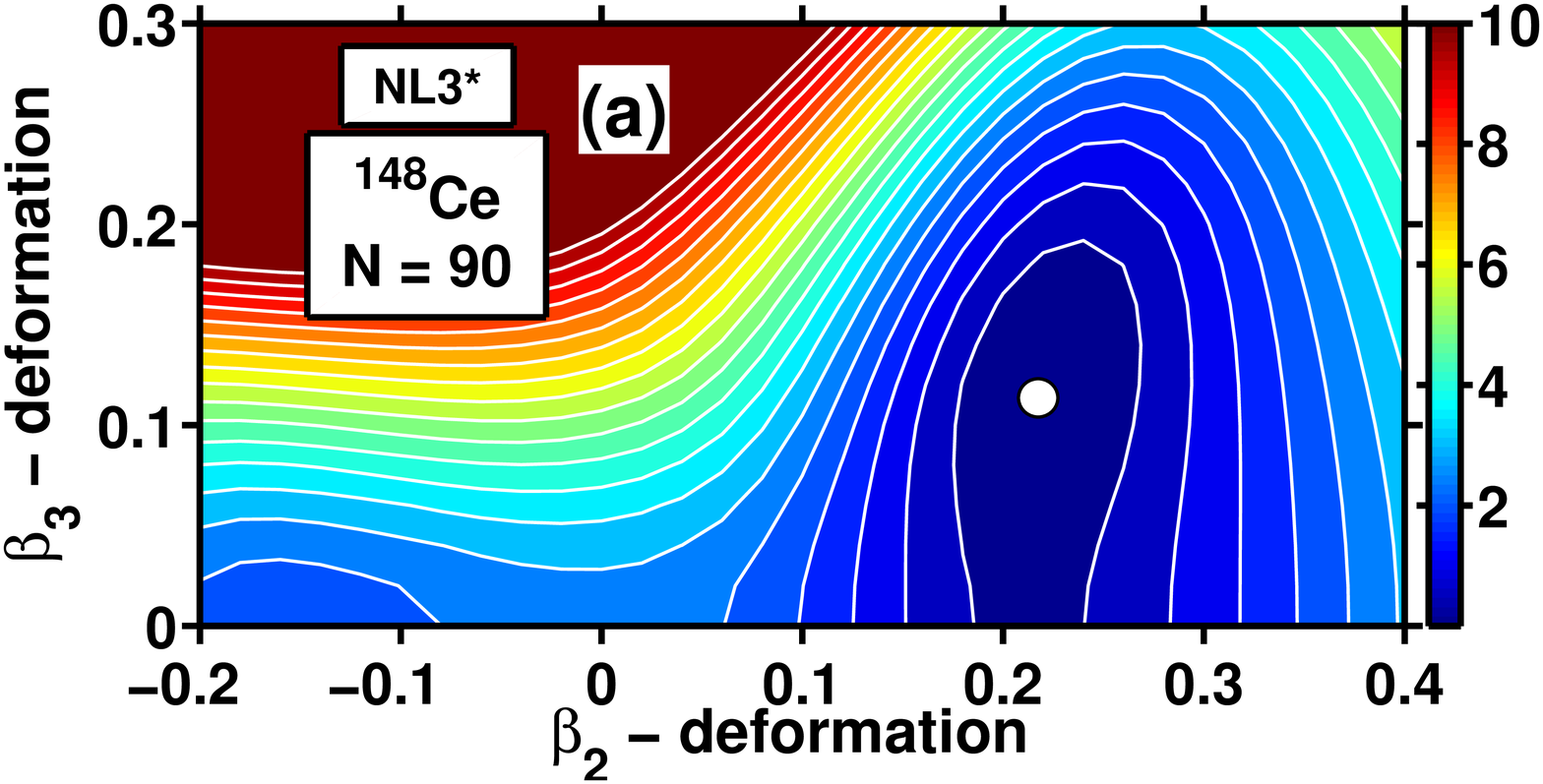}
  \includegraphics[angle=0,width=5.9cm]{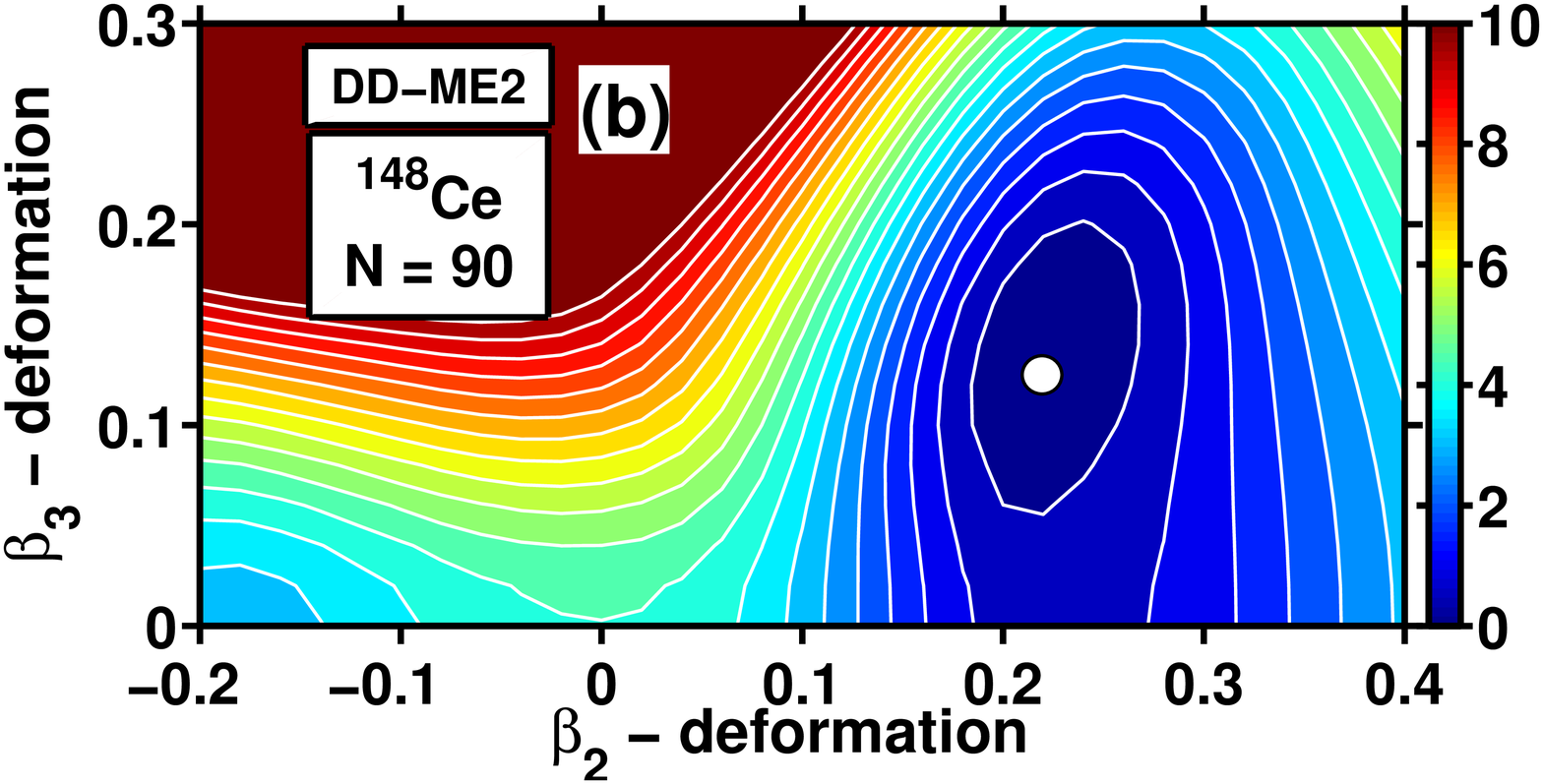}
  \includegraphics[angle=0,width=5.9cm]{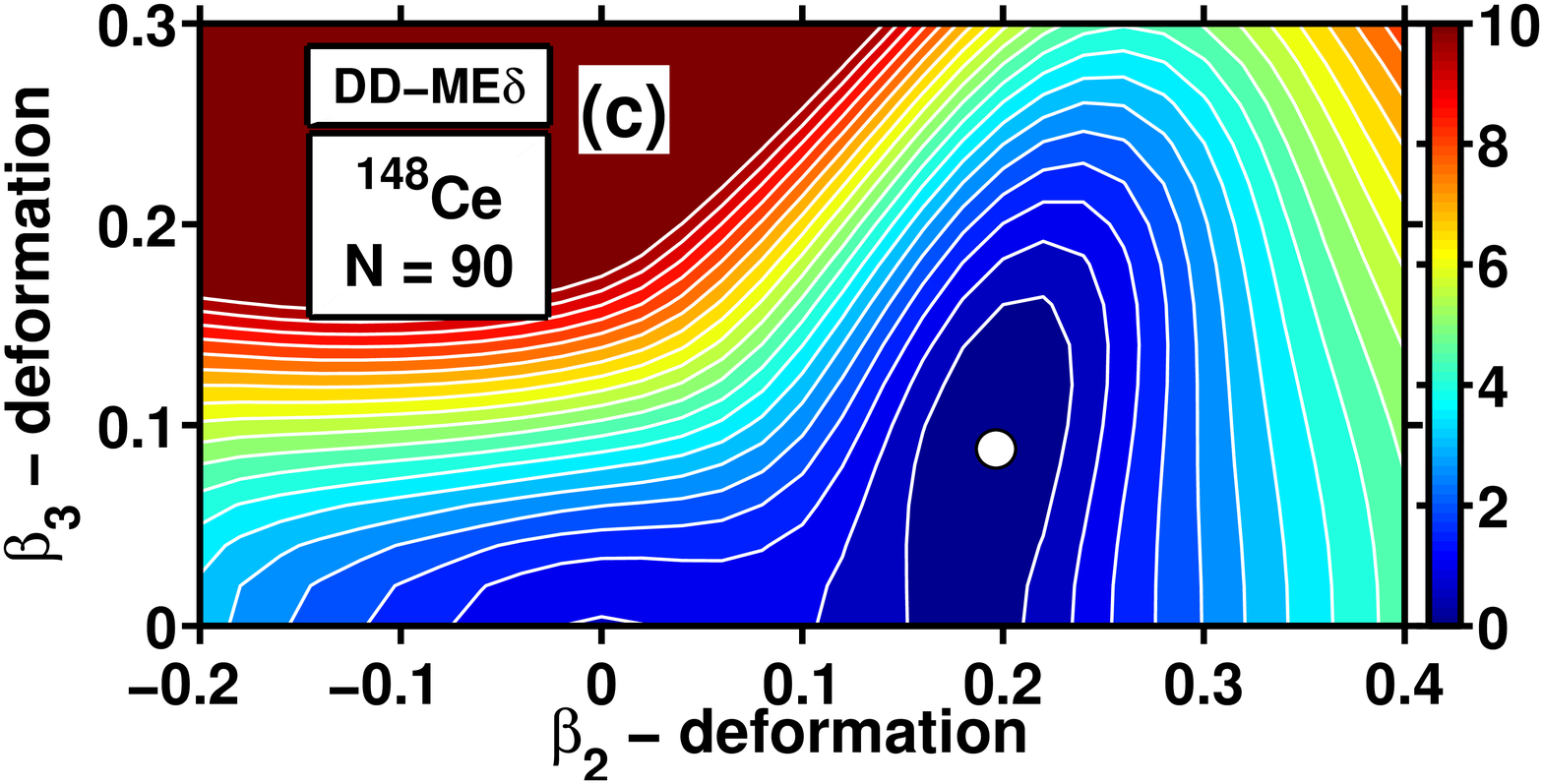}
  \includegraphics[angle=0,width=5.9cm]{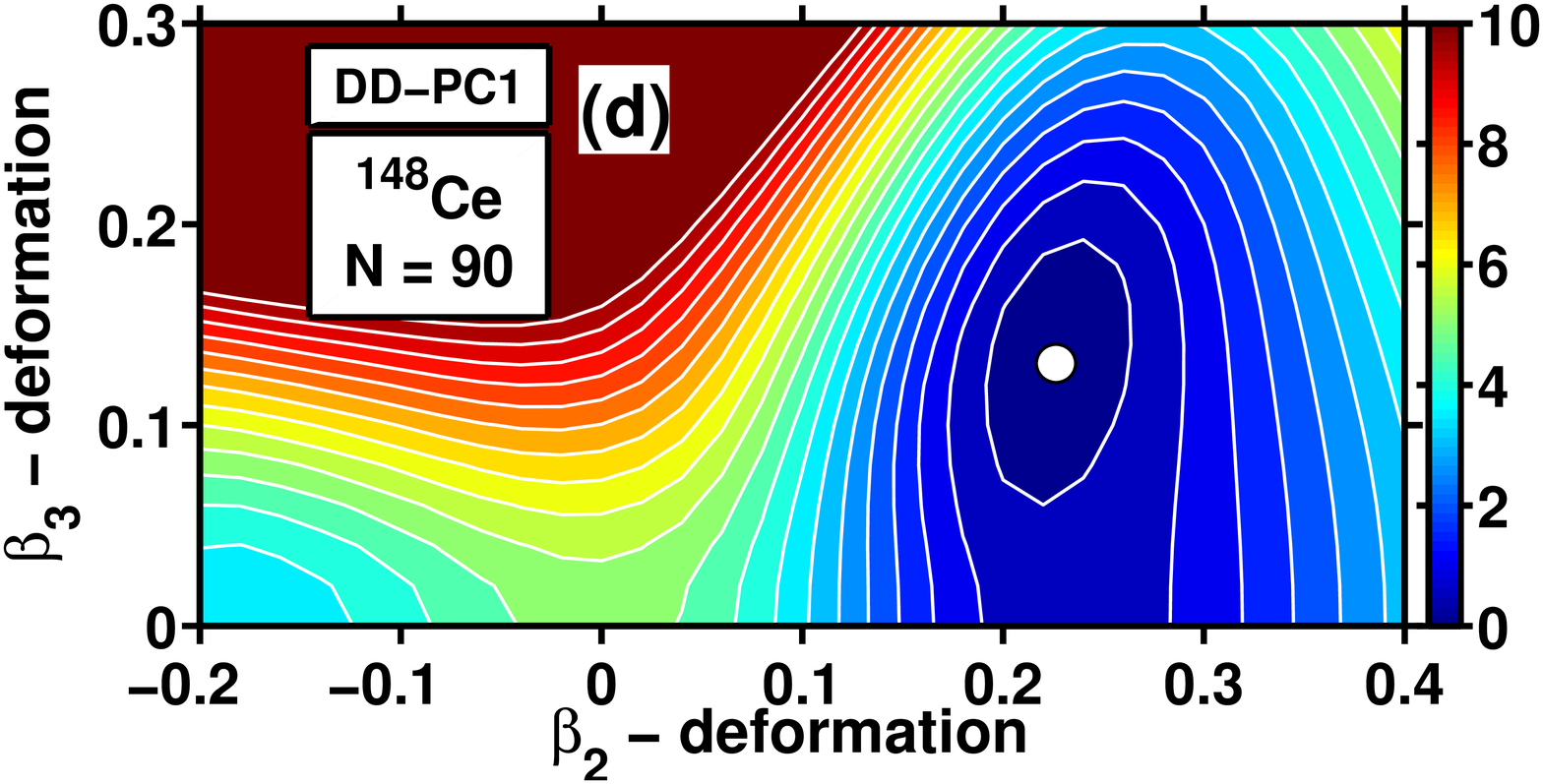}
  \includegraphics[angle=0,width=5.9cm]{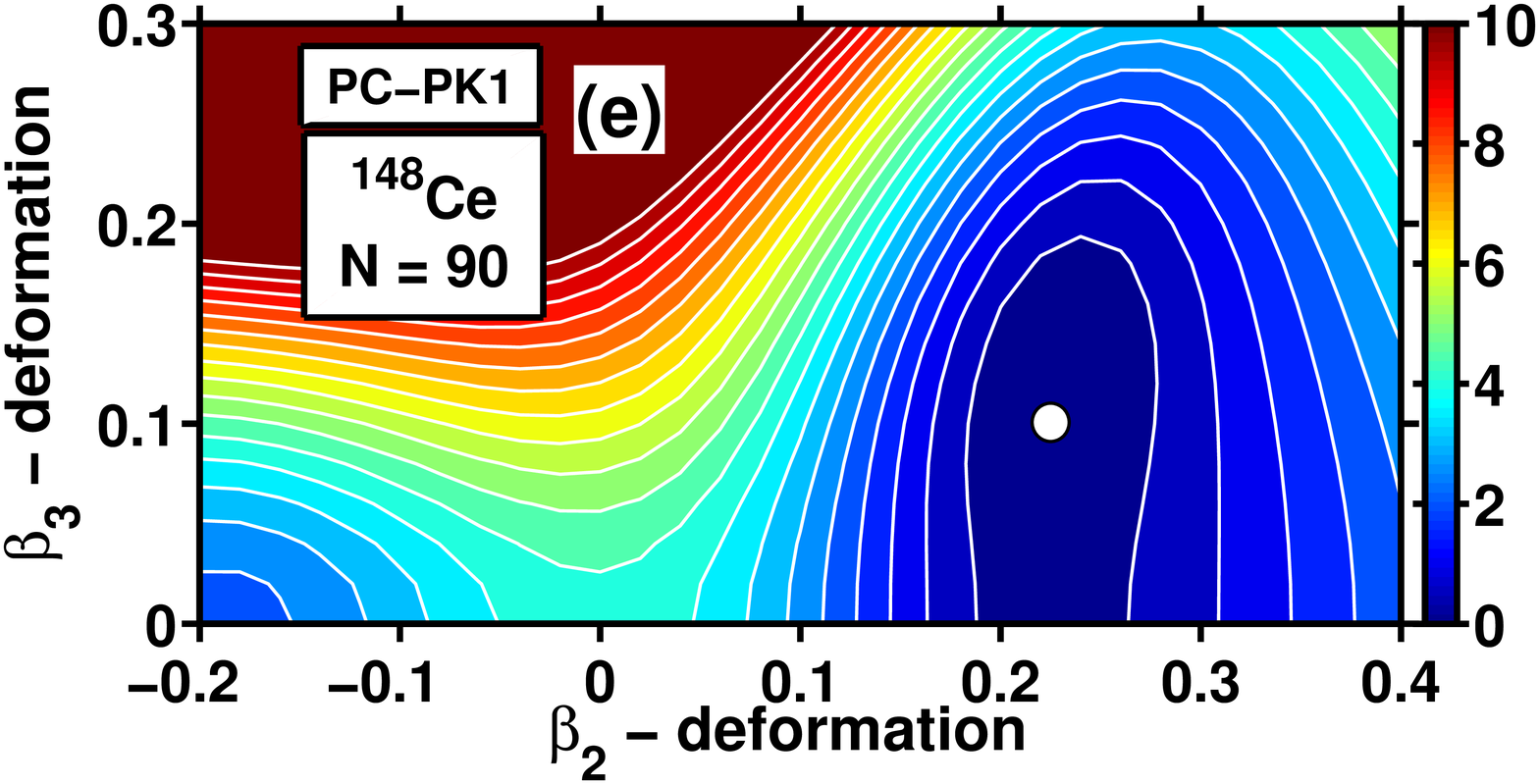}
  \caption{(Color online) The same as Fig.\ \ref{Rn_DD-PC1},
           but for the $^{148}$Ce obtained with the indicated CEDFs}
\label{Ce148-5-cedf}
\end{figure*}

  Second, the results obtained with DD-ME$\delta$ still differ from
the ones obtained with other functionals. However, the differences are
less pronounced as compared with the actinides where the RHB results
obtained with this functional contradict drastically to
experimental data and the results of other functionals. One can also
see in Fig.\ \ref{Ce148-5-cedf} that the topologies of the PESs obtained
with the five employed functionals are similar;  the only difference
is the fact that octupole minimum is somewhat deeper in the DD-ME2 and
DD-PC1 CEDFs as compared with other functionals. Note that we will not
discuss the details of the results obtained with CEDF DD-ME$\delta$ 
in the following.

 In this mass region we focus on the presentation of the RHB results and
their comparison with non-relativistic ones. In general, the island of octupole
deformation predicted in the RHB calculations is close to the ones obtained
in non-relativistic calculations. Moreover, it is close to the one extracted
from experimental data indicating either octupole deformation or enhanced
octupole correlations (see Ref.\ \cite{BN.96} for details). However, a detailed
interpretation of experimental data in this mass region at the mean field level
is complicated by the fact that PES are extremely soft in the  octupole direction
which favors the fluctuations and vibrations in this degree of freedom. 
For example, expected parity doublets in odd-mass nuclei, which are clear 
fingerprints of static octupole deformation \cite{BN.96}, are frequently not 
observed even near the center of the island of octupole deformation in the 
lanthanides \cite{Nd147.15,Ba147.13,Ba145.12}.

\subsection{Xe isotopes.}

  Our RHB calculations (including those with the CEDFs DD-ME2,
PC-PK1 and DD-ME$\delta$, not shown in Table \ref{table-global}) do not
predict non-zero octupole deformation in the $N\sim 88$ nuclei (Table
\ref{table-global}). 
On the contrary, the minimum in the PESs of the nuclei $^{142,144}$Xe with $N=86, 88$ 
is characterized by
$\beta_3 \sim 0.06$ in the MM calculations of Ref.\ \cite{MBCOISI.08} with
a folded Yukawa potential. A non-zero $\beta_3$ deformation is also seen in the
nucleus $^{144}$Xe $N=90$ in the MM calculations with a Woods-Saxon potential
\cite{NT.92}. The HF+BCS calculations with the D1S Gogny force predict 
non-zero octupole deformation in $^{142,144}$Xe \cite{MR.93}. Experimental data 
on $^{142,144}$Xe \cite{Eval-data} do not suggest the stabilization of octupole 
deformation at the ground state.

\subsection{Ba isotopes.}

\begin{figure*}
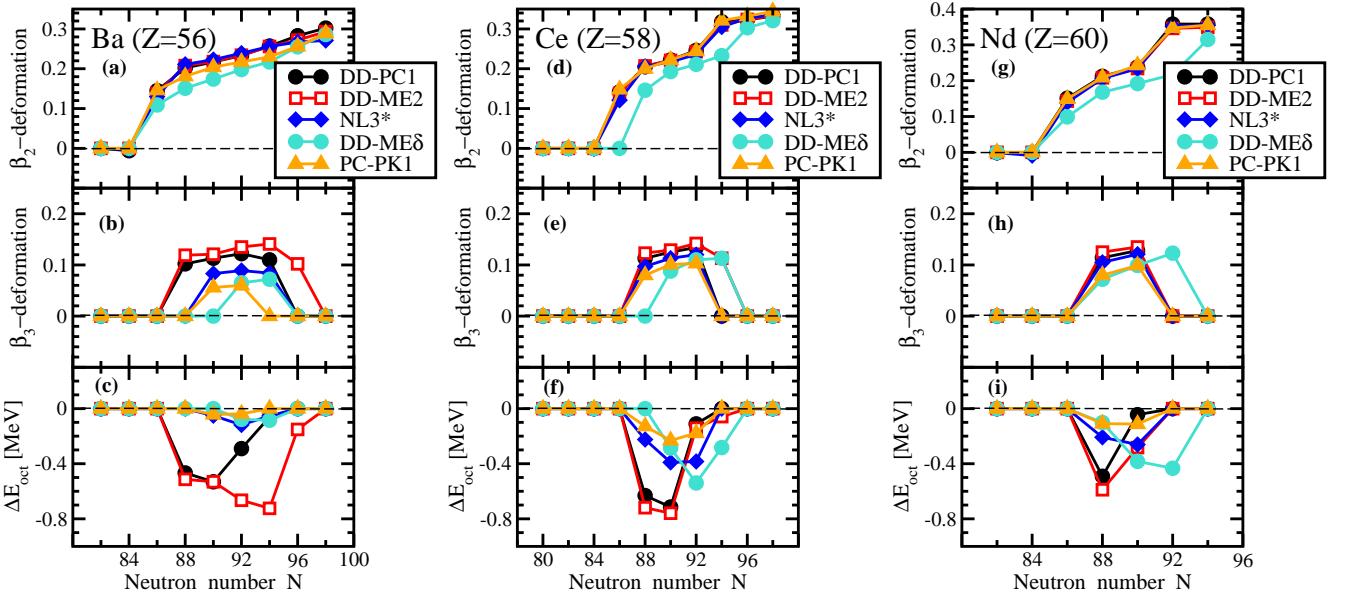

  \includegraphics[angle=0,width=5.8cm]{fig-13-a-new.eps}
  \includegraphics[angle=0,width=5.8cm]{fig-13-b-new.eps}
  \includegraphics[angle=0,width=5.8cm]{fig-13-c-new.eps}
  \caption{(Color online) The same as Fig.\ \ref{massdependence},
           but for the Ba, Ce, and Nd isotopes.}
\label{massdependence-Ba-etc}
\end{figure*}

 A non-zero octupole deformation is predicted for the $N=88-94$ $^{144-150}$Ba 
isotopes  in calculations with DD-PC1\footnote{The same nuclei were predicted
to have non-zero octupole deformation in the RHB calculations with DD-PC1 in
Ref.\ \cite{NVNL.14}.}, for the $N=88-96$ $^{144-152}$Ba isotopes with DD-ME2 
and NL3*, and for the $N=90-92$ $^{146-148}$Ba isotopes with PC-PK1
(Fig.\ \ref{massdependence-Ba-etc}). The maximum gain in binding due to octupole 
deformation takes place at $N=90$ for DD-PC1 and PC-PK1, at $N=92$ for NL3* 
and at $N=94$ for DD-ME2. The RMF+BCS calculations with PK1 CEDF of Ref.\ 
\cite{ZLZ.10} predict a finite octupole deformation in the $N=88-98$ Ba isotopes 
with a maximum octupole deformation around $N=92-94$.  On the contrary, in the 
MM calculations with a folded Yukawa potential (Ref.\ \cite{MBCOISI.08} and 
Table \ref{table-global}) only the $N=86-90$ isotopes possess non-zero octupole 
deformation. The MM results of Ref.\ \cite{NT.92} based on a Woods-Saxon potential 
show non-zero octupole deformation only in the $N=88-90$ nuclei. The HF+BCS 
calculations with Gogny D1S force of Ref.\ \cite{ER.92} predict non-zero octupole 
deformation in the $N=88-92$ $^{142-148}$Ba nuclei with a maximum gain of binding 
due to octupole deformation at the nucleus $^{144}$Ba with $N=90$.

\subsection{Ce isotopes.}

  The $N=88-92$ $^{146-150}$Ce nuclei have $\beta_3\neq 0$ in the RHB calculations
with the CEDFs DD-PC1, NL3* and PC-PK1. The DD-ME2 functional provides one extra nucleus
($N=94$ $^{152}$Ce) with non-zero octupole deformation. In all functionals, the
maximum of $|\Delta E^{oct}|$ is reached at $N=90$. The MM calculations with folded
Yukawa \cite{MBCOISI.08} (see also Table \ref{table-global}) and Woods-Saxon \cite{NT.92}
potentials predict non-zero octupole deformation in the  $N=86-90$ $^{144-148}$Ce and
$N=86-88$ $^{144,146}$Ce nuclei, respectively. The HF+BCS calculations with the Gogny D1S
force of Ref.\ \cite{ER.92} predict octupole deformation in the $N=84-90$ $^{142-148}$Ce
nuclei.

\subsection{Nd isotopes.}

  The $N=88-90$ $^{148-150}$Nd nuclei are predicted to be octupole deformed
with four CEDFs. The maximum gain in binding due to octupole deformation
is reached at $N=88$ for DD-PC1 and DD-ME2 and at $N=90$ for NL3* and PC-PK1.
The $N=86-88$ $^{146,148}$Nd nuclei are predicted to be octupole deformed in
the MM calculations with a folded Yukawa potential (see Table \ref{table-global}).
On the contrary, the MM calculations with a Woods-Saxon potential do not predict
octupole deformed Nd nuclei \cite{NT.92}. The $N=86-88$ $^{146-148}$Nd isotopes
are predicted to be octupole deformed in the HF+BCS calculations with the Gogny
force D1S \cite{ER.92}.

\subsection{Sm isotopes.}

  Our calculations predict that only in the nucleus $^{150}$Sm a very shallow
octupole minimum  (with $|\Delta E^{oct}|=0.25$ MeV for DD-PC1 and
$|\Delta E^{oct}|=0.09$ MeV for NL3*) is formed. On the contrary,
the RMF+BCS calculations with PK1 presented in Ref.\ \cite{ZLZM.10}
predict non-zero octupole deformation in the nuclei $^{146-152}$Sm. The maximum
gain in binding due to octupole deformation ($|\Delta E^{oct}|=1.36$
MeV) takes place in the $N=88$ $^{150}$Sm nucleus. This large $|\Delta E^{oct}|$
value would suggest a stabilization of octupole deformation in the ground state.
However, the experimental data do not support such a possibility \cite{BN.96,Eval-data}.
Non-zero octupole deformation has also been seen in RHB calculations with
DD-PC1 in $^{148,150,156}$Sm \cite{NVNL.14}. However, the PESs
are extremely soft in octupole direction in the $^{148,156}$Sm nuclei, so that  a
slightly stronger pairing (as compared with Ref.\ \cite{NVNL.14})
could easily drive the system to reflection symmetry. A somewhat deeper octupole
minimum is seen in $^{150}$Sm in Ref.\ \cite{NVNL.14}. However, in this case the
gain in binding due to octupole deformation is comparable with the one presented
in Table \ref{table-global} if the difference in pairing is taken into account.
The MM calculations with a Woods-Saxon potential of Ref.\ \cite{NT.92} show that
$^{150}$Sm is reflection symmetric in the ground state, while the ones with
a folded Yukawa potential (Ref.\ \cite{MBCOISI.08} and Table \ref{table-global})
suggests that this nucleus is only soft in octupole direction ($|\Delta E^{oct}| = 20$
keV). The HF+BCS calculations with the Gogny force D1S of Ref.\ \cite{ER.92} predict
octupole deformation in the $N=86-88$ $^{148-150}$Sm nuclei. On the contrary, the
HFB calculations with D1S and D1M forces predict octupole deformation only
in $^{150}$Sm \cite{RRS.12}. However, the gain in binding due to octupole 
deformation is small (204 keV for D1S and 43 keV for D1M).

\section{The impact of pairing on the relative energies of minima 
         with and without octupole deformation.}
\label{sec-pairing-impact}

 The extrapolation beyond  the known region of nuclei is always accompanied
with a number of uncertainties related to the theoretical description of
finite nuclei. For example, some of them are connected with the uncertainties
in the theoretical description of single-particle energies
\cite{AS.11,A.14-jpg,AARR.14,AARR.15}; they emerge from the particle-hole
channel of the density functional theories (DFTs). To some degree, these 
uncertainties can be estimated
by using a set of different functionals as it is done in the present manuscript. In
addition, there are the uncertainties in the particle-particle (pairing)
channel which become especially large in the vicinity of two-neutron drip
line (see Refs.\ \cite{PMSV.13,AARR.15}).  It is also expected that they can
affect the relative energies of the minima with and without octupole deformation 
and possibly
the topology of  the potential energy surfaces in the cases of very soft PESs. For
example, it is well known that the selection of the pairing force and the pairing
strength affects the potential energy surfaces of fissioning nuclei and their
fission barriers (see Ref.\ \cite{KALR.10} and references therein).  Moreover,
as shown in this reference, fission barrier heights decrease with increasing
pairing strength.

\begin{figure*}
  \includegraphics[angle=0,width=8.8cm]{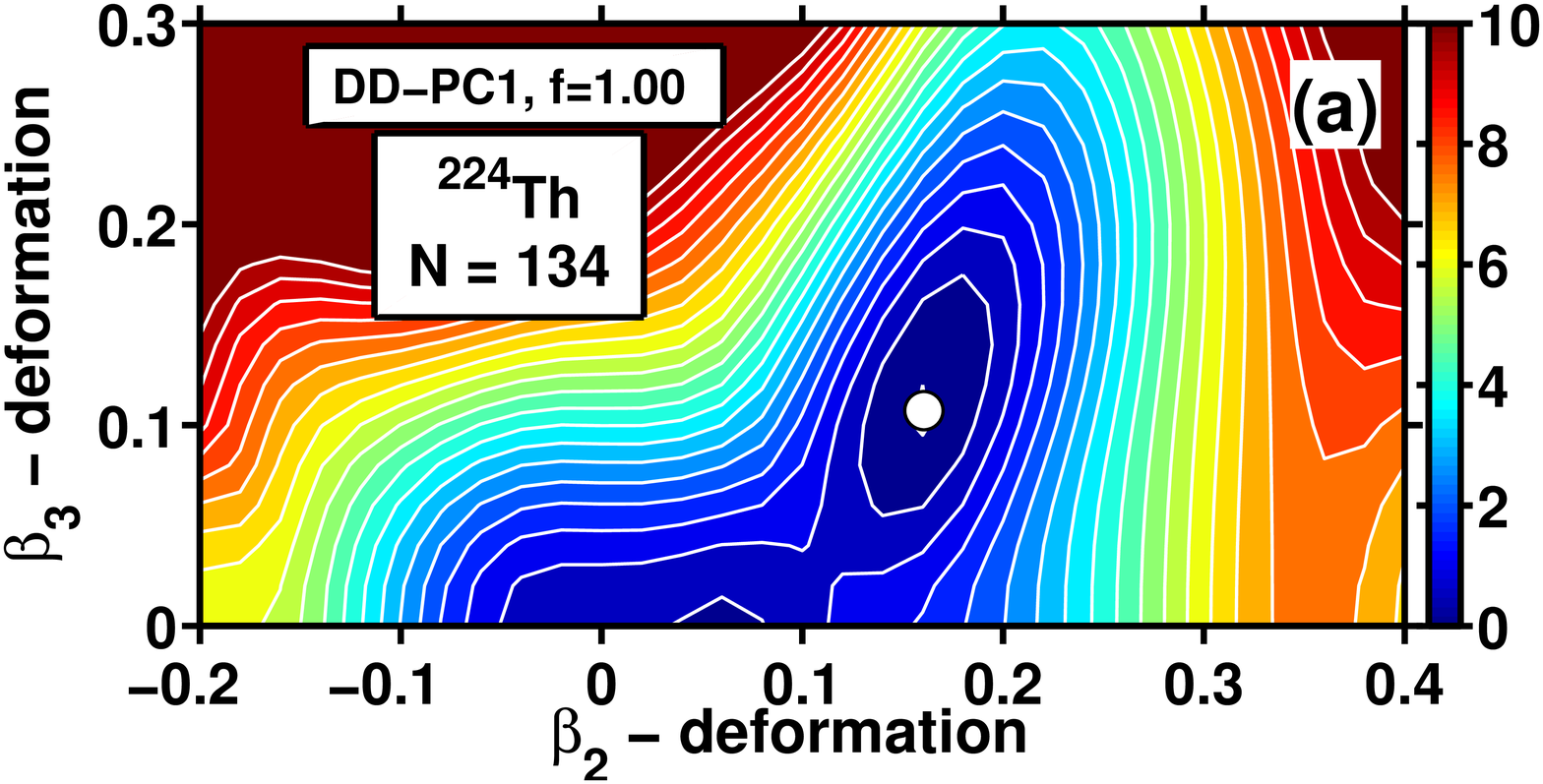}
  \includegraphics[angle=0,width=8.8cm]{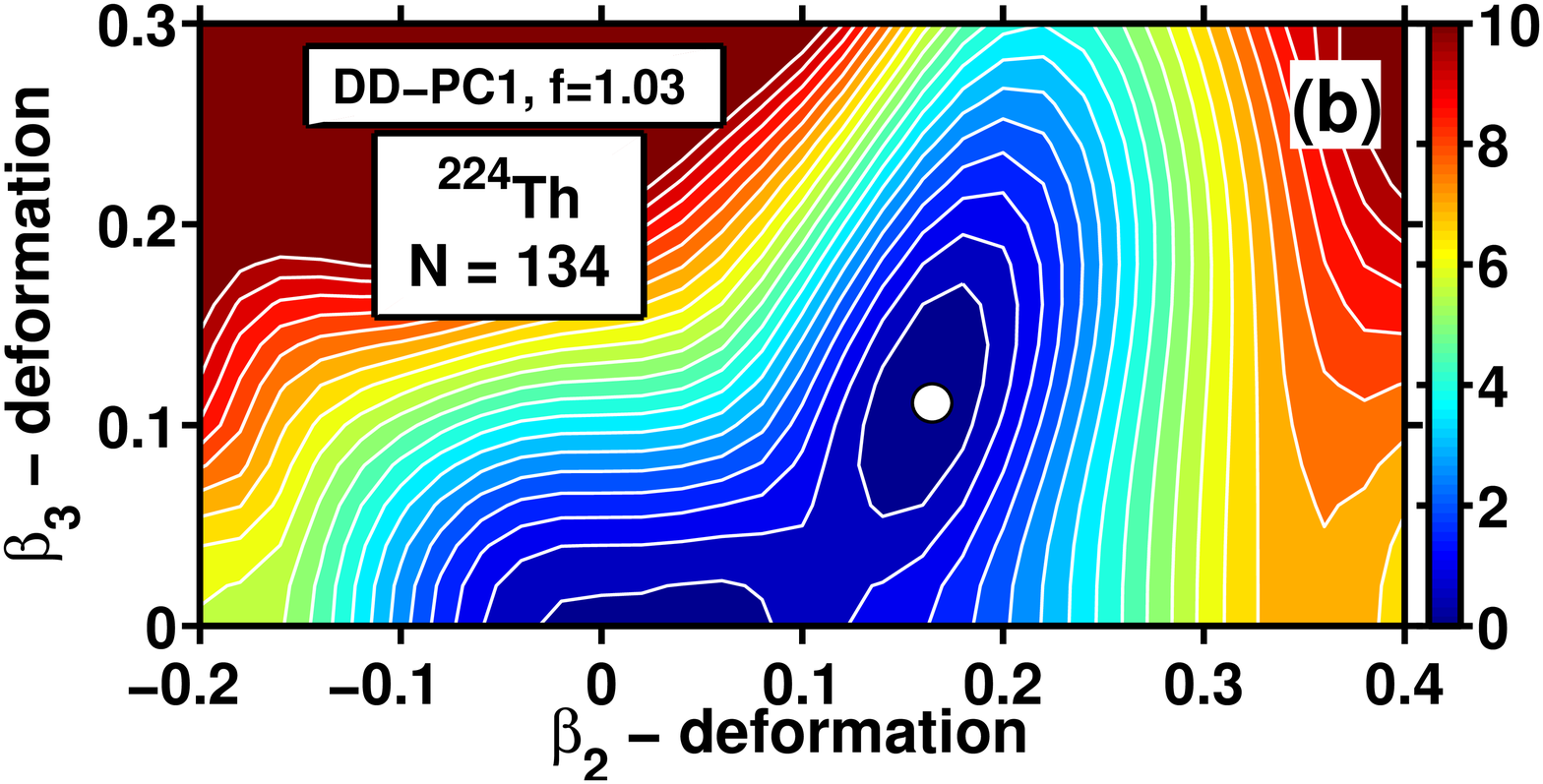}
  \includegraphics[angle=0,width=8.8cm]{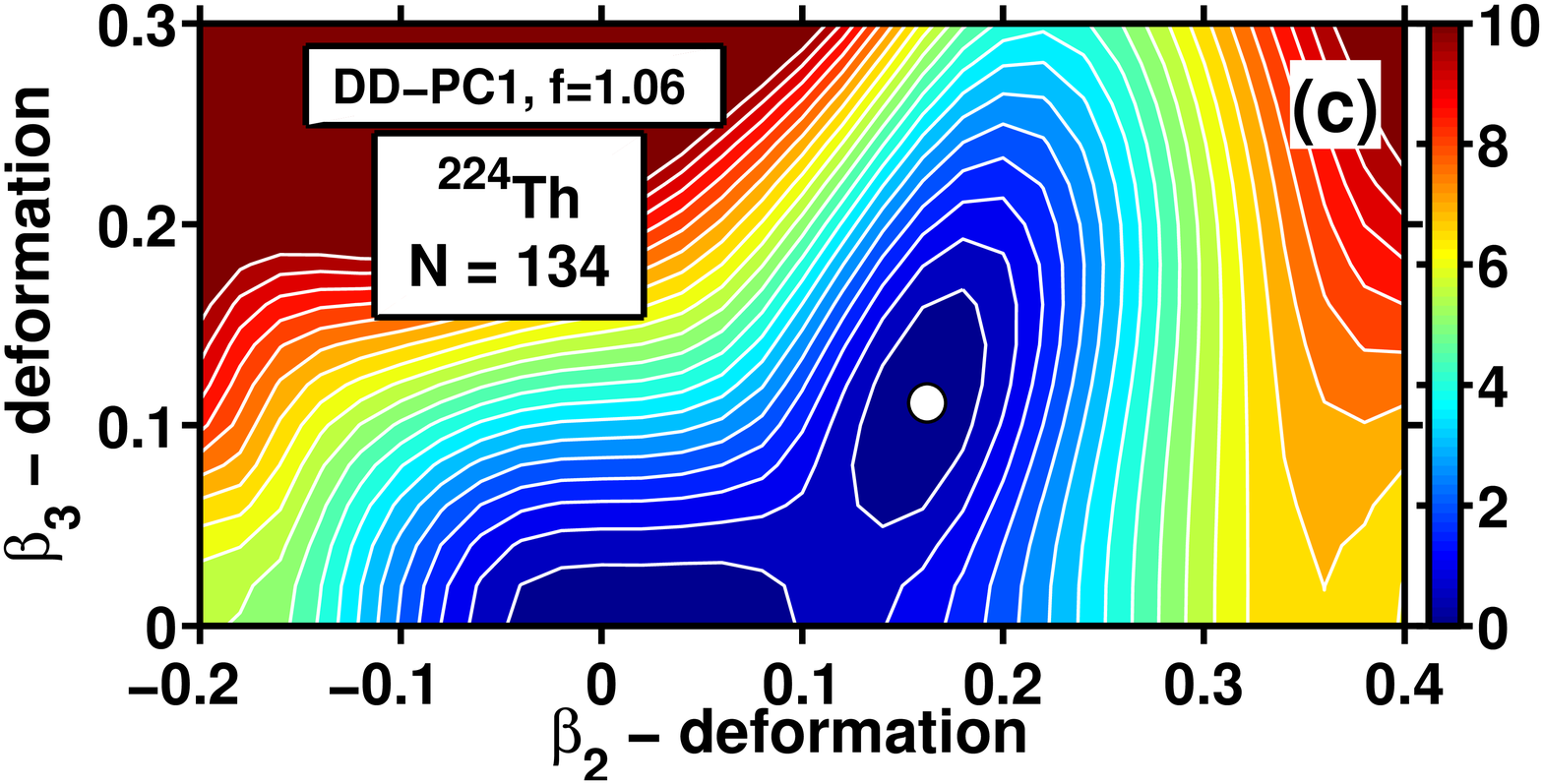}
  \includegraphics[angle=0,width=8.8cm]{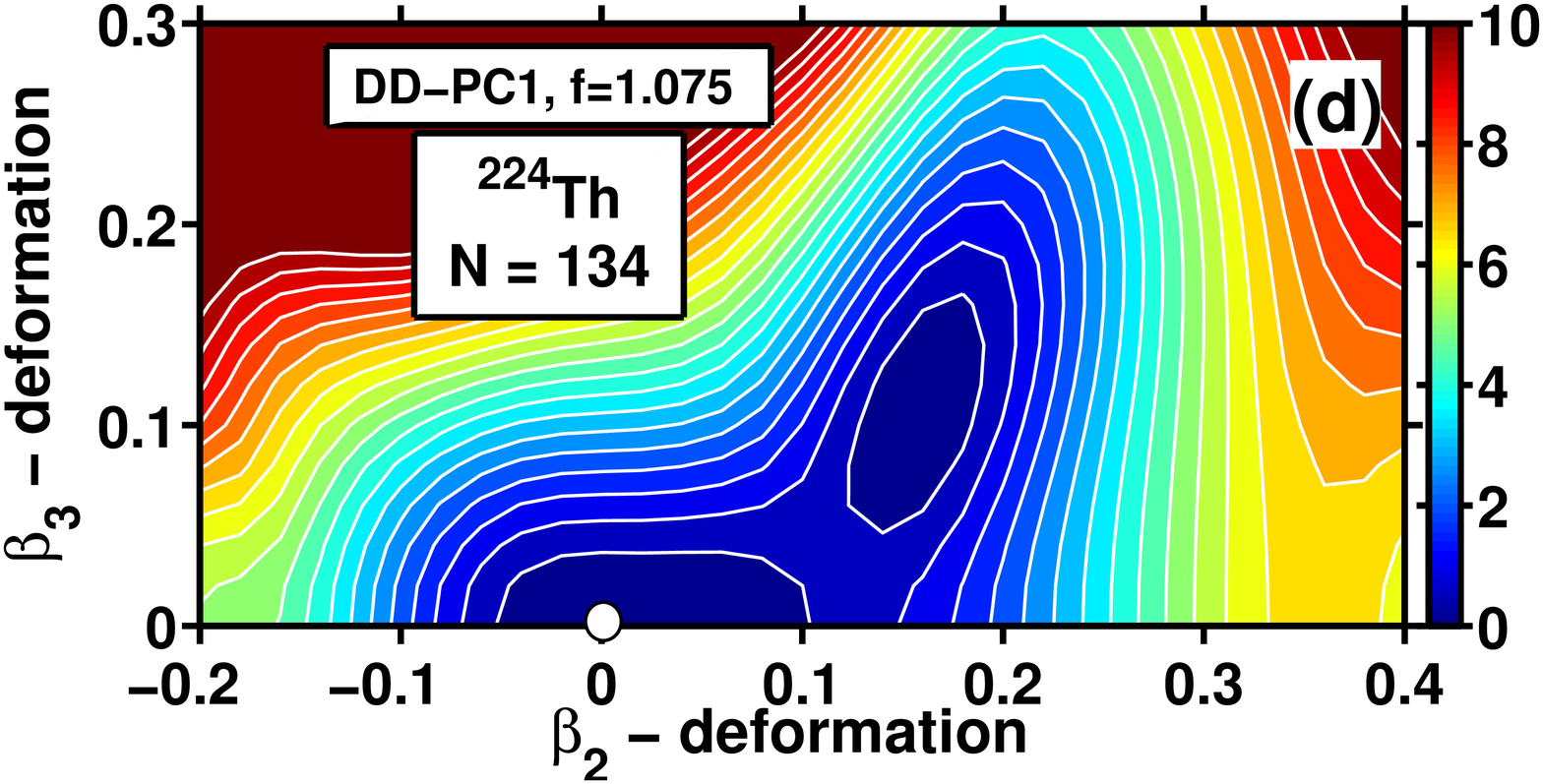}
 \caption{(Color online) Potential energy surfaces of  $^{224}$Th in the
          $(\beta_2,\beta_3)$ plane
          calculated with the CEDF DD-PC1 for different values
          of scaling factor $f$ of the pairing strength. White
          circle indicates the global minimum. Equipotential
          lines are shown in steps of 0.5 MeV.}
\label{Th_DD-PC1_diff_vfac}
\end{figure*}

\begin{figure}
  \includegraphics[angle=0,width=8.0cm]{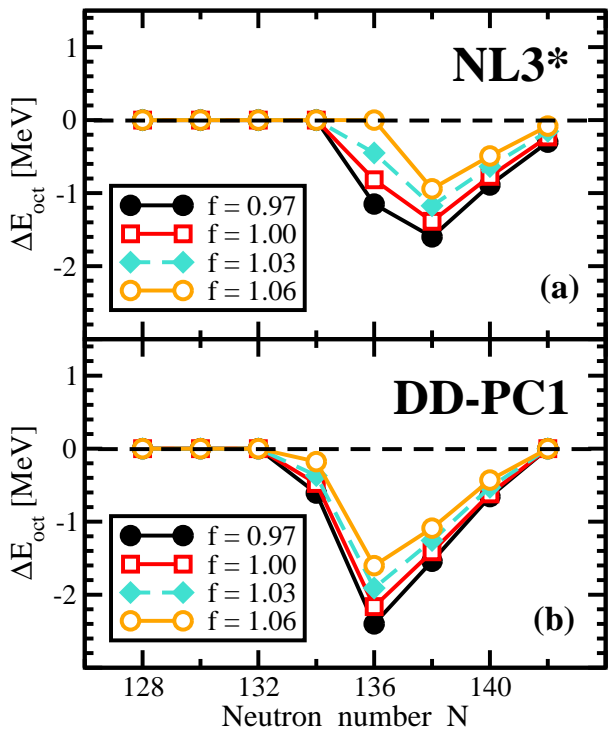}
  \caption{(Color online). The impact of the variation of the scaling
           factor $f$ of the pairing force (Eq.(\ref{TMR})) on the
           $\Delta E_{oct}$ quantity in the $^{218-232}$Th isotopes. The 
           results obtained with CEDFs NL3* and DD-PC1 and scaling 
           factors $f=0.97, 1.00, 1.03$
           and 1.06 are presented.}
\label{pairing}
\end{figure}

 In order to understand how the variation of pairing strength affects
the potential energy surfaces and relative energies of minima with and 
without octupole deformation in octupole soft nuclei we have performed 
RHB calculations with different values of the scaling factor $f$ in 
Eq.\ (\ref{TMR}) for the pairing force. The results of these calculations 
are summarized in Figs.\ \ref{Th_DD-PC1_diff_vfac} and \ref{pairing}.

  The impact of the pairing strength on the topology of the PESs
is shown in Fig.\ \ref{Th_DD-PC1_diff_vfac}. Two local minima
with $\beta_2 \sim 0.05, \beta_3=0.0$ (further on called quadrupole
minimum) and $\beta_2 \sim 0.15, \beta_3 \sim 0.12$ (further on called octupole
minimum) exist for all values of the scaling factor $f$. Although in
this case the topology of the PES is not strongly affected by the change of
$f$, two important features are seen.  First, similar to fissioning nuclei
(Ref.\ \cite{KALR.10}) the barrier between quadrupole and octupole minima
decreases with the increase of pairing strength. Second, the increase of
pairing strength changes the relative energies of octupole and quadrupole 
minima. The octupole minimum is the lowest in energy for the values of
$f=1.00, 1.03$ and 1.06. However, the energy difference $|\Delta E^{oct}|$
between these two minima decreases with increasing scaling factor $f$. It is 
well known that strong pairing favors spherical configurations. This leads to 
the fact that at $f=1.075$ pairing is so strong that the quadrupole minimum 
becomes spherical; this minimum is also the lowest in energy.

 Similar effects are seen in more systematic investigations presented
in Fig.\ \ref{pairing} for the chain of the Th isotopes. The largest values
for $|\Delta E^{oct}|$ are always observed for the weakest pairing. This result
does not depend on the functional under consideration. Thus, one can conclude that
in general pairing counteracts the shell effects and favors the shapes
with no octupole deformation. Or vise versa, the strongest impact of the
octupole deformation (as quantified by $\Delta E^{oct}$) is expected in
the systems with no pairing. Note also that the variation of pairing
strength typically does not affect the neutron number at which the
maximum gain due to octupole deformation takes place (at $N=136$ for
DD-PC1 and at $N=138$ in NL3*).

   Note that the simple analysis within the random phase approximation
presented in Sec.\ IIIA of Ref.\ \cite{BN.96} also indicates that
pairing has a tendency to make the system less octupole deformed.

\section{Consequences for rotational nuclei}
\label{rotation}

In Ref.\ \cite{AO.13} a systematic investigation of rotational properties of
the actinides within the cranked relativistic
Hartree-Bogoliubov approach with approximate particle
number projection by means of the Lipkin-Nogami method
(further CRHB+LN) has revealed 
in light even-even actinides with neutron number $N\leq 146$ ranging from
$^{230}$Th up to $^{240}$Pu  a paired band crossing
leading to an upbend in the kinematic moment of inertia $J^{(1)}$
at a rotational frequency $\Omega_x \sim 0.2$ MeV (see Figs.\ 9 and 10 in Ref.\
\cite{AO.13}). However, this upbend in $J^{(1)}$ is absent in the
experiment. Note that such problems with the description of
rotational properties of light actinides exists in all cranking
calculations (see Sec.\ IV of Ref.\ \cite{AO.13} for details).
On the contrary, no such problems exist in the description of
the band crossings in even-even actinides with neutron number
$N\geq 148$ within the CRHB+LN approach (Ref.\ \cite{AO.13});
for these nuclei this approach has also a good predictive power
(Refs.\ \cite{A.14,240U}).

  The problem in the description of rotational properties
of the $N\leq 146$ actinides in Ref.\ \cite{AO.13} is most likely related to the
stabilization of octupole deformation at high spin which is
not taken into account in these model calculations. The
arguments in favor of such an interpretation have been
reviewed in Sec. IV of Ref.\ \cite{AO.13}. In particular,
stable octupole deformation has been shown to delay
alignment processes \cite{FP.84} and this may explain
the differences between theory and experiment in light
actinides.

\begin{figure*}
  \includegraphics[angle=0,width=5.9cm]{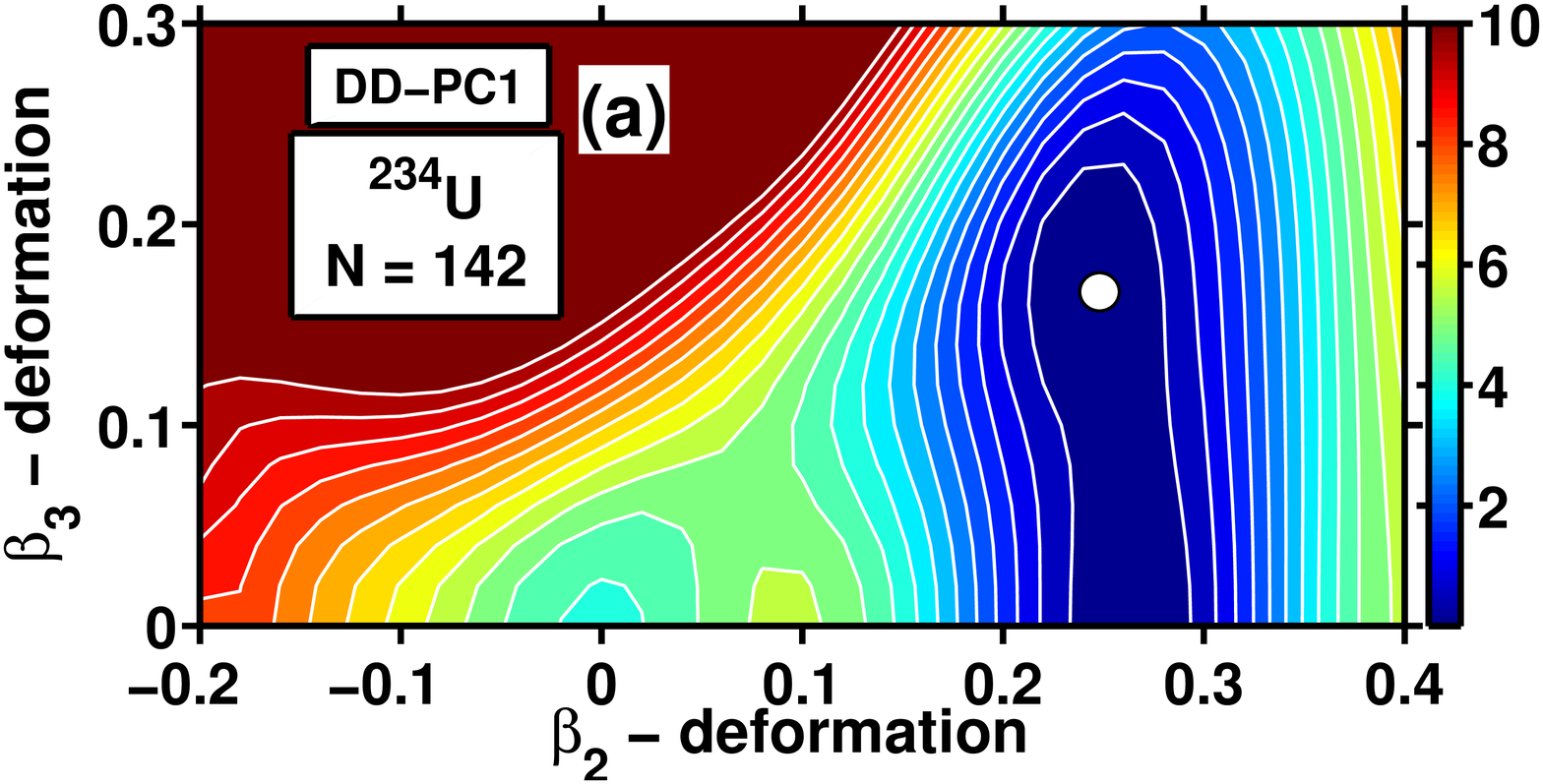}
  \includegraphics[angle=0,width=5.9cm]{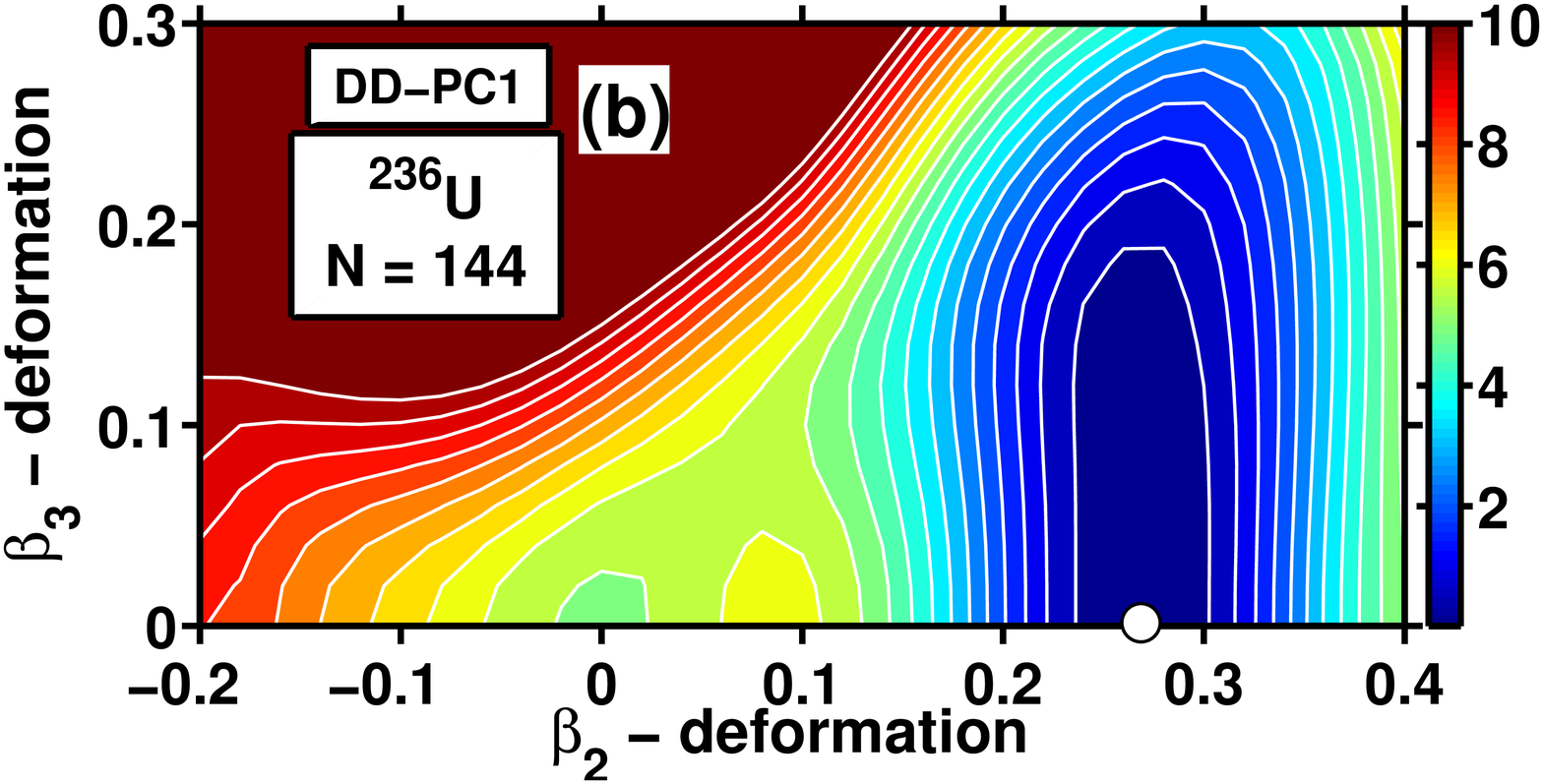}
  \includegraphics[angle=0,width=5.9cm]{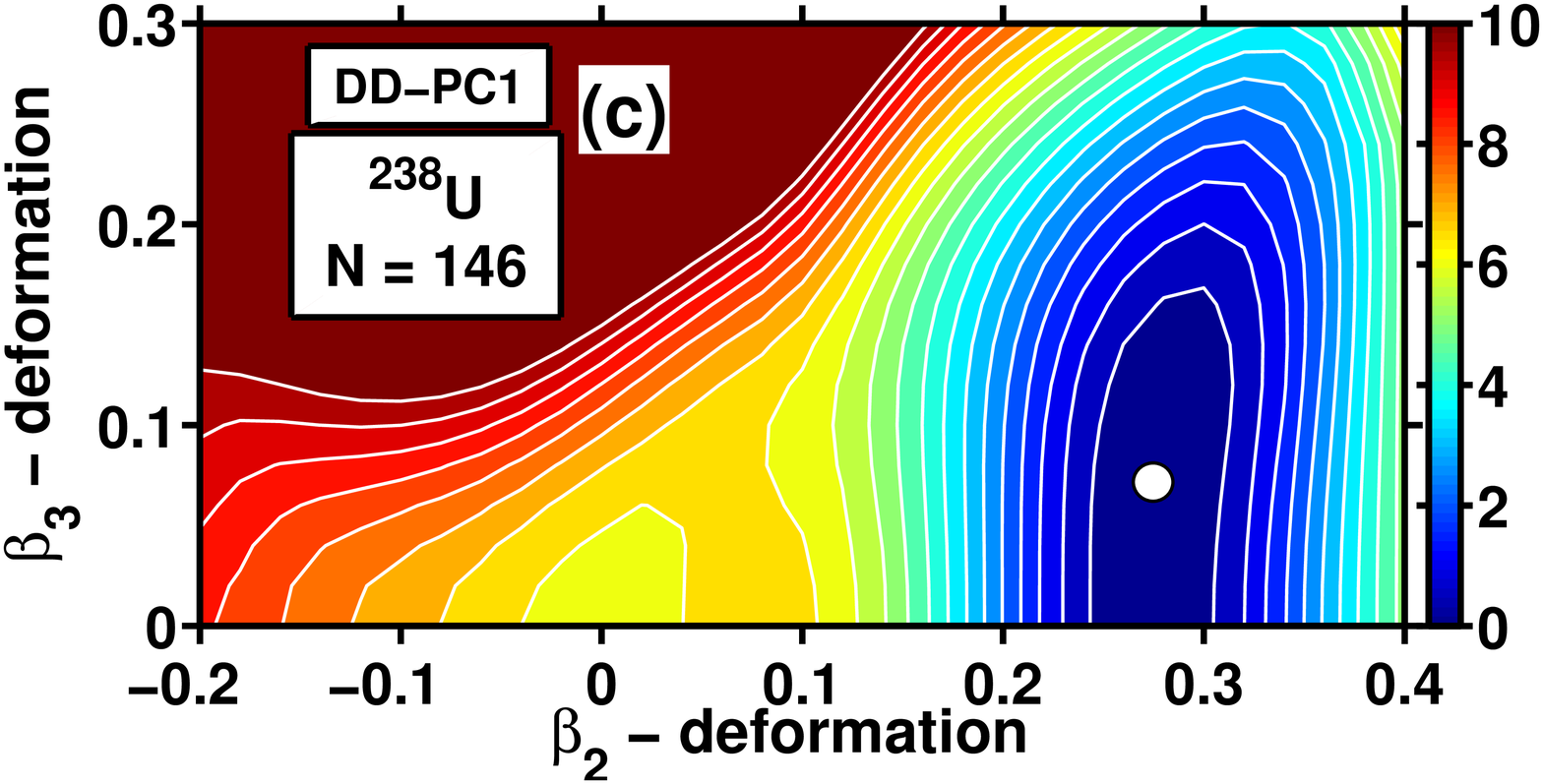}
  \includegraphics[angle=0,width=5.9cm]{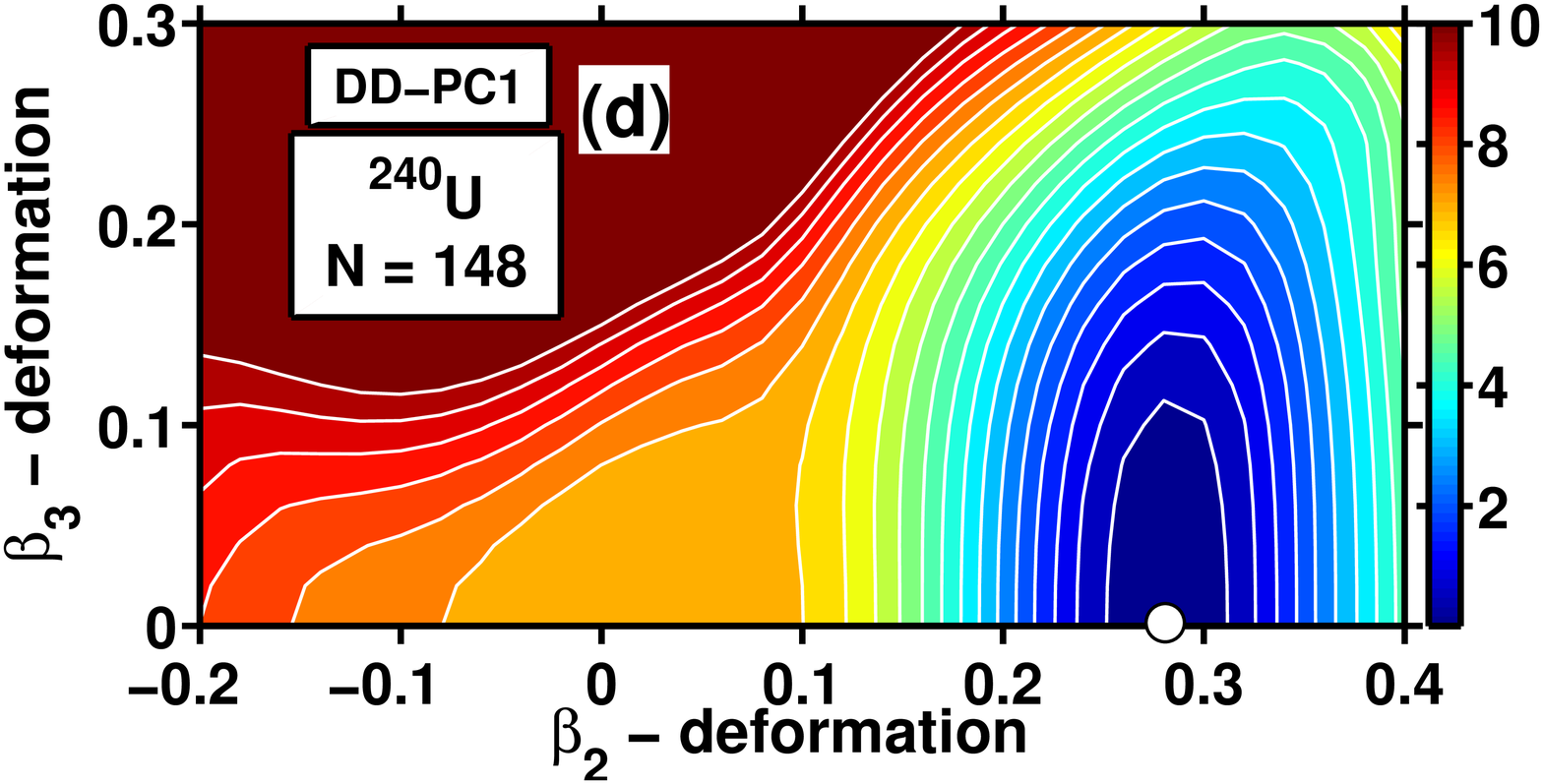}
  \includegraphics[angle=0,width=5.9cm]{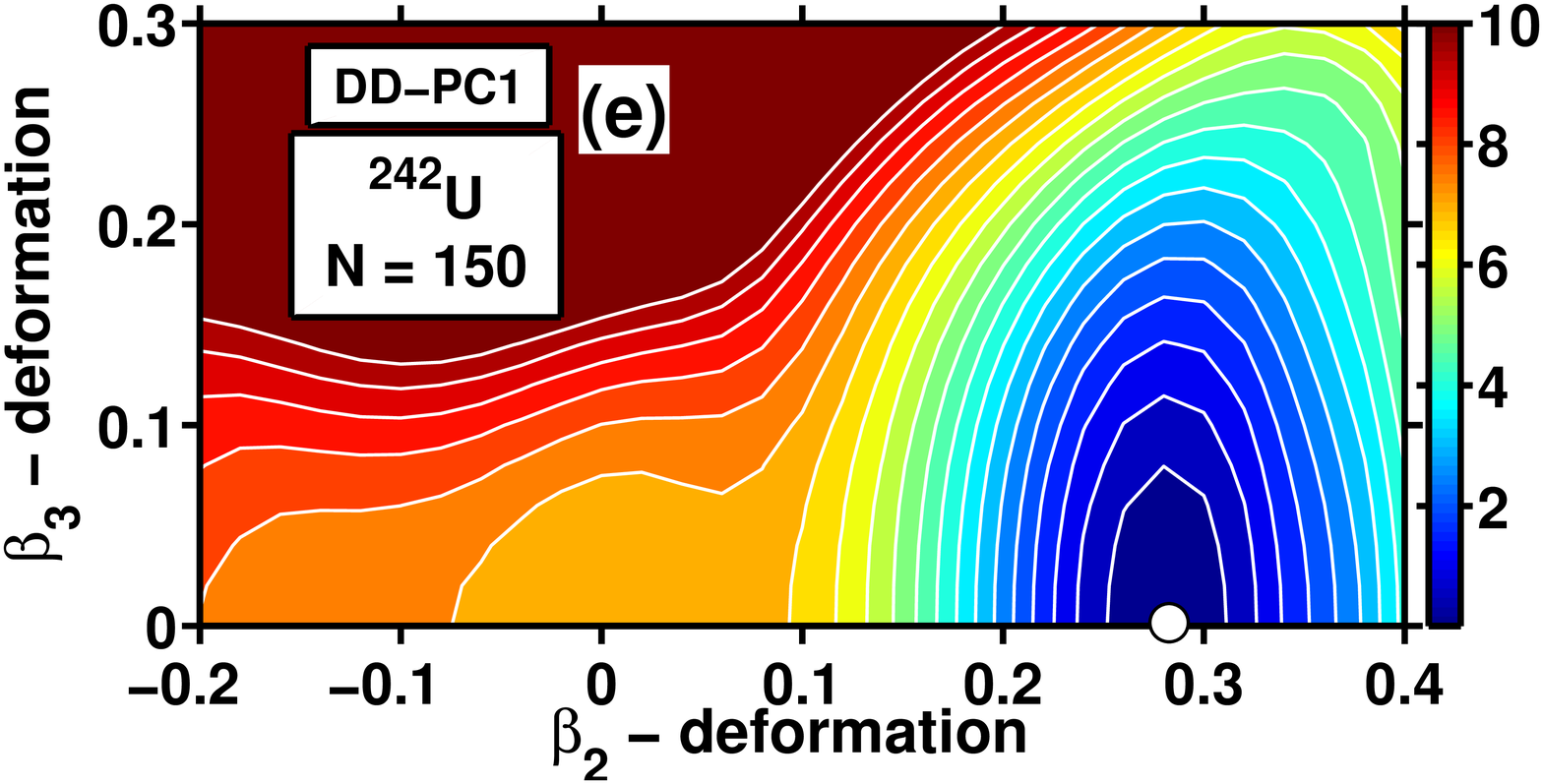}
  \caption{(Color online) The same as Fig.\ \ref{Rn_DD-PC1},
           but for the U isotopes with $N=142-150$ calculated
           with the DD-PC1 functional.}
\label{U_DD-PC1_rot}
\end{figure*}

\begin{figure*}
  \includegraphics[angle=0,width=5.9cm]{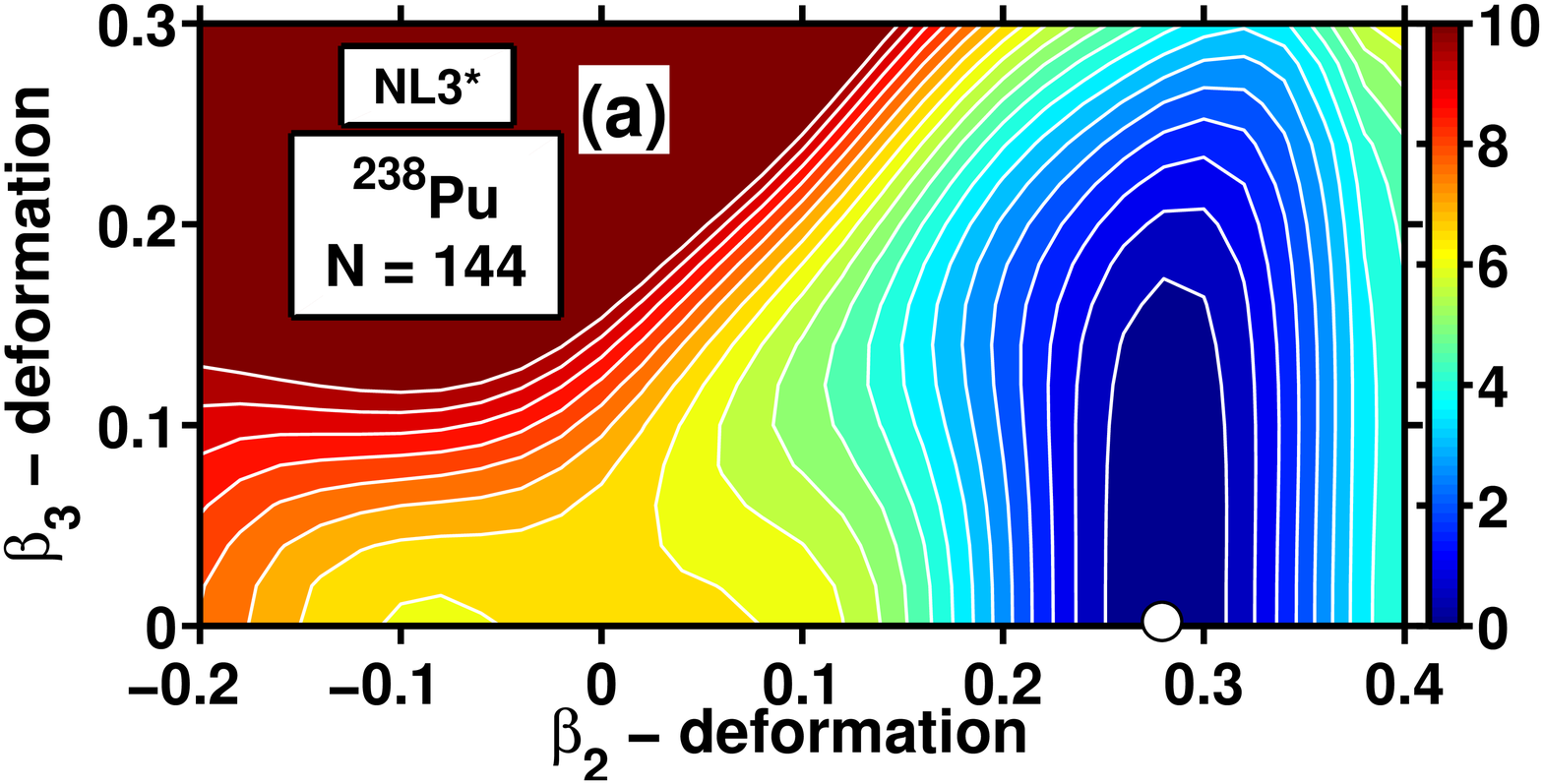}
  \includegraphics[angle=0,width=5.9cm]{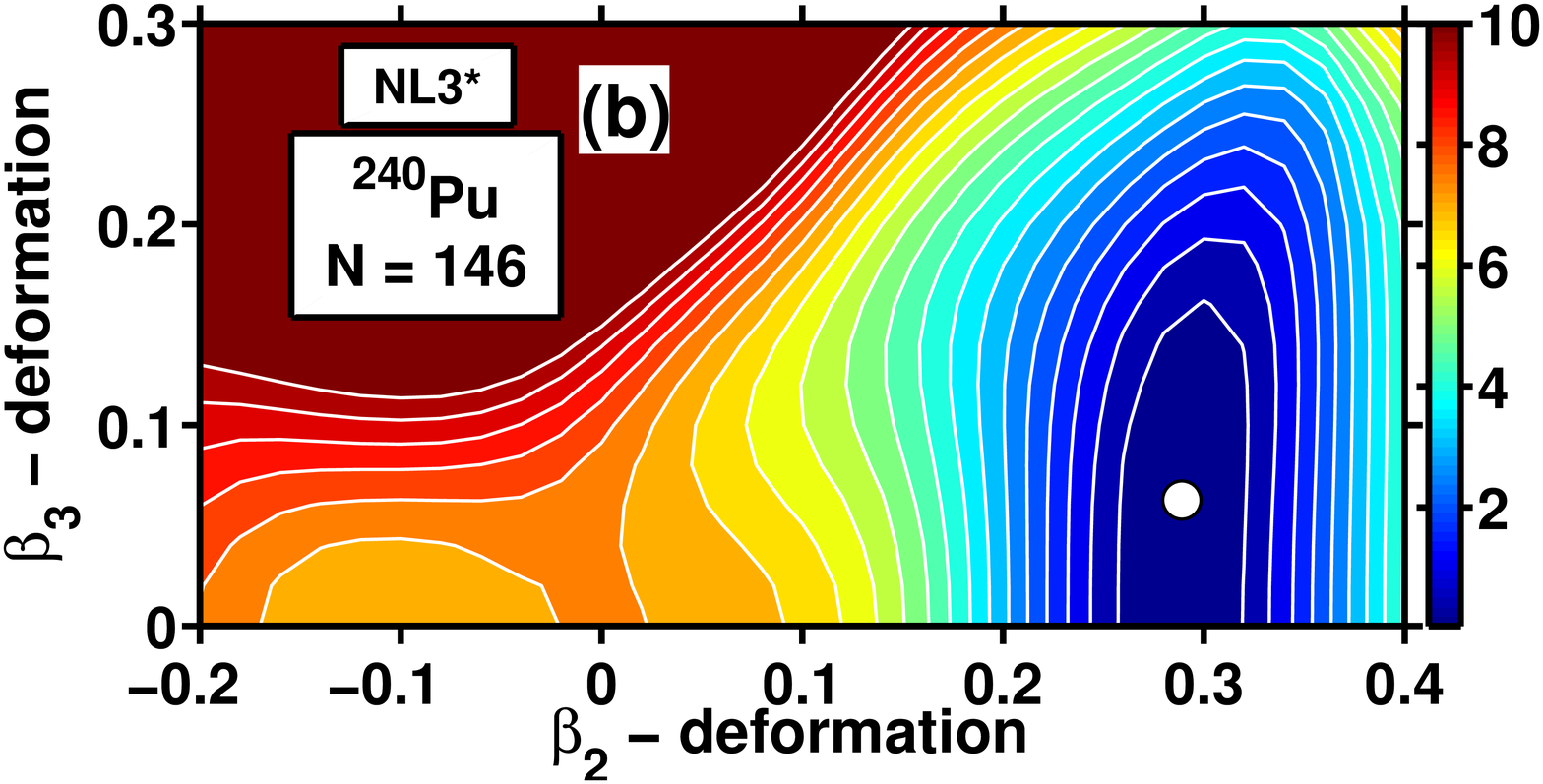}
  \includegraphics[angle=0,width=5.9cm]{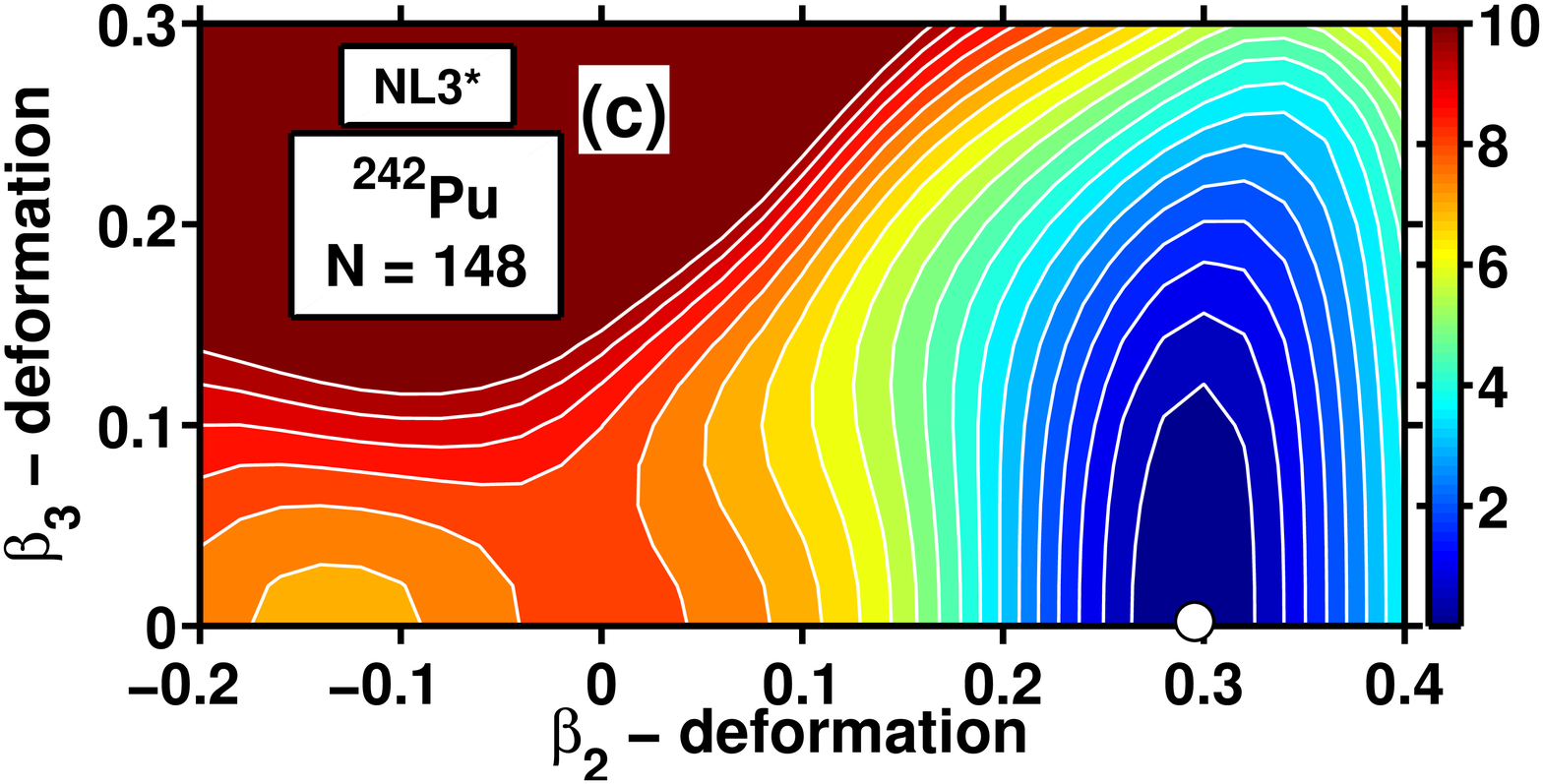}
  \includegraphics[angle=0,width=5.9cm]{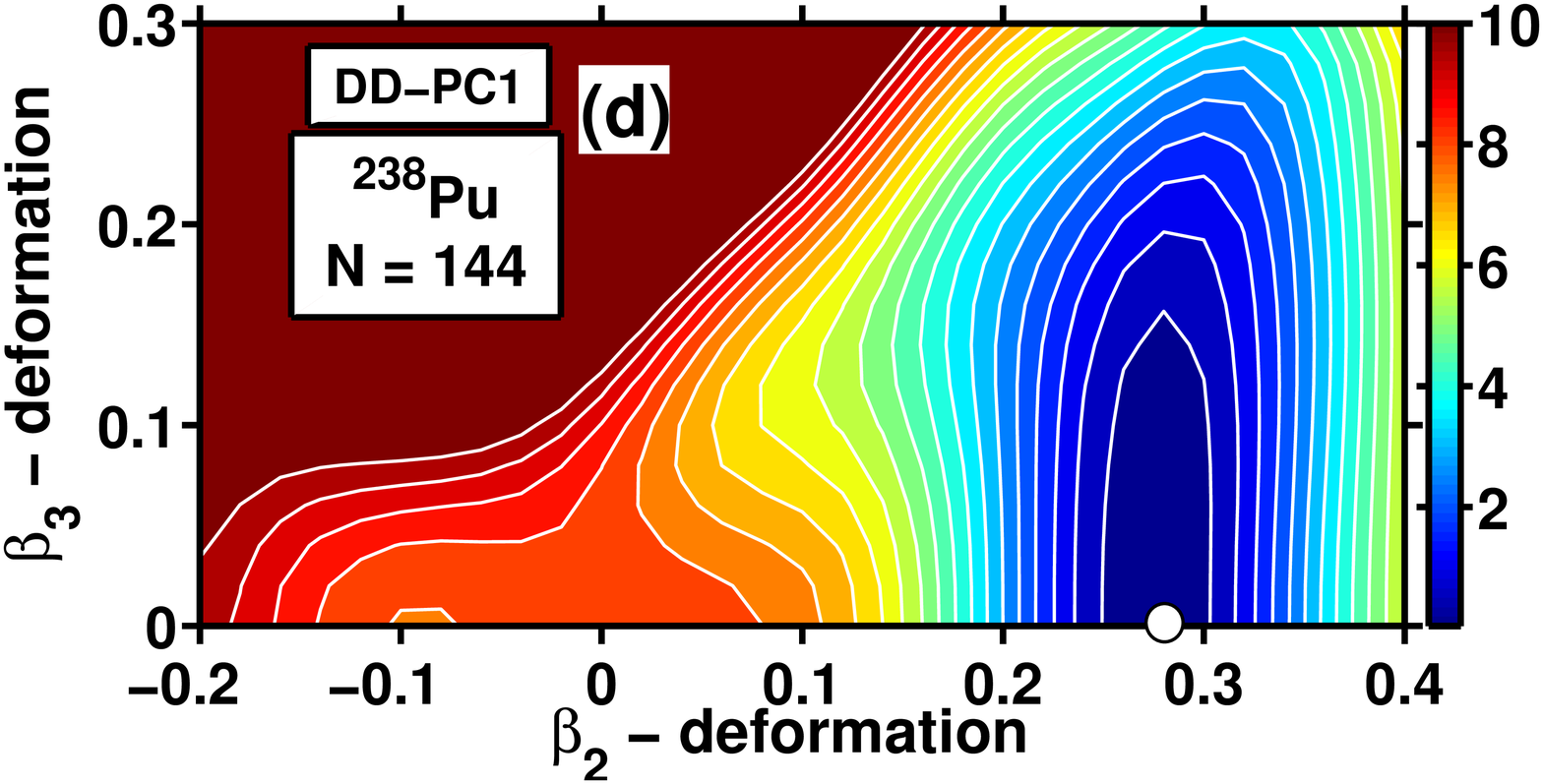}
  \includegraphics[angle=0,width=5.9cm]{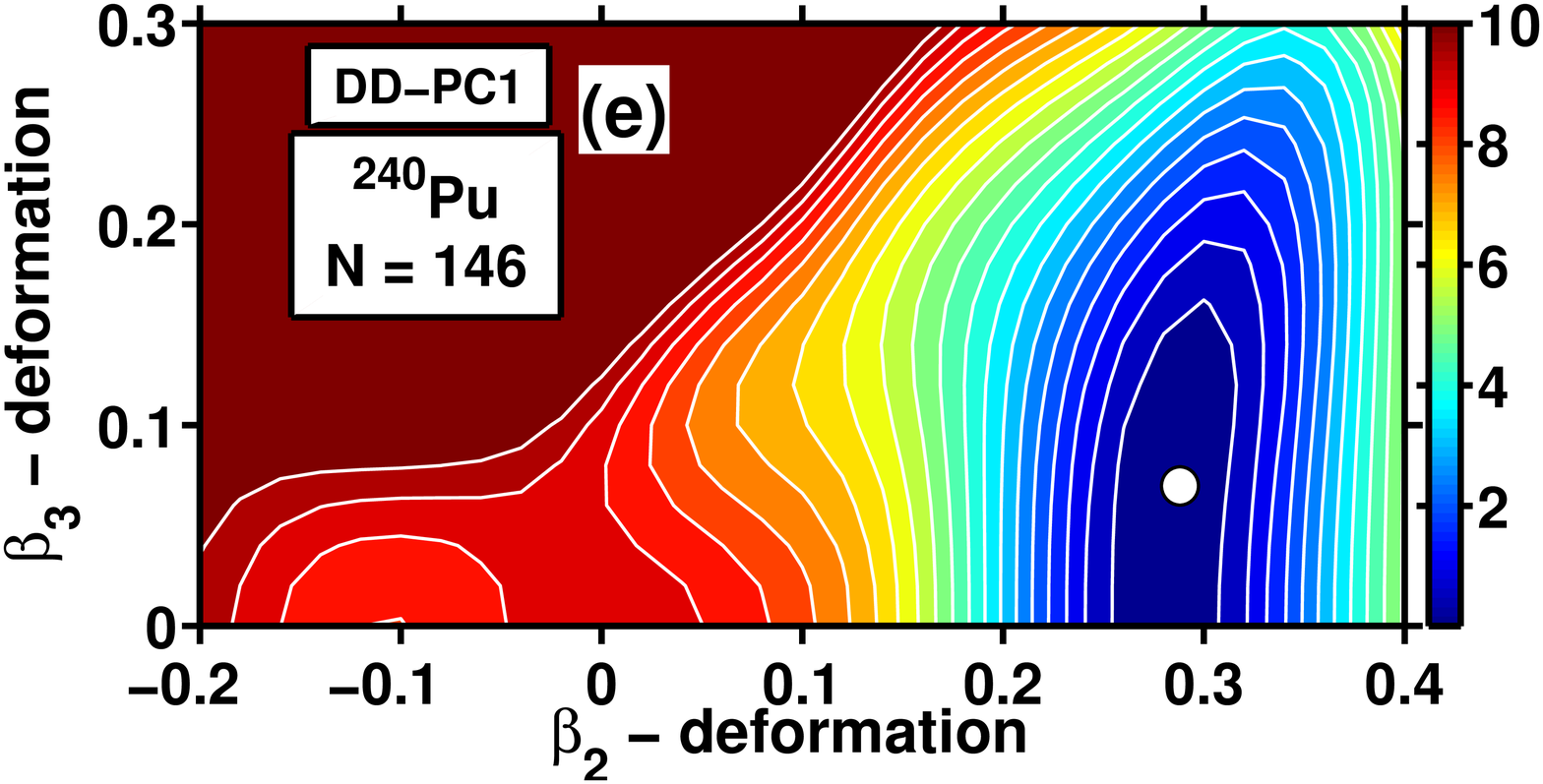}
  \includegraphics[angle=0,width=5.9cm]{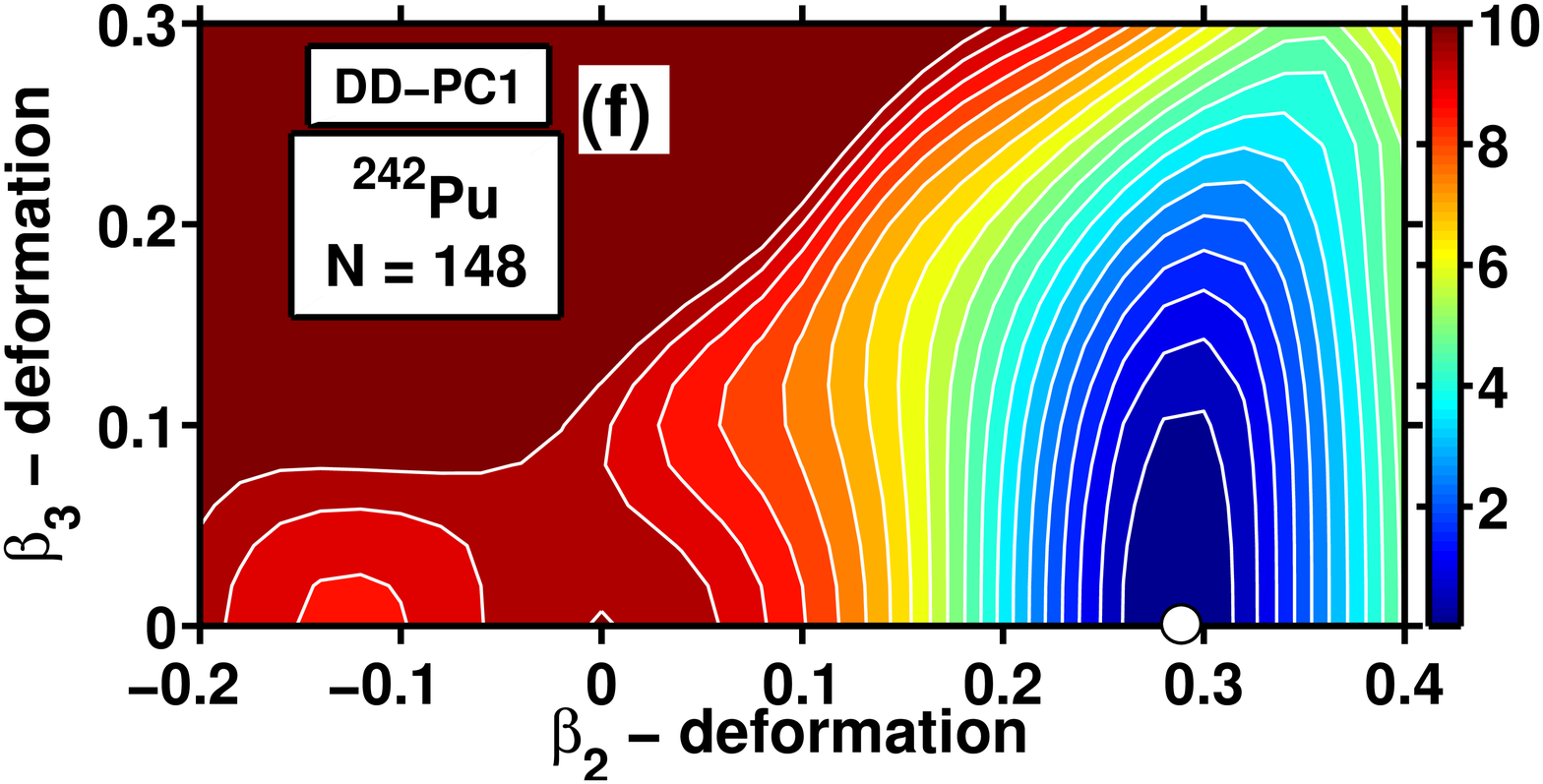}
  \caption{(Color online) The same as Fig.\ \ref{Rn_DD-PC1}, but
           for the  $N=144-148$ Pu isotopes calculated with the NL3*
           (upper panels) and DD-PC1 (bottom panels)
           functionals.}
\label{Pu_DD-PC1_NL3s}
\end{figure*}

 The analysis of the PES of the nuclei with $N\sim 146$ indicates
that such a scenario is possible. This is illustrated by the
examples of the U and Pu isotopes
presented in Figs.\ \ref{U_DD-PC1_rot} and \ref{Pu_DD-PC1_NL3s}.
One can see that the PESs of the $N\leq 146$ U
isotopes are very soft in $\beta_3$ direction. The
octupole deformed solution with $\beta_3\neq 0.0$ is even
the lowest in energy in $^{238}$U. However, the gain of binding
due to octupole deformation $|\Delta E^{oct}|$ in this nuclei
is very small - only 94 keV (see Table \ref{table-global}),
and, thus, this nucleus remains in the octupole vibrational
regime. Dependent on the underlying single-particle structure
and its evolution with spin, the rotation of  these octupole
soft nuclei may lead to a stabilization of octupole deformation
at high spin \cite{NLD.87,NT.92,GER.98} and the analysis of experimental
data on the $N\leq 146$ actinides (see Sec. IV in Ref.\ \cite{AO.13})
strongly points to such a possibility. The stiffness of the PES in the
direction of  octupole deformation increases with the increase of
neutron number above $N=146$ (see panels for $^{240}$U and $^{242}$U
in Fig.\ \ref{U_DD-PC1_rot}).  As a result, the stabilization of
octupole deformation at high spin due to rotation in these
two nuclei is not likely. Indeed, the predictions of the
CRHB+LN calculations with no octupole deformation \cite{AO.13}
for rotational properties of the ground state band in $^{240}$U almost
coincide with recent experimental data \cite{240U}.

  The same transition from octupole soft to octupole stiff potential
energy surfaces is observed between $N=146$ and $N=148$ also in Pu
(see Fig.\ \ref{Pu_DD-PC1_NL3s}), Th and Cm isotopes. Fig.\
\ref{Pu_DD-PC1_NL3s} also illustrates that the results almost
do not depend on the functional since the PES
obtained with DD-PC1 and NL3* are very similar. The potential
energy surfaces of the $N=144$ isotones (Figs.\ \ref{U_DD-PC1_rot} and
\ref{Pu_DD-PC1_NL3s}) are soft in $\beta_3$ direction but the minimum
of the PES is located at $\beta_3=0$ (see also Table
\ref{table-global}). The same softness is observed in the $N=146$ isotopes
(Figs.\ \ref{U_DD-PC1_rot} and \ref{Pu_DD-PC1_NL3s}) but the local minimum is
located at $\beta_3 \neq 0$ (see Table \ref{table-global}). However, the
binding energy gains due to octupole deformation remain small (around 100 keV,
see Table \ref{table-global}) so that the $N=146$
Th, U, Pu and Cm isotopes remain in the octupole vibrational regime. On the
contrary, the $N=148$ Th, U, Pu and Cm isotones are characterized by 
PES which are stiff in $\beta_3$-deformation
(see, for example, Figs.\ \ref{U_DD-PC1_rot} and \ref{Pu_DD-PC1_NL3s}).

A  detailed investigation of the impact of octupole deformation
on rotational properties requires the development of a symmetry
unrestricted cranked RHB code which is definitely beyond the scope
of the present manuscript. However, the analysis of octupole
softness in the ground states establish a clear correlation with the
CRHB+LN results presented in Ref.\ \cite{AO.13}. The $N\geq 148$ actinides are
characterized by PESs which are stiff in the
direction of octupole deformation. As a result, no stabilization of octupole
deformation due to rotation is expected in these nuclei and the
experimental data on ground state rotational bands are well
described in the CRHB+LN calculations with no octupole deformation
\cite{AO.13,A.14,240U}. On the contrary, the PESs of the $N\leq 146$
actinides are soft in octupole
deformation.  Thus, at low spin these nuclei are in the octupole
vibrational regime in which the CRHB+LN calculations with no
octupole deformation describe well the moments of inertia
(see Figs.\ 9 and 10 in Ref.\ \cite{AO.13}). However, with
increasing spin the transition to static octupole deformation
(or to the aligned vibrational limit) quite likely takes place
and the CRHB+LN calculations of Ref.\ \cite{AO.13} with no
octupole deformation could not properly describe this process;
they predict  upbends in the kinematic moments of inertia $J^{(1)}$
which are not observed in experiment.

\section{Global analysis}
\label{glob-an}

We have carried out a  global search for octupole deformation, which covers
all even-even $Z\leq 106$ nuclei between the two-proton and
two-neutron lines, with CEDFs DD-PC1 and
NL3*. The selection of these two functionals is guided by the
following reasons. First, the DD-ME$\delta$
functional is omitted from global studies since it does not
reproduce the experimental situation in octupole deformed actinides
(Sec.\ \ref{actin}) and provides unrealistically low fission barriers
in superheavy nuclei (see Ref.\ \cite{AANR.16}). Second, the systematic studies
with all four functionals are numerically too time-consuming to
be undertaken. Thus, among the remaining functionals we selected the two
functionals DD-PC1 and NL3*, which show the largest spread
of the predictions not only for octupole-deformed nuclei (Sec.\
\ref{actin}) but also for two-neutron drip lines \cite{AARR.13,AARR.14},
superheavy nuclei \cite{AANR.15} and the evolution of pairing with
isospin in neutron-rich nuclei \cite{AARR.14}.

 The results of this search are summarized in Table \ref{table-global}
and in Fig.\ \ref{fig-global}. The RHB results are also compared with the
MM results of Ref.\ \cite{MBCOISI.08}. Note that
the MM results cover only the part of the nuclear chart with neutron
numbers $N\leq 160$. There are also the HFB calculations with the Gogny forces D1S,
D1M and D1N \cite{RB.11} which cover a region of nuclei 
not extending far beyond the known nuclei (see  Fig.\ 9 in Ref.\  \cite{RB.11} for
details). However, these results are not added to Table \ref{table-global}
since the equilibrium octupole deformations $\beta_3$ are not properly
recorded in the supplemental material to this {publication; they are 
always given as the multiplies of 0.02 (0.025) for the D1M (D1N and D1S) functionals 
which indicates that they do not correspond to the $\beta_3$ values of the energy 
minimum.

  One can see in Fig.\ \ref{fig-global} that in addition to the Ba, Ce
and Nd isotopes as well as the actinides discussed above there are several
regions of octupole deformed nuclei. These are nuclei
around $^{80}$Zr, $^{110}$Zr and $^{200}$Dy which are octupole soft. Since
the gain of binding due to octupole deformation is quite small, no
stabilization of octupole deformation is expected in these nuclei. Indeed,
the interpretation of experimental data on nuclei around $^{80}$Zr does not
require the involvement of stable octupole deformation \cite{NDBBR.85}.
Note that the HFB calculations with the Gogny forces also indicate octupole softness
of the nuclei around $^{80}$Zr (see  Fig.\ 9 in Ref.\  \cite{RB.11}). However,
in the MM calculations of Ref.\ \cite{RB.11}, these nuclei do not have octupole 
deformation  (Table \ref{table-global}).

In the RHB calculations with DD-PC1 there exists a region of octupole soft Gd, Dy, and Er 
nuclei with $N\sim 136$ and  $A\sim 200$ (Fig. \ref{fig-global} and
Table \ref{table-global}). However, in the RHB calculations with NL3*
octupole softness is seen only in $^{200}$Dy. This difference is quite likely due
to the fact that pairing correlations, which counteract octupole deformation (see
Sec.\ \ref{sec-pairing-impact}), are substantially stronger in neutron-rich
nuclei for the NL3* functional as compared with DD-PC1 (Ref.\ \cite{AARR.15}).
This region of nuclei will not be accessible with future facilities like FRIB
since it is located beyond the expected reach of FRIB (Fig.\ \ref{fig-global}).
As compared with our results, the MM calculations of Ref.\ \cite{MBCOISI.08}
predict a much broader $Z\sim 60, N\sim 132$ region of nuclei with non-zero
octupole deformation (see Fig.\ 3 in Ref.\ \cite{MBCOISI.08}). Note that the HFB
calculations with Gogny forces of Ref.\ \cite{RB.11} do not cover the Gd-Dy-Er
region around $A\sim 200$.

  In addition, octupole deformation is predicted in the ground states
of the actinides and light superheavy nuclei with neutron number around
$N\sim 196$ (Table \ref{table-global} and Fig.\ \ref{fig-global}). To
our knowledge, the existence of this region of octupole deformation,
centered around $Z\sim 98, N\sim  196$, has not been  predicted before.
In many respects, it  is similar to the one located in the $Z\sim 90,
N\sim 136$ actinides. For example, the gains in binding due to octupole
deformation are similar in both regions and the size of
these regions in the $(Z,N)$ plane are comparable. In the center of the
$Z\sim 98, N=196$ region, $|\Delta E^{oct}|$ is around 1.5 MeV in the 
calculations with DD-PC1 and around of 1.2 MeV in the calculations with 
NL3*. As a result, some of the $N\sim 196$ actinides could have a stable 
octupole deformation
in the ground state.  This difference in $|\Delta E^{oct}|$ could be due
to the fact that pairing correlations, which counteract octupole deformation
(see Sec.\ \ref{sec-pairing-impact}), are substantially stronger in
neutron-rich nuclei in the NL3* functional as compared with DD-PC1 one
(Ref.\ \cite{AARR.15}).

  The maximum gain in binding due to octupole deformation is changing
from $N=198$ ($N=200$) to $N=192$ ($N=196$) on going from Th ($Z=90$)
to Fm ($Z=100$) nuclei in the calculations with DD-PC1 (NL3*)
(see Table \ref{table-global}). While the predictions for the location of
octupole deformed nuclei in the Th-Fm region are more or less similar in
the calculations with DD-PC1 and NL3*, they diverge for the No, Rf
and Sg isotopes (Fig.\ \ref{fig-global}). However, this difference
exists in octupole soft nuclei which have only relatively small (less than
0.6 MeV) gain in binding due to octupole deformation $|\Delta E^{oct}|$. It is
definitely caused by the differences in underlying single-particle structure
(see  Ref.\ \cite{AANR.15}) which also leads to different predictions for the
ground  state properties of superheavy nuclei (see Ref.\ \cite{AANR.15} for
details).

Note that this region of nuclei will not be accessible with future
facilities like FRIB since it is located beyond the expected reach of FRIB
(Fig.\ \ref{fig-global}). However, the accounting of octupole deformation
in the ground states of these nuclei is important for the modeling of fission
recycling in neutron star mergers \cite{GBJ.11} since the gain in binding of
the ground states due to octupole deformation  will increase the fission 
barrier heights as compared with the case when octupole deformation is 
neglected.

  The presence of octupole deformation in these nuclei is due to the
interaction of the normal-parity 2$h_{11/2}$ (from the $N=7$ shell) and the intruder
1$k_{17/2}$ (from the $N=8$ shell) neutron orbitals. These orbitals,
located above the $N=184$ shell gap, are very close in energy at the Fermi
level in very heavy and superheavy nuclei in many nuclear potentials (see,
for example, Fig.\ 6.9 in Ref.\ \cite{NilRag-book} and Fig.\ 1 in Ref.\
\cite{AANR.15}). The octupole coupling  between proton 1$i_{13/2}$ and
2$f_{7/2}$ orbitals is still active in these nuclei but its maximum is around
$Z=98$. On the contrary, it has a maximum around $Z=92$ in the $A\sim 230$
octupole deformed actinides. This change in the position of the maximum of
octupole interaction in the proton subsystem is due to the increase of the 
neutron excess on going from $N\sim 136$ to $N\sim 196$ and related modifications 
of the properties of the proton potential.

\begin{figure*}
\includegraphics[angle=-90,width=10.0cm]{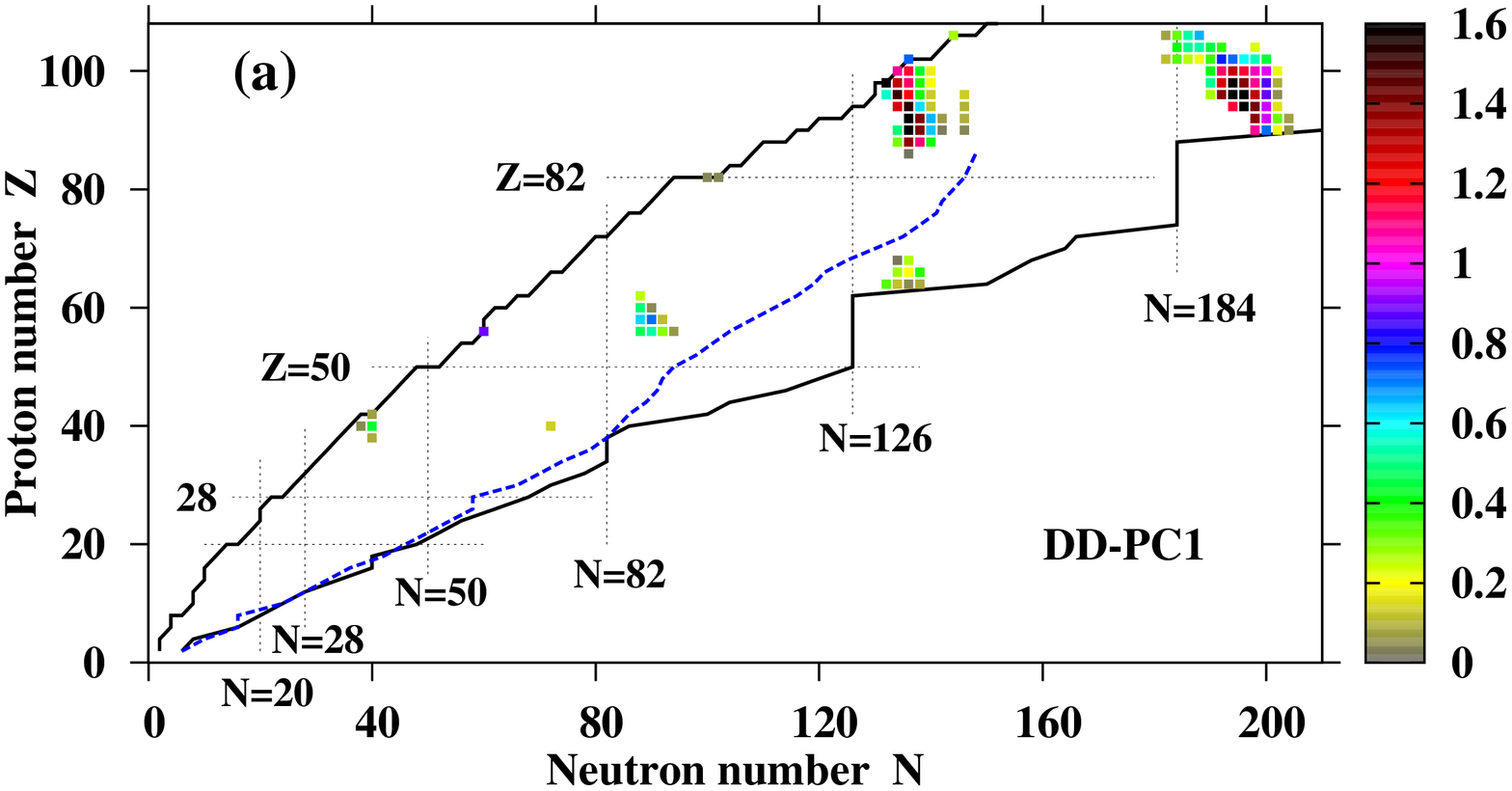}
\includegraphics[angle=-90,width=10.0cm]{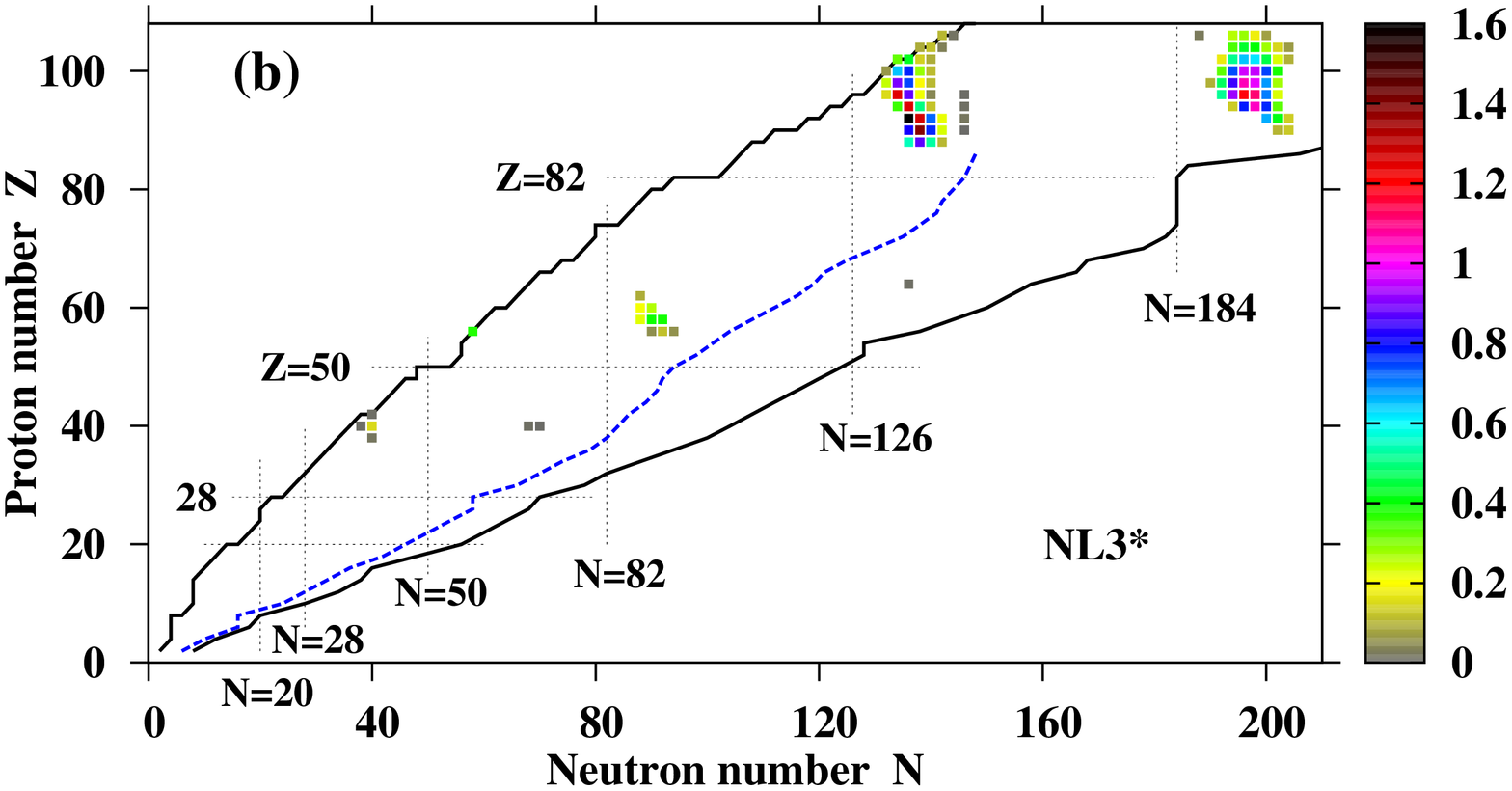}
  \caption{(Color online) Octupole deformed nuclei
           in the nuclear chart. Only nuclei with non vanishing
           $\Delta E^{oct}$ are shown by squares; the colors of the
           squares represent the values of $|\Delta E^{oct}|$ (see colormap).
           Top and bottom panels display the results obtained with the 
           CEDFs DD-PC1
           and NL3*, respectively. The blue dashed line shows the
           limits of the nuclear chart (defined as fission yield greater
           than 10$^{−6}$) which may be achieved with dedicated existence
           measurements at FRIB \cite{S-priv.14}. The two-proton and two-neutron
           drip lines are displayed by solid black lines.}
\label{fig-global}
\end{figure*}

\section{Conclusions}
\label{conclusions}

 A global search for octupole deformation has been
performed within covariant density functional theory employing
the DD-PC1 and NL3* functionals; this search covers
all even-even nuclei with $Z\leq 106$ located between the two-neutron and 
two-proton
drip lines. In the regions of octupole
deformed light lanthanides and actinides, additional
studies have been performed with the CEDFs DD-ME2, DD-ME$\delta$,
and PC-PK1  in order to establish the sensitivity
of the results to the choice of the functional and to estimate
theoretical uncertainties. The main
results can be summarized as follows:

\begin{itemize}

\item
  The RHB calculations with the DD-PC1, PC-PK1 and DD-ME2
functionals correctly predict the islands of  octupole
deformation in the light lanthanides and actinides which
in general agrees with available experimental data. The
NL3* tends to place the centers (in the $(Z,N)$ plane)
of these two islands by two neutrons higher than in above
mentioned functionals. The DD-ME$\delta$ functional fails to
describe experimental data in the actinides.

\item
  The gain in binding due to octupole deformation $|\Delta E^{oct}|$
is the quantity which defines the location and the extend of the islands
of octupole deformation. If one excludes the DD-ME$\delta$ functional,
theoretical uncertainties in its prediction are typically around
0.5 MeV; however, in some nuclei they reach 1 MeV. This leads to the
differences in the predictions of the islands of octupole deformation.
The most important source of these uncertainties is the difference
in the prediction of underlying single-particle structure (see Ref.\
\cite{DABRS.15} for comparison of different DFTs). This is clearly seen for
example in light lanthanides where the maximum gain in binding due to
octupole deformation is located mostly at $N=88$ and $N=90$ in
non-relativistic and relativistic models, respectively (see Sec.\
\ref{lanth}). Another example is the  differences in the predictions
of the borderlines of the island of octupole deformed nuclei in the U, Pu,
Cm, Cf and Fm isotopes in relativistic and non-relativistic theories
(see Sec.\ \ref{actin}).

\item
A new region of octupole deformation, centered around $Z\sim 98, N\sim  196$,
has been predicted for the first time. In terms of the size
in the $(Z,N)$ plane and the impact of octupole deformation on binding energies
this region is similar to the known $Z\sim 90, N\sim 136$ region of octupole
deformation in actinides.  The presence of octupole deformation in these
nuclei is due to the interaction of the 2$h_{11/2}$ and  1$k_{17/2}$ neutron
orbitals and of the 1$i_{13/2}$ and 2$f_{7/2}$ proton orbitals. Note that the
maximum of the interaction of proton orbitals occurs at a higher proton
number $Z$ as compared with the well known $A\sim 230$ region of octupole 
deformation in actinides. 

\item
Important correlations between the softness of the potential
energy surfaces in octupole deformation and the behavior
of ground state rotational bands of the $N\geq 140$ even-even
nuclei have been revealed. These nuclei do not possess stable
octupole deformation in the ground states. The rotational
properties of the $N\geq 148$ nuclei with stiff potential energy
surfaces with respect to octupole deformation are well
described in the CRHB+LN approach neglecting octupole deformation
\cite{AO.13}. In addition, for these nuclei this approach has a
predictive power as illustrated in Refs.\ \cite{A.14,240U}. The
moments of inertia of lighter nuclei are also well described at
low and medium spins but the CRHB+LN approach predicts paired
band crossings at rotational frequency $\Omega_x \sim 0.2$ MeV
which are not observed in experiment. This discrepancy between
theory and experiment is most likely due to stabilization of
octupole deformation at high spin which becomes possible because the
potential energy surfaces of the $N\leq 146$ actinides are soft in
octupole deformation at the ground states.

\item
  The impact of pairing correlations on the properties of
octupole deformed nuclei has been studied. In general,
pairing counteracts shell effects and favors shapes
with no octupole deformation. Thus, the strongest impact of
octupole deformation is expected in systems with no
or weak pairing. The barrier between quadrupole and octupole
local minima is at maximum for no pairing and decreases with
increasing pairing strength.

\item
  Comparing different functionals, one can see that the
results obtained with the covariant energy density functional
DD-ME$\delta$ differ substantially from the results of other
functionals. This effect is especially pronounced in
actinides where DD-ME$\delta$ does not lead to octupole
deformation of the nuclei which are known to be octupole
deformed. In addition, the heights of the inner fission barriers in
superheavy nuclei with $Z=112-116$ obtained in this functional
are significantly lower than the experimental estimates and
the values calculated in all other models \cite{AANR.16}. This
functional is different from all the other functionals used
here, because it has been adjusted in Ref.\ \cite{DD-MEdelta}
using only four phenomenological parameters in addition to
some input from ab initio calculations
\cite{Baldo2004_NPA736-241,VanDalen2007_EPJA31-29}.
All these facts suggest that either the ab initio input for
this functional is not precise enough or the number of only
four phenomenological parameters (fitted to masses of spherical
nuclei) is too small to provide a proper description of the
details of the single-particle structure. Thus, this functional
is not recommended for future investigations, in spite of the
fact that it provides a good description of masses and some
other ground-state observables \cite{AARR.14}.

\end{itemize}

\section{Acknowledgements}

  This material is based upon work supported by the U.S. Department
of Energy, Office of Science, Office of Nuclear Physics under Award
Number DE-SC0013037 and by the DFG cluster of excellence \textquotedblleft
Origin and Structure of the Universe\textquotedblright\
(www.universe-cluster.de).

\begin{longtable*}{|c|c|c|c|c|c|c|c|c|c|c|c|c|c|c|}
 \caption{Calculated effect of reflection asymmetry on nuclear ground state properties.
          For each functional or model, the equilibrium quadrupole ($\beta_2$) and octupole
          ($\beta_3$) deformations as well as the gains in binding energy due to octupole
          deformation $|\Delta E^{oct}|$ are given. The results are presented only in the case
          when the octupole deformed minimum is the lowest in energy. Note that $\epsilon_{2}$ and
          $\epsilon_{3}$ are the quadrupole and octupole deformations (in the Nilsson
          perturbed-spheroid parametrization) obtained in the MM approach of Ref.\
          \cite{MBCOISI.08}.
\label{table-global}} \\
\hline\hline
\multicolumn{3}{|c|}{} &&  \multicolumn{3}{c|}{\textbf{DD-PC1}} & &
\multicolumn{3}{c|}{\textbf{NL3*}} & &
\multicolumn{3}{c|}{\textbf{mic+mac}} \\ \hline
$Z$(Nucleus) & $N$ & $A$ && $\beta_{2}$ & $\beta_{3}$ & $|\Delta{E}^{oct}|$ && $\beta_{2}$ &
$\beta_{3}$ & $|\Delta{E}^{oct}|$ && $\epsilon_{2}$ & $\epsilon_{3}$ & $|\Delta{E}^{oct}|$ \\
\hline
\endfirsthead
\multicolumn{14}{c}%
{\tablename\ \thetable\ -- \textit{Continued from previous page}} \\
\hline\hline
\multicolumn{3}{|c|}{} &&  \multicolumn{3}{c|}{\textbf{DD-PC1}} & &
\multicolumn{3}{c|}{\textbf{NL3*}} & &
\multicolumn{3}{c|}{\textbf{mic+mac}} \\
\hline\hline
$Z$(Nucleus) & $N$ & $A$ && $\beta_{2}$ & $\beta_{3}$ & $|\Delta{E}^{oct}|$ && $\beta_{2}$ &
$\beta_{3}$ & $|\Delta{E}^{oct}|$ & $$ & $\epsilon_{2}$ & $\epsilon_{3}$ & $|\Delta{E}^{oct}|$ \\
\hline
\endhead
\hline \multicolumn{14}{r}{\textit{Continued on next page}} \\
\endfoot
\hline
\endlastfoot
 20(Ca) & 36  & 56  &&       &       &       & &       &       &       & &-0.07 & 0.02 & 0.04 \\
        & 40  & 60  &&       &       &       & &       &       &       & & 0.00 & 0.07 & 0.03 \\
        &     &     &&       &       &       & &       &       &       & &      &      &      \\
 38(Sr) & 40  & 78  && 0.005 & 0.084 & 0.089 & & 0.005 & 0.078 & 0.019 & &      &      &      \\
        &     &     &&       &       &       & &       &       &       & &      &      &      \\
 40(Zr) & 38  & 78  && 0.003 & 0.068 & 0.043 & & 0.003 & 0.060 & 0.005 & &      &      &      \\
        & 40  & 80  && 0.008 & 0.145 & 0.439 & & 0.007 & 0.139 & 0.149 & &      &      &      \\
        &     &     &&       &       &       & &       &       &       & &      &      &      \\
        & 68  & 108 &&       &       &       & & 0.002 & 0.060 & 0.009 & &      &      &      \\
        & 70  & 110 &&       &       &       & & 0.001 & 0.053 & 0.004 & &      &      &      \\
        & 72  & 112 &&-0.003 & 0.094 & 0.133 & &       &       &       & &      &      &      \\
        &     &     &&       &       &       & &       &       &       & &      &      &      \\
 42(Mo) & 40  & 82  &&-0.001 & 0.078 & 0.070 & &-0.001 & 0.064 & 0.007 & &      &      &      \\
        &     &     &&       &       &       & &       &       &       & &      &      &      \\
 48(Cd) & 42  & 90  &&       &       &       & &       &       &       & &-0.01 & 0.04 & 0.04 \\
        &     &     &&       &       &       & &       &       &       & &      &      &      \\
 54(Xe) & 54  & 108 &&       &       &       & &       &       &       & & 0.15 & 0.05 & 0.05 \\
        & 56  & 110 &&       &       &       & &       &       &       & & 0.16 & 0.07 & 0.20 \\
        & 58  & 112 &&       &       &       & &       &       &       & & 0.18 & 0.07 & 0.14 \\
        &     &     &&       &       &       & &       &       &       & &      &      &      \\
        & 88  & 142 &&       &       &       & &       &       &       & & 0.13 & 0.06 & 0.11 \\
        & 90  & 144 &&       &       &       & &       &       &       & & 0.15 & 0.07 & 0.11 \\
        &     &     &&       &       &       & &       &       &       & &      &      &      \\
 56(Ba) & 52  & 108 &&       &       &       & &       &       &       & & 0.13 & 0.05 & 0.05 \\
        & 54  & 110 &&       &       &       & &       &       &       & & 0.17 & 0.09 & 0.34 \\
        & 56  & 112 && 0.244 & 0.114 & 0.284 & & 0.274 & 0.188 & 0.792 & & 0.18 & 0.10 & 0.48 \\
        & 58  & 114 && 0.252 & 0.097 & 0.157 & & 0.267 & 0.155 & 0.374 & & 0.20 & 0.09 & 0.31 \\
        & 60  & 116 && 0.275 & 0.074 & 0.888 & &       &       &       & &      &      &      \\
        &     &     &&       &       &       & &       &       &       & &      &      &      \\
        & 86  & 142 &&       &       &       & &       &       &       & & 0.12 & 0.06 & 0.14 \\
        & 88  & 144 && 0.201 & 0.101 & 0.467 & &       &       &       & & 0.15 & 0.09 & 0.49 \\
        & 90  & 146 && 0.216 & 0.112 & 0.531 & & 0.202 & 0.083 & 0.051 & & 0.16 & 0.09 & 0.47 \\
        & 92  & 148 && 0.232 & 0.122 & 0.290 & & 0.216 & 0.089 & 0.118 & &      &      &      \\
        & 94  & 150 && 0.254 & 0.114 & 0.061 & & 0.230 & 0.084 & 0.053 & &      &      &      \\
        &     &     &&       &       &       & &       &       &       & &      &      &      \\
 58(Ce) & 56  & 114 && 0.254 & 0.100 & 0.166 & & 0.286 & 0.161 & 0.396 & & 0.21 & 0.08 & 0.21 \\
        &     &     &&       &       &       & &       &       &       & &      &      &      \\
        & 86  & 144 &&       &       &       & &       &       &       & & 0.13 & 0.07 & 0.22 \\
        & 88  & 146 && 0.205 & 0.113 & 0.631 & & 0.194 & 0.097 & 0.224 & & 0.16 & 0.09 & 0.46 \\
        & 90  & 148 && 0.222 & 0.125 & 0.714 & & 0.215 & 0.113 & 0.390 & & 0.19 & 0.07 & 0.02 \\
        & 92  & 150 && 0.246 & 0.134 & 0.111 & & 0.236 & 0.120 & 0.384 & &      &      &      \\
        &     &     &&       &       &       & &       &       &       & &      &      &      \\
 60(Nd) & 86  & 146 &&       &       &       & &       &       &       & & 0.14 & 0.06 & 0.08 \\
        & 88  & 148 && 0.206 & 0.114 & 0.491 & & 0.198 & 0.105 & 0.208 & & 0.18 & 0.06 & 0.09 \\
        & 90  & 150 && 0.235 & 0.128 & 0.044 & & 0.231 & 0.121 & 0.261 & &      &      &      \\
        &     &     &&       &       &       & &       &       &       & &      &      &      \\
 62(Sm) & 88  & 150 && 0.211 & 0.098 & 0.253 & & 0.206 & 0.091 & 0.091 & & 0.19 & 0.04 & 0.02 \\
        &     &     &&       &       &       & &       &       &       & &      &      &      \\
 64(Gd) & 132 & 196 && 0.136 & 0.062 & 0.335 & &       &       &       & &      &      &      \\
        & 134 & 198 && 0.167 & 0.090 & 0.117 & &       &       &       & &      &      &      \\
        & 136 & 200 && 0.192 & 0.119 & 0.046 & & 0.182 & 0.003 & 0.008 & &      &      &      \\
        & 138 & 202 && 0.217 & 0.142 & 0.088 & &       &       &       & &      &      &      \\
        &     &     &&       &       &       & &       &       &       & &      &      &      \\
 66(Dy) & 134 & 200 && 0.176 & 0.049 & 0.274 & &       &       &       & &      &      &      \\
        & 136 & 202 && 0.202 & 0.090 & 0.200 & &       &       &       & &      &      &      \\
        & 138 & 204 && 0.231 & 0.106 & 0.368 & &       &       &       & &      &      &      \\
        &     &     &&       &       &       & &       &       &       & &      &      &      \\
 68(Er) & 130 & 198 &&       &       &       & &       &       &       & & 0.06 & 0.05 & 0.10 \\
        & 132 & 200 &&       &       &       & &       &       &       & & 0.11 & 0.04 & 0.05 \\
        & 134 & 202 && 0.170 & 0.004 & 0.017 & &       &       &       & & 0.11 & 0.06 & 0.04 \\
        & 136 & 204 && 0.200 & 0.065 & 0.265 & &       &       &       & &      &      &      \\
        &     &     &&       &       &       & &       &       &       & &      &      &      \\
 70(Yb) & 134 &     &&       &       &       & &       &       &       & & 0.11 & 0.04 & 0.04 \\
        &     &     &&       &       &       & &       &       &       & &      &      &      \\
 76(Os) & 134 &     &&       &       &       & &       &       &       & & 0.09 & 0.02 & 0.02 \\
        &     &     &&       &       &       & &       &       &       & &      &      &      \\
 78(Pt) & 136 &     &&       &       &       & &       &       &       & & 0.09 & 0.03 & 0.03 \\
        &     &     &&       &       &       & &       &       &       & &      &      &      \\
 80(Hg) & 136 &     &&       &       &       & &       &       &       & & 0.06 & 0.05 & 0.02 \\
        & 138 &     &&       &       &       & &       &       &       & & 0.08 & 0.05 & 0.14 \\
        &     &     &&       &       &       & &       &       &       & &      &      &      \\
 82(Pb) & 98  & 180 &&       &       &       & &       &       &       & & 0.00 & 0.03 & 0.02 \\
        & 100 & 182 && 0.004 & 0.041 & 0.038 & &       &       &       & & 0.01 & 0.02 & 0.08 \\
        & 102 & 184 && 0.002 & 0.041 & 0.038 & &       &       &       & & 0.00 & 0.02 & 0.04 \\
        &     &     &&       &       &       & &       &       &       & &      &      &      \\
        & 134 & 216 &&       &       &       & &       &       &       & & 0.01 & 0.04 & 0.02 \\
        & 136 & 218 &&       &       &       & &       &       &       & & 0.01 & 0.06 & 0.16 \\
        & 138 & 220 &&       &       &       & &       &       &       & & 0.01 & 0.07 & 0.23 \\
        & 140 & 222 &&       &       &       & &       &       &       & & 0.01 & 0.07 & 0.26 \\
        &     &     &&       &       &       & &       &       &       & &      &      &      \\
 84(Po) & 134 & 218 &&       &       &       & &       &       &       & & 0.05 & 0.09 & 0.44 \\
        & 136 & 220 &&       &       &       & &       &       &       & & 0.09 & 0.09 & 0.42 \\
        & 138 & 222 &&       &       &       & &       &       &       & & 0.10 & 0.08 & 0.16 \\
        &     &     &&       &       &       & &       &       &       & &      &      &      \\
 86(Rn) & 132 & 218 &&       &       &       & &       &       &       & & 0.07 & 0.10 & 0.67 \\
        & 134 & 220 &&       &       &       & &       &       &       & & 0.10 & 0.09 & 0.85 \\
        & 136 & 222 &&       &       &       & &       &       &       & & 0.10 & 0.09 & 0.64 \\
        & 138 & 224 &&       &       &       & &       &       &       & & 0.13 & 0.08 & 0.29 \\
        & 140 & 226 &&       &       &       & &       &       &       & & 0.15 & 0.04 & 0.09 \\
        & 146 & 232 &&       &       &       & &       &       &       & & 0.21 & 0.02 & 0.02 \\
        &     &     &&       &       &       & &       &       &       & &      &      &      \\
 88(Ra) & 130 & 218 &&       &       &       & &       &       &       & & 0.07 & 0.09 & 0.59 \\
        & 132 & 220 &&       &       &       & &       &       &       & & 0.10 & 0.09 & 1.20 \\
        & 134 & 222 && 0.160 & 0.104 & 0.310 & &       &       &       & & 0.11 & 0.10 & 1.27 \\
        & 136 & 224 && 0.177 & 0.125 & 1.370 & & 0.178 & 0.124 & 0.547 & & 0.13 & 0.10 & 0.91 \\
        & 138 & 226 && 0.196 & 0.133 & 1.110 & & 0.197 & 0.134 & 0.874 & & 0.15 & 0.08 & 0.40 \\
        & 140 & 228 && 0.208 & 0.123 & 0.385 & & 0.208 & 0.126 & 0.526 & & 0.16 & 0.06 & 0.08 \\
        & 142 & 230 &&       &       &       & & 0.225 & 0.098 & 0.105 & &      &      &      \\
        &     &     &&       &       &       & &       &       &       & &      &      &      \\
 90(Th) & 130 & 220 &&       &       &       & &       &       &       & & 0.08 & 0.10 & 1.33 \\
        & 132 & 222 &&       &       &       & &       &       &       & & 0.10 & 0.10 & 1.35 \\
        & 134 & 224 && 0.167 & 0.112 & 0.491 & &       &       &       & & 0.13 & 0.11 & 1.22 \\
        & 136 & 226 && 0.186 & 0.137 & 1.999 & & 0.187 & 0.134 & 0.814 & & 0.14 & 0.10 & 0.50 \\
        & 138 & 228 && 0.214 & 0.154 & 1.402 & & 0.212 & 0.150 & 1.387 & & 0.16 & 0.08 & 0.08 \\
        & 140 & 230 && 0.224 & 0.152 & 0.642 & & 0.223 & 0.149 & 0.770 & &      &      &      \\
        & 142 & 232 && 0.234 & 0.141 & 0.025 & & 0.236 & 0.138 & 0.231 & &      &      &      \\
        & 146 & 236 && 0.261 & 0.054 & 0.039 & & 0.274 & 0.041 & 0.002 & &      &      &      \\
        &     &     &&       &       &       & &       &       &       & &      &      &      \\
        & 198 & 288 && 0.176 & 0.127 & 1.084 & &       &       &       & &      &      &      \\
        & 200 & 290 && 0.189 & 0.135 & 0.716 & &       &       &       & &      &      &      \\
        & 202 & 292 && 0.205 & 0.113 & 0.216 & & 0.182 & 0.095 & 0.102 & &      &      &      \\
        & 204 & 294 && 0.221 & 0.065 & 0.051 & & 0.198 & 0.090 & 0.126 & &      &      &      \\
        &     &     &&       &       &       & &       &       &       & &      &      &      \\
 92(U)  & 128 & 220 &&       &       &       & &       &       &       & & 0.05 & 0.08 & 0.08 \\
        & 130 & 222 &&       &       &       & &       &       &       & & 0.09 & 0.10 & 1.21 \\
        & 132 & 224 &&       &       &       & &       &       &       & & 0.12 & 0.10 & 1.22 \\
        & 134 & 226 &&       &       &       & &       &       &       & & 0.13 & 0.10 & 0.60 \\
        & 136 & 228 && 0.201 & 0.155 & 1.724 & & 0.201 & 0.151 & 1.813 & &      &      &      \\
        & 138 & 230 && 0.229 & 0.170 & 1.399 & & 0.228 & 0.165 & 1.264 & &      &      &      \\
        & 140 & 232 && 0.238 & 0.169 & 0.659 & & 0.238 & 0.166 & 0.721 & &      &      &      \\
        & 142 & 234 && 0.245 & 0.170 & 0.067 & & 0.247 & 0.162 & 0.217 & &      &      &      \\
        & 146 & 238 && 0.275 & 0.078 & 0.094 & & 0.284 & 0.068 & 0.019 & &      &      &      \\
        &     &     &&       &       &       & &       &       &       & &      &      &      \\
        & 198 & 290 && 0.181 & 0.140 & 1.378 & &       &       &       & &      &      &      \\
        & 200 & 292 && 0.196 & 0.151 & 0.969 & & 0.183 & 0.124 & 0.664 & &      &      &      \\
        & 202 & 294 && 0.214 & 0.137 & 0.319 & & 0.200 & 0.127 & 0.416 & &      &      &      \\
        & 204 & 296 && 0.233 & 0.082 & 0.074 & & 0.220 & 0.117 & 0.133 & &      &      &      \\
        &     &     &&       &       &       & &       &       &       & &      &      &      \\
 94(Pu) & 128 & 222 &&       &       &       & &       &       &       & & 0.05 & 0.08 & 0.35 \\
        & 130 & 224 &&       &       &       & &       &       &       & & 0.09 & 0.10 & 1.09 \\
        & 132 & 226 &&       &       &       & &       &       &       & & 0.12 & 0.10 & 0.59 \\
        & 134 & 228 && 0.170 & 0.134 & 1.260 & & 0.167 & 0.129 & 0.354 & & 0.14 & 0.10 & 0.04 \\
        & 136 & 230 && 0.197 & 0.155 & 1.535 & & 0.196 & 0.152 & 1.251 & &      &      &      \\
        & 138 & 232 && 0.246 & 0.161 & 0.622 & & 0.240 & 0.159 & 0.501 & &      &      &      \\
        & 140 & 234 && 0.263 & 0.133 & 0.125 & & 0.261 & 0.142 & 0.121 & &      &      &      \\
        & 146 & 240 && 0.284 & 0.066 & 0.099 & & 0.290 & 0.054 & 0.010 & &      &      &      \\
        &     &     &&       &       &       & &       &       &       & &      &      &      \\
        & 194 & 288 && 0.131 & 0.108 & 1.156 & & 0.119 & 0.089 & 0.132 & &      &      &      \\
        & 196 & 290 && 0.156 & 0.131 & 1.774 & & 0.151 & 0.118 & 0.780 & &      &      &      \\
        & 198 & 292 && 0.176 & 0.146 & 1.419 & & 0.171 & 0.135 & 1.046 & &      &      &      \\
        & 200 & 294 && 0.192 & 0.158 & 0.965 & & 0.187 & 0.142 & 0.770 & &      &      &      \\
        & 202 & 296 && 0.216 & 0.141 & 0.162 & & 0.206 & 0.143 & 0.327 & &      &      &      \\
        &     &     &&       &       &       & &       &       &       & &      &      &      \\
 96(Cm) & 128 & 224 &&       &       &       & &       &       &       & & 0.04 & 0.08 & 0.52 \\
        & 130 & 226 &&       &       &       & &       &       &       & & 0.08 & 0.10 & 0.84 \\
        & 132 & 228 && 0.134 & 0.115 & 0.562 & & 0.130 & 0.111 & 0.152 & & 0.14 & 0.08 & 0.02 \\
        & 134 & 230 && 0.162 & 0.135 & 1.511 & & 0.159 & 0.132 & 1.248 & &      &      &      \\
        & 136 & 232 && 0.195 & 0.158 & 1.190 & & 0.194 & 0.154 & 0.877 & &      &      &      \\
        & 138 & 234 && 0.252 & 0.142 & 0.387 & & 0.249 & 0.145 & 0.212 & &      &      &      \\
        & 140 & 236 && 0.274 & 0.098 & 0.131 & & 0.275 & 0.096 & 0.041 & &      &      &      \\
        & 146 & 242 && 0.295 & 0.063 & 0.132 & & 0.298 & 0.044 & 0.005 & &      &      &      \\
        &     &     &&       &       &       & &       &       &       & &      &      &      \\
        & 190 & 286 && 0.095 & 0.105 & 0.271 & &       &       &       & &      &      &      \\
        & 192 & 288 && 0.115 & 0.116 & 1.394 & & 0.116 & 0.102 & 0.516 & &      &      &      \\
        & 194 & 290 && 0.131 & 0.126 & 1.994 & & 0.135 & 0.119 & 0.923 & &      &      &      \\
        & 196 & 292 && 0.150 & 0.137 & 1.790 & & 0.155 & 0.136 & 1.191 & &      &      &      \\
        & 198 & 294 && 0.170 & 0.151 & 1.294 & & 0.172 & 0.147 & 1.110 & &      &      &      \\
        & 200 & 296 && 0.190 & 0.164 & 0.878 & & 0.189 & 0.154 & 0.701 & &      &      &      \\
        & 202 & 298 && 0.217 & 0.142 & 0.049 & & 0.211 & 0.153 & 0.229 & &      &      &      \\
        &     &     &&       &       &       & &       &       &       & &      &      &      \\
 98(Cf) & 126 & 224 && 0.008 & 0.065 & 0.151 & &       &       &       & & 0.00 & 0.05 & 0.10 \\
        & 128 & 226 && 0.015 & 0.049 & 0.032 & &       &       &       & & 0.03 & 0.08 & 0.56 \\
        & 130 & 228 &&       &       &       & &       &       &       & & 0.07 & 0.10 & 0.06 \\
        & 132 & 230 && 0.146 & 0.111 & 0.427 & & 0.142 & 0.112 & 0.247 & &      &      &      \\
        & 134 & 232 && 0.172 & 0.141 & 1.292 & & 0.170 & 0.138 & 0.895 & &      &      &      \\
        & 136 & 234 && 0.198 & 0.164 & 1.122 & & 0.198 & 0.160 & 0.747 & &      &      &      \\
        & 138 & 236 && 0.245 & 0.146 & 0.379 & & 0.244 & 0.146 & 0.195 & &      &      &      \\
        & 140 & 238 && 0.266 & 0.114 & 0.197 & & 0.266 & 0.114 & 0.107 & &      &      &      \\
        &     &     &&       &       &       & &       &       &       & &      &      &      \\
        & 190 & 288 && 0.106 & 0.123 & 0.515 & & 0.102 & 0.090 & 0.095 & &      &      &      \\
        & 192 & 290 && 0.131 & 0.131 & 1.259 & & 0.132 & 0.114 & 0.473 & &      &      &      \\
        & 194 & 292 && 0.146 & 0.140 & 1.632 & & 0.153 & 0.134 & 0.848 & &      &      &      \\
        & 196 & 294 && 0.164 & 0.150 & 1.388 & & 0.171 & 0.152 & 1.056 & &      &      &      \\
        & 198 & 296 && 0.180 & 0.162 & 1.087 & & 0.185 & 0.162 & 0.983 & &      &      &      \\
        & 200 & 298 && 0.196 & 0.173 & 0.868 & & 0.199 & 0.168 & 0.680 & &      &      &      \\
        & 202 & 300 && 0.218 & 0.147 & 0.094 & & 0.217 & 0.164 & 0.273 & &      &      &      \\
        &     &     &&       &       &       & &       &       &       & &      &      &      \\
100(Fm) & 126 & 226 && 0.011 & 0.089 & 0.490 & & 0.014 & 0.085 & 0.207 & & 0.00 & 0.06 & 0.12 \\
        & 128 & 228 && 0.014 & 0.079 & 0.149 & & 0.023 & 0.073 & 0.092 & & 0.02 & 0.08 & 0.52 \\
        & 132 & 232 && 0.161 & 0.085 & 0.152 & & 0.164 & 0.095 & 0.061 & &      &      &      \\
        & 134 & 234 && 0.187 & 0.146 & 1.084 & & 0.188 & 0.145 & 0.646 & &      &      &      \\
        & 136 & 236 && 0.201 & 0.172 & 1.156 & & 0.202 & 0.166 & 0.774 & &      &      &      \\
        & 138 & 238 && 0.226 & 0.158 & 0.434 & & 0.223 & 0.160 & 0.282 & &      &      &      \\
        & 140 & 240 && 0.258 & 0.110 & 0.177 & & 0.253 & 0.122 & 0.110 & &      &      &      \\
        &     &     &&       &       &       & &       &       &       & &      &      &      \\
        & 190 & 290 && 0.137 & 0.129 & 0.403 & &       &       &       & &      &      &      \\
        & 192 & 292 && 0.162 & 0.149 & 1.119 & & 0.147 & 0.124 & 0.327 & &      &      &      \\
        & 194 & 294 && 0.170 & 0.159 & 1.309 & & 0.170 & 0.151 & 0.740 & &      &      &      \\
        & 196 & 296 && 0.183 & 0.169 & 1.160 & & 0.187 & 0.167 & 0.964 & &      &      &      \\
        & 198 & 298 && 0.195 & 0.177 & 1.056 & & 0.199 & 0.176 & 0.954 & &      &      &      \\
        & 200 & 300 && 0.204 & 0.184 & 0.976 & & 0.208 & 0.180 & 0.749 & &      &      &      \\
        & 202 & 302 && 0.216 & 0.157 & 0.218 & & 0.220 & 0.174 & 0.383 & &      &      &      \\
        &     &     &&       &       &       & &       &       &       & &      &      &      \\
102(No) & 134 & 236 &&       &       &       & & 0.197 & 0.120 & 0.321 & &      &      &      \\
        & 136 & 238 && 0.205 & 0.145 & 0.725 & & 0.206 & 0.144 & 0.424 & &      &      &      \\
        & 138 & 240 &&       &       &       & & 0.227 & 0.134 & 0.138 & &      &      &      \\
        & 140 & 242 &&       &       &       & & 0.251 & 0.113 & 0.091 & &      &      &      \\
        &     &     &&       &       &       & &       &       &       & &      &      &      \\
        & 182 & 284 && 0.014 & 0.065 & 0.115 & &       &       &       & &      &      &      \\
        & 184 & 286 && 0.007 & 0.085 & 0.334 & &       &       &       & &      &      &      \\
        & 186 & 288 &&-0.004 & 0.085 & 0.257 & &       &       &       & &      &      &      \\
        & 188 & 290 &&-0.033 & 0.082 & 0.214 & &       &       &       & &      &      &      \\
        & 190 & 292 && 0.176 & 0.122 & 0.357 & &       &       &       & &      &      &      \\
        & 192 & 294 && 0.172 & 0.139 & 0.776 & & 0.153 & 0.110 & 0.173 & &      &      &      \\
        & 194 & 296 && 0.181 & 0.145 & 0.717 & & 0.174 & 0.136 & 0.512 & &      &      &      \\
        & 196 & 298 && 0.194 & 0.152 & 0.602 & & 0.192 & 0.148 & 0.646 & &      &      &      \\
        & 198 & 300 && 0.206 & 0.156 & 0.550 & & 0.206 & 0.156 & 0.619 & &      &      &      \\
        & 200 & 302 && 0.214 & 0.159 & 0.445 & & 0.219 & 0.157 & 0.454 & &      &      &      \\
        & 202 & 304 &&       &       &       & & 0.235 & 0.138 & 0.230 & &      &      &      \\
        & 204 & 306 &&       &       &       & & 0.248 & 0.105 & 0.095 & &      &      &      \\
        &     &     &&       &       &       & &       &       &       & &      &      &      \\
104(Rf) & 138 & 242 &&       &       &       & & 0.241 & 0.094 & 0.090 & &      &      &      \\
        & 140 & 244 &&       &       &       & & 0.254 & 0.102 & 0.117 & &      &      &      \\
        & 142 & 246 &&       &       &       & & 0.264 & 0.094 & 0.032 & &      &      &      \\
        &     &     &&       &       &       & &       &       &       & &      &      &      \\
        & 184 & 288 && 0.002 & 0.090 & 0.407 & &       &       &       & &      &      &      \\
        & 186 & 290 &&-0.025 & 0.101 & 0.513 & &       &       &       & &      &      &      \\
        & 188 & 292 &&-0.039 & 0.100 & 0.536 & &       &       &       & &      &      &      \\
        & 190 & 294 && 0.195 & 0.105 & 0.454 & &       &       &       & &      &      &      \\
        & 192 & 296 && 0.201 & 0.116 & 0.366 & &       &       &       & &      &      &      \\
        & 194 & 298 &&       &       &       & & 0.179 & 0.115 & 0.340 & &      &      &      \\
        & 196 & 300 && 0.230 & 0.126 & 0.226 & & 0.218 & 0.134 & 0.402 & &      &      &      \\
        & 200 & 304 &&       &       &       & & 0.232 & 0.131 & 0.276 & &      &      &      \\
        & 202 & 306 &&       &       &       & & 0.246 & 0.111 & 0.145 & &      &      &      \\
        & 204 & 308 &&       &       &       & & 0.256 & 0.086 & 0.067 & &      &      &      \\
        &     &     &&       &       &       & &       &       &       & &      &      &      \\
106(Sg) & 142 & 248 &&       &       &       & & 0.264 & 0.098 & 0.104 & &      &      &      \\
        & 144 & 250 && 0.266 & 0.052 & 0.271 & & 0.272 & 0.060 & 0.013 & &      &      &      \\
        &     &     &&       &       &       & &       &       &       & &      &      &      \\
        & 182 & 288 &&-0.003 & 0.055 & 0.076 & &       &       &       & &      &      &      \\
        & 184 & 290 &&-0.003 & 0.084 & 0.320 & &       &       &       & &      &      &      \\
        & 186 & 292 &&-0.032 & 0.102 & 0.530 & &       &       &       & &      &      &      \\
        & 188 & 294 &&-0.046 & 0.103 & 0.660 & &-0.033 & 0.060 & 0.024 & &      &      &      \\
        & 194 & 300 &&       &       &       & & 0.182 & 0.097 & 0.257 & &      &      &      \\
        & 196 & 302 &&       &       &       & & 0.211 & 0.109 & 0.282 & &      &      &      \\
        & 198 & 304 &&       &       &       & & 0.230 & 0.111 & 0.179 & &      &      &      \\
        & 200 & 306 &&       &       &       & & 0.246 & 0.098 & 0.073 & &      &      &      \\
        &     &     &&       &       &       & &       &       &       & &      &      &      \\
\end{longtable*}

\bibliography{references13}

\end{document}